\documentclass[fleqn,usenatbib]{mnras}

\usepackage{newtxtext,newtxmath}

\usepackage[T1]{fontenc}

\DeclareRobustCommand{\VAN}[3]{#2}
\let\VANthebibliography\thebibliography
\def\thebibliography{\DeclareRobustCommand{\VAN}[3]{##3}\VANthebibliography}

\usepackage{graphicx}	
\usepackage{amsmath}	
\usepackage{multicol}        
\usepackage{bm}		
\usepackage{pdflscape}	
\usepackage{longtable}  
\usepackage{caption} 
\usepackage{subfigure} 
\usepackage[flushleft]{threeparttable}
\usepackage{multirow}   
\usepackage{xfrac}   
\usepackage{xcolor} 
\usepackage{todonotes} 
\usepackage{outlines} 
\usepackage{ulem} 

\newcommand{\dm}{~pc~cm$^{-3}$} 

\def\sigproc{\mbox{\texttt{SIGPROC}}}
\def\psrchive{\mbox{\textsc{psrchive}}}

\def\mosaic{\mbox{\textsc{mosaic}}}

\def\seekat{\mbox{S\textsc{ee}KAT}}
\def\rratsolve{\mbox{\textsc{rratsolve}}}
\def\mtcutils{\mbox{\textsc{mtcutils}}}
\def\spyden{\mbox{\textsc{spyden}}}
\def\tempo{\mbox{\textsc{tempo}2}}
\def\scatfit{\mbox{\textsc{scatfit}}}
\def\dmphase{\mbox{\textsc{dm\_phase}}}
\def\iqrm{\mbox{\textsc{iqrm\_apollo}}}
\def\salsa{\mbox{\textsc{psrsalsa}}}

\def\dspsr{\mbox{\texttt{DSPSR}}}

\def\fourteen{\mbox{PSR J1911$-$2020}}
\def\fifteen{\mbox{PSR J2237+2828}}
\def\sixteen{\mbox{PSR J1525$-$2322}}
\def\seventeen{\mbox{PSR J0402$-$6542}}

\def\twenty{\mbox{PSR J1930$-$1856}}
\def\twentyone{\mbox{PSR J0219$-$06}}
\def\twentythree{\mbox{PSR J1319$-$4536}}

\def\twentyeight{\mbox{PSR J1540$-$5821}}

\def\thirtyone{{\mbox{PSR J0917$-$4245}}}

\def\thirtyfour{\mbox{PSR J1108$-$5946}}

\def\thirtynine{\mbox{PSR J1533$-$5609}}

\def\fortythree{\mbox{PSR J1649$-$4230}}
\def\fortyfour{\mbox{PSR J2218+2902}}
\def\fortyfive{\mbox{PSR J1531$-$5557}}

\title[26 new MeerTRAP transients]{Discovery of 26 new Galactic radio transients by MeerTRAP}

\author[J. D. Turner et al.]{J. D. Turner,$^{1}$\thanks{E-mail: james.turner-13@postgrad.manchester.ac.uk}
B. W. Stappers,$^{1}$
J. Tian,$^{1}$
M.\ C.\ Bezuidenhout,$^{2,3}$
M. Caleb,$^{4}$
L. N. Driessen,$^{4}$
F.~Jankowski,$^{5}$\newauthor
I. Pastor-Marazuela,$^{1}$
K. M. Rajwade,$^{6}$
M. Surnis,$^{7}$
M. Kramer, $^{8,1}$ 
E. D. Barr,$^{8}$
M. Berezina,$^{8,9}$\\
$^{1}$ Jodrell Bank Centre for Astrophysics, Department of Physics and Astronomy, The University of Manchester, Manchester M13 9PL, UK\\
$^{2}$Department of Mathematical Sciences, University of South Africa, Cnr Christiaan de Wet Rd and Pioneer Avenue, Florida Park, 1709, Roodepoort, South Africa\\
$^{3}$Centre for Space Research, Potchefstroom Campus, North-West University, Potchefstroom 2520, South Africa\\
$^{4}$ Sydney Institute for Astronomy, School of Physics, The University of Sydney, New South Wales 2006, Australia\\
$^{5}$ LPC2E, OSUC, Univ Orleans, CNRS, CNES, Observatoire de Paris, F-45071 Orleans, France\\
$^{6}$ Astrophysics, The University of Oxford, Denys Wilkinson Building, Keble Road, Oxford OX1 3RH, UK\\
$^{7}$ Department of Physics, IISER Bhopal, Bhauri Bypass Road, Bhopal, 462066, India\\
$^{8}$ Max-Planck-Institut f\"{u}r Radioastronomie, Auf dem H\"{u}gel 69, D-53121 Bonn, Germany\\
$^{9}$ Landessternwarte, Universit\"{a}t Heidelberg, K\"{o}nigstuhl 12, D-69117 Heidelberg, Germany 
}

\date{Accepted XXX. Received YYY; in original form ZZZ}

\pubyear{2025}

\begin{document}
\label{firstpage}
\pagerange{\pageref{firstpage}--\pageref{lastpage}}
\maketitle

\begin{abstract}
Radio searches for single pulses provide the opportunity to discover one-off events, fast transients and some pulsars that might otherwise be missed by conventional periodicity searches.
The MeerTRAP real-time search pipeline operates commensally to observations with the MeerKAT telescope. Here, we report on 26 new Galactic radio transients, mostly rotating radio transients (RRATs) and also the detection of one RRAT and two pulsars that were independently discovered by other surveys. The dispersion measures of two of the new sources marginally exceed the Galactic contribution depending on the electron density model used.
Using a simple method of fitting a Gaussian function to individual pulses, and obtaining positions of arcsecond accuracy from image-based localisations using channelised voltage data from our transient buffer, we have derived timing solutions spanning multiple years for five sources. The timing parameters imply ages of several Myr and low surface magnetic field strengths which is characteristic of RRATs. We were able to measure spin periods for eight more transients, including one source which appears to rotate every 17.5\,seconds. A majority of the sources have only been seen in one observation, sometimes despite multiple return visits to the field. Some sources exhibit complex emission features like component switching and periodic microstructure.
\end{abstract}




\section{Introduction}\label{intro}
Radio-emitting neutron stars (NSs) are known as pulsars if they appear to flash or pulse as their magnetic axis sweeps across the line of sight to an observer. Most pulsars have been discovered using Fourier-domain periodicity searches. However, these searches disfavour the discovery of intermittent sources like Rotating RAdio Transients \citep[RRATs;\footnote{RRATs are generally treated as a separate class from nulling pulsars: RRATs tend to be more transient and are characterised by a dramatic variability in brightness, while nullers tend to spend time between a bright and non-visible state.}][]{Keane2011b}, as they do not have significant spectral power. Currently, 155 out of the 211 RRATs in the ATNF Pulsar catalogue\footnote{\href{https://www.atnf.csiro.au/research/pulsar/psrcat/}{https://www.atnf.csiro.au/research/pulsar/psrcat/} v2.2.0} \citep{Manchester2005} have a period measurement, and 44 of those also have a period derivative. After their discovery \citep{McLaughlin2006}, efforts to reprocess survey data to find single dispersed pulses \citep[e.g.][]{Keane2010, Michilli2018} led to the discovery of many new RRATs. As a result, it has become the standard practice to include a single pulse search pipeline parallel to pulsar searches. Multibeam receivers provide a larger field-of-view (FoV) which boosts the chances for discoveries during those searches \citep{Burke-Spolaor2010, Burke-Spolaor2011, Keane2018, Karako-Argaman2015, Deneva2016, Patel2018, Tyul'bashev2018, Zhou2023}.
Interferometric real-time searches can potentially access a larger FoV, offer more sensitivity and provide instantaneous localisations, e.g. the Canadian Hydrogen Intensity Mapping Experiment \citep[CHIME;][]{CHIME2018} and the Commensal Realtime ASKAP Fast Transient COherent searches \citep[CRACO;][]{Wang2024b}. The continued discovery and timing of RRATs by these surveys provides an improved opportunity to constrain their emission states \citep[e.g.][]{Zhou2023}, constrain the Galactic NS birth rate \citep{Keane2008} and inform how future transient searching surveys should operate. \\
Aside from RRATs, these surveys are sensitive to very long period pulsars that lie beyond the pulsar death line. 
The 76-second pulsar discovered by \citet{Caleb2022} is so far from canonical pulsars in rotational phase-space that it could belong to the hypothesised ultra-long period magnetar class of NS \citep{Beniamini2023}. The 421-second transient discovered by CHIME \citep{Dong2024} is a member of the long-period transient (LPT) class of very slow rotators, potentially NSs, capable of producing beamed coherent radio emission. At the same time, fast image-domain searches have discovered even slower LPTs, for example the 18-minute GLEAM-X J162759.5$-$523504.3 \citep{Hurley-Walker2022}, the 21-minute GPM J1839$-$10 \citep{Hurley-Walker2023} and the 54-minute ASKAP J193505.1+214841.0 \citep{Caleb2024}. The slowest pulsars and LPTs pose challenges to our understanding of radio emission from a slowly rotating dipole and provide clues as to how NSs are born and evolve. 

The MeerKAT telescope \citep{Jonas2018, Camilo2018a} in South Africa is the most sensitive telescope in the Southern hemisphere. MeerTRAP \citep[more TRAnsients and Pulsars with MeerKAT, for project details see][]{Sanidas2018} has since September 2020 been searching for fast radio transients in real-time by piggy-backing on a significant portion of MeerKAT observations. This has resulted in the discovery of 14 Galactic transients \citep{Caleb2022, Bezuidenhout2022, Surnis2023}, excluding those presented here, and a few dozen fast radio bursts (FRBs). FRBs are short duration, extra-galactic radio pulses of unknown physical origin \cite[see][for a review]{Cordes2019, Petroff2022}. Due to the large data volumes of $\approx$\,21\,GB/s handled by the MeerTRAP transient search backend, data must be irreversibly reduced in frequency or time resolution to be searched. The role of the transient buffer, which has been in operation since June 2022, is to retain a full, or near-full resolution copy of the data to be deposited on demand for offline analysis. This has proven essential in rapidly localising and following-up MeerTRAP discoveries, including FRBs \citep{Rajwade2024}. Section \ref{obs} briefly describes the key components of the real-time search pipeline and the transient buffer, and details the offline processing of the data. In Section \ref{results} we present the newly discovered sources and their properties. In Section \ref{disc} we interpret the discoveries and the method for calculating pulse arrival times and in Section \ref{conc} we summarise our findings and conclude.

New sources are given the MeerTRAP name `\textit{MTPXXXX}', which is based on the chronology of discoveries. A standard J2000 name is also given for sources where we have measured a spin period. We report all detections of these sources up to 2024 July 25. Throughout the article, bracketed values that follow results refer to the 1-sigma uncertainty on the final digit. All timestamps provided are topocentric and are dispersion-corrected to the highest frequency of the observing band, unless stated otherwise.

\section{Observations and Data Reduction}\label{obs}
\subsection{MeerKAT data processing}
\subsubsection*{Real-time pulse searches}
MeerTRAP piggybacks many MeerKAT observations, of which a significant fraction have been during time dedicated to Large Survey Projects (LSPs). LSPs tend to focus on a particular class of sources or region of the sky. Consequently, MeerTRAP does not uniformly sample the Southern sky, as some fields are frequented often and others rarely or not at all \citep[see][]{Bezuidenhout2022}. MeerTRAP's frequency coverage depends on the MeerKAT receiver in use. These data here were captured by the L--band receiver \citep[856-1712 MHz;][]{Lehmensiek2012} or the Ultra High Frequency (UHF)--band receiver \citep[544-1088 MHz;][]{Lehmensiek2014a}. Two observing modes are operated simultaneously; in the incoherent mode the signals from the receivers of up to 64 telescopes are combined to form the incoherent beam (IB), and in the coherent mode the signals are coherently summed with respect to the phase centre of the array. 
Up to 768 coherent beams (CBs) are instantaneously formed on the sky as computed by the Filterbanking Beamformer User Supplied Equipment \citep[FBFUSE;][]{Barr2018, Chen2021} a dedicated high-performance cluster installed on-site. 
Usually, the innermost 40 telescopes with a 1~km baseline are used instead of the full array to balance sky coverage and system gain. The central CB is therefore a factor of around 40/$\sqrt{64}\approx$5 times more sensitive than the IB \citep[e.g.,][]{Rajwade2022}. To ensure the sensitivity of the coherent FoV exceeds that of the IB, the beam overlap is chosen to be at 25 per cent of the peak sensitivity level. \\
Channelised data from FBFUSE are transferred to the Transient User Supplied Equipment (TUSE) cluster then downsampled and processed by the real-time search pipeline. The specifications of the data are provided in \autoref{tab:specs}. The IB area\footnote{In some previous MeerTRAP papers the FoV was given as 1.27 deg$^{2}$ which was calculated assuming a central frequency of about 1120 MHz.} is computed by \cite{Jankowski2023} at the half-power width, and the CB widths refer to the width at the overlap level. The typical smallest half-power width of a CB is $\sim$80\,arcsec at 816\,MHz and $\sim$50\,arcsec at 1284\,MHz. The CB shape strongly depends on the available baselines and the source elevation, so \autoref{tab:specs} provides the central 90 per cent range of widths. \\
The MeerTRAP search pipeline was introduced in \citet{Malenta2020} and later described in detail in \citet{Rajwade2021, Rajwade2022} and \citet{Bezuidenhout2022}. \autoref{tab:specs} lists the search parameters and excision thresholds applied to the candidates.
Radio frequency interference (RFI) is mitigated with the zero-DM technique \citep{Eatough2009, Men2019} and with IQRM\footnote{\href{https://github.com/v-morello/iqrm}{https://github.com/v-morello/iqrm}} \citep{Morello2022}. Broadband RFI is further mitigated by rejecting candidates below a dispersion measure (DM) of 20\dm. The maximum DM trials were previously 1480\dm{} (UHF) and 5000\dm{} (L--band), but were set to 2100\dm{} and 3600\dm{} respectively in October 2023. Candidates above the signal-to-noise ratio (S/N) threshold are then sifted and classified \citep[see][for further information]{Rajwade2022} using \textsc{frbid}\footnote{\href{https://github.com/Zafiirah13/FRBID}{https://github.com/Zafiirah13/FRBID} by Zafiirah Hosenie} \citep{Hosenie2021}. Initially, the S/N threshold was varied between 7-8 depending on RFI prevalence, but has been set at 8 since October 2021. Candidates are written as a filterbank in \sigproc{}\footnote{\href{https://sigproc.sourceforge.net/}{https://sigproc.sourceforge.net/}} format and padded with 0.5\,s of data either side of the pulse. For any candidate of S/N $<$ 8, the data are checked using the MeerTRAP candidate inspection tool \mtcutils\footnote{\href{https://bitbucket.org/vmorello/mtcutils}{https://bitbucket.org/vmorello/mtcutils} by Vincent Morello} to see if the DM can be optimised to extract enough S/N to claim a detection. During observations, a JSON file is written to TUSE that contains information about the user's observation, such as the target, the telescopes and the beam positions. These are created every 10\,minutes or every new target, whichever is sooner.

\begin{table}
\begin{center}
\caption{Specification summary of MeerTRAP data and the search pipeline. Beam widths are given at the central frequency of the observing band. These values are true for the majority of searches; details of how the thresholds have been varied are given in the text. The CB widths are the beam widths at the 0.25 level overlap.}\label{tab:specs}
\begin{tabular}[\textwidth]{lcc}
\hline
\multicolumn{3}{c}{Data specifications} \\
\hline
& UHF & L--band \\
\hline
Central frequency, MHz \dotfill     & 816                       & 1284     \\
Bandwidth, MHz \dotfill             & 544                       & 856      \\
Sampling time, $\upmu$s \dotfill    & 482                       & 306      \\
Number of channels \dotfill         & \multicolumn{2}{c}{1024}             \\
Number of polarisations \dotfill    & \multicolumn{2}{c}{2}                \\
CB overlap level \dotfill           & \multicolumn{2}{c}{0.25}             \\
90\% CB width range, arcsec \dotfill & 76-228 & 46-100 \\
Half-power IB area, deg$^{2}$ \dotfill        & 2.4  & 1.0 \\
\hline
\multicolumn{3}{c}{Pipeline thresholds} \\
\hline
Min. DM, \dm \dotfill               & \multicolumn{2}{c}{20} \\
Max. DM, \dm \dotfill               & 1480-2100 & 3600-5000 \\
Min. boxcar width, s \dotfill       & 0.000482 & 0.000306               \\
Max. boxcar width, s \dotfill       & 0.25 & 0.67               \\
S/N cutoff \dotfill                 & \multicolumn{2}{c}{7-8}                \\
IQRM maximum lag, samples \dotfill    & \multicolumn{2}{c}{100}              \\
IQRM threshold, $\upsigma$ \dotfill            & \multicolumn{2}{c}{3.0}              \\
\hline
\end{tabular}
\end{center}
\end{table}

\subsubsection*{Localisation}\label{image}
We seek to rapidly localise our new discoveries so that when the field is reobserved, we can strategically form a coherent beam at the best position to achieve a better sensitivity and detect more pulses. We refer to this setup as a targeted observation. A detection in a single CB indicates the pulse originated within that region, provided the beam does not lie on the edge of the tiling. If a pulse is detected in three or more CBs, we can use spatial S/N variation to constrain the position, usually to within a few arcseconds. To do this, we use the multibeam localisation tool \seekat{}\footnote{\href{https://github.com/BezuidenhoutMC/SeeKAT}{https://github.com/BezuidenhoutMC/SeeKAT} by Mechiel Bezuidenhout} \citep{Bezuidenhout2023}. We provide \seekat{} with a coherent point spread function (PSF) simulated by the multibeam simulation package \mosaic\footnote{\href{https://github.com/wchenastro/Mosaic}{https://github.com/wchenastro/Mosaic}{ by Weiwei Chen}} described in \citet{Chen2021}.

We can also use voltage data captured by the transient buffer (TB) \citep{Malenta2020, Jankowski2022}, in operation since June 2022. Several of the new source triggered the TB, allowing image-based localisations. A detailed description of the TB and the formation of measurement sets (MSs) from the voltage data using \textsc{casa} \citep{McMullin2007} is given in \citet{Rajwade2024}. The system does not distinguish FRBs from other transients, thus we follow the same data reduction process described in \citet{Rajwade2024} and \citet{Tian2024}. Usually, data from $\sim$\,60 telescopes is saved, thus providing more sensitivity than the CB or IB detections in the time domain.
Subbanded MSs are cleaned with \texttt{WSClean} \citep{Offringa2014} then frequency-averaged `on-' and `off-pulse' images, separated by a few dozen milliseconds, are compared. Once identified, the best position and uncertainty of the emission is found using \textsc{python} Blob Detector and Source Finder, \texttt{pyBDSF}\footnote{\href{https://www.astron.nl/citt/pybdsf/}{https://www.astron.nl/citt/pybdsf/}}. The position is corrected by performing absolute astrometry using catalogued sources in the field with position of sub-arcsecond accuracy. These measurements follow the same methodology as is described in \citet{Driessen2022, Driessen2024}. All matched sources used to solve for a transformation matrix to shift and rotate the MeerKAT sources were Rapid ASKAP Continuum Survey (RACS) source positions\footnote{The code for performing the astrometric correction can be found on GitHub: \url{https://github.com/AstroLaura/MeerKAT_Source_Matching}}. The total uncertainty is typically about 1 arcsec, and has three contributions: the uncertainty from the fit in {\sc pybdsf}, the absolute systematic astrometric uncertainty from the RACS positions \citep{Hale2021, Duchesne2024} and the median offset of the positions after the astrometric correction.

\subsection{Determining the dispersion measure}
For each source we report either the DM that optimises the S/N of the pulse or the DM that optimises for frequency-averaged profile structure. For all sources, we measure the DM of the brightest pulse and assume it is the correct DM for all other pulses. 
Structure-optimised dispersion correction is preferred over optimising S/N for pulses with complex profiles and frequency-time substructure \citep{Hessels2019}. As has been the case for many one-off and repeating FRBs \citep[e.g.][]{Sand2024}, this method can help reveal underlying substructure in our source, so we used this method for as many sources as possible. To measure the structure-optimised DM we use \dmphase{}\footnote{\href{https://github.com/InesPM/DM\_phase}{https://github.com/InesPM/DM\_phase} by Andrew Seymour, Daniele Michilli and Ziggy Pleunis, modified by In\'{e}s Pastor-Marazuela} \citep{Seymour2019}. \dmphase{} does not clean the data of RFI, so we do this before hand using \iqrm\footnote{\href{https://gitlab.com/kmrajwade/iqrm\_apollo}{https://gitlab.com/kmrajwade/iqrm\_apollo}} (\citealt{Morello2023}; also see \citealt{Morello2022} for more details). If \dmphase{} is not suitable due to low S/N or only single component pulses are seen, we use \scatfit\footnote{\href{https://github.com/fjankowsk/scatfit}{https://github.com/fjankowsk/scatfit} by Fabian Jankowski} \citep{Jankowski2022b}, which models subbanded pulse profiles as an exponentially-modified Gaussian pulse that is broadened due to the dispersive and scattering action of the interstellar medium and smearing due to intra-channel dispersion and the time resolution. \scatfit{} uses a Bayesian analysis by Markov Chain Monte Carlo sampling to provide a refined DM and uncertainty. More details can be found in \citet{Jankowski2023}. Using the brightest single-component pulse for each source, we fit a scattered pulse model. Due to the low S/N of many of these pulses, we trial the fit for data that are downsampled in time by a factor of 2 and 4, and in frequency to 4, 8 and 16 subbands, and choose the DM with the best Bayesian information criterion (BIC) and reduced-$\chi^{2}$ closest to unity. Subbands with a S/N value below 3.0 are not included in the fit. If a scattering timescale cannot be measured across all subbands, we instead fit an unscattered model. We did this instead of allowing the scattering timescale to fall to zero to avoid it  correlating with the smearing timescale.\footnote{\scatfit{} contains models that consider DM smearing between channels, though we can ignore this as the smearing timescale is less than the intrinsic pulse width at the bottom of the band for all the pulses we inspected.} We ultimately did not find a scattered profile to be better than the unscattered model for any of the pulses. The results from the fits are provided in \autoref{tab:fitstats}. \\
\subsection{Properties and timing of single pulses}\label{timing}
For some sources we detected a cluster of pulses in a single day from which we could use their separation in time to measure the spin period, $P$. Of these sources, there were some with detections spanning many days from which we could measure the spin-down rate, $\dot{P}$ and constrain the position using pulsar timing. In order to do this we calculated the time of arrival (TOA) of the signal for each detection. We must do this to each pulse individually, as only a portion of the rotation phase is contained in the data stored for each detection. The TOAs are calculated using \texttt{make\_toas}\footnote{Available at: \href{https://bitbucket.org/meertrap-ipm/mtcutils/src/jturner-timing/}{https://bitbucket.org/meertrap-ipm/mtcutils/src/jturner-timing/}}, an implementation of \mtcutils. For each pulse (the brightest if there are multibeam detections), \texttt{make\_toas} finds the width and phase alignment that maximises the S/N by using \spyden\footnote{\href{https://bitbucket.org/vmorello/spyden}{https://bitbucket.org/vmorello/spyden} by Vincent Morello} to convolve single-Gaussian templates of various widths with the time series. This is done on the full time resolution data dedispersed at the best DM to the top of the frequency band. The dispersion delay-corrected timestamp of the first bin is added to the time of the peak bin of the best Gaussian to get the TOA. The uncertainty on the TOA is calculated to be the width of the pulse divided by the S/N. With this method, the uncertainty due to pulse jitter is worse than conventional pulsar timing, because we fit to the feature in intensity that best resembles a Gaussian. Additionally, we may inadvertently fit to substructure or subpulses that would not otherwise be a dominant pulse component. However, the width of any subpulse is always smaller than the full pulse width, so our TOA value will not be greatly affected by this. \\
To make an initial period estimate of some pulses that are closely spaced in time, we use \rratsolve\footnote{\href{https://github.com/v-morello/rratsolve}{https://github.com/v-morello/rratsolve} by Vincent Morello} to find the lowest common period connecting the TOAs. The period returned is the largest value that connects the TOAs, such that P/N where N is an integer is also a valid solution. If we have pulses over multiple days, we use the pulsar timing program \tempo\footnote{\href{https://bitbucket.org/psrsoft/tempo2/src/master/}{https://bitbucket.org/psrsoft/tempo2/src/master/}} \citep{Hobbs2006} to find a phase-connected timing solution. Our initial timing model is an ephemeris containing the period from \rratsolve{} and a position obtained either from \seekat{} or from imaging. Then using \tempo{}, we find the barycentric rotational parameters and, if possible, a constrained position that best predicts the TOAs. The rapid and accurate localisations improve our ability to obtain a timing solutions even with a low number of TOAs. \\
The timestamps recorded in the JSONs from TUSE are used to calculate the total observation time and analyse the detection statistics of the new sources. Specifically, we use the total time spent observing any targets during which a new source was detected; we henceforth refer to these as primary targets (e.g. the primary target for observations of the new source MTP0014 is the gain calibrator J1911$-$2006). For the case where detections are, or would be, serendipitous we define a non-targeted observing time, $T_{\text{n}}$. For targeted observations (see Section \ref{image} `Localisation'), we separately calculate a targeted observing time, $T_{\text{t}}$. We make calculations of the average detection rate for several sources using $T_{\text{n}}$ or, where possible, $T_{\text{t}}$. Due to how the MeerTRAP system operates, the real-time search does not include approximately 25\,s of data per new pointing. The amount of unsearched data varies, so we provide an uncertainty on $T_{\text{n}}$ and $T_{\text{t}}$ that assumes the 25\,s time loss has an error of 50 per cent. The fractional uncertainty is thus lower where integration times are longer.\\
$T_{\text{n}}$ must be considered a lower limit as there may be observations of other sources in the field of our discoveries where detections were not made. We assume in such cases that if there are no detections, then either the observing time was not significant or the source was not covered by CBs. With this caveat in mind, we make two statements about our calculations. Firstly, the average non-targeted detection rate must be considered an upper limit if $T_{\text{n}}$ is a lower limit. Secondly, the targeted detection rate is much more reliable, given the unchanging beam position and the extra sensitivity. Although the full sensitivity provided by MeerKAT's coherent observing mode allows us to sample fainter pulses, the rates we find are ultimately flux-limited. Detailed analyses of pulse energy distributions are beyond the scope of this paper. The pulse statistics will also be biased by the nature of the primary targets. For example, if a source is in the field of a calibrator, then those observations will tend to sample short wait times between pulses.

\begin{table*}
\centering
\caption{Detection and observation information for all of the reported transients, starting with their names and the date, observing mode and candidate S/N of their discovery pulses. Two detection counts are provided; the non-targeted number, $N_{\text{det,n}}$ is the count during the time $T_{\text{n}}$ spent observing any identified nearby primary target(s) at L--band and UHF combined, and $N_{\text{det,t}}$ is for the time, $T_{\text{t}}$ when the source was targeted. The total number of non-targeted and targeted observations, $N_{\text{obs}}$ is given, and the frequency band at which detections have been made. The typical length of observations is approximately $T$/$N_{\text{obs}}$, but more detail is provided in the text for each source.}
\label{tab:detections}
\begin{tabular}{llclrrrrrrl}
\hline
MTP name & PSR J2000 name & Discov. MJD & Discov. mode & Discov. S/N & $N_{\text{det,n}}$ & $T_{\text{n}}$ (s) & $N_{\text{det,t}}$ & $T_{\text{t}}$ (s) & $N_{\text{obs}}$ & Band \\
\hline
\multicolumn{11}{l}{Published sources} \\
\hline
MTP0015 & J2237+2828 & 59177.713493 & CB & 37.8 & 1 & 271(13) & 2 & 6009(124) & 99 & L \\
MTP0028 & J1540$-$5821 & 59436.852030 & CB & 10.1 & 3 & 572(13) & $-$ & 0 & 1 & L \\
MTP0043 & J1649$-$4230 & 59579.482282 & CB & 10.0 & 2 & 553(13) & $-$ & 0 & 1 & L \\
\hline
\multicolumn{11}{l}{New sources} \\
\hline
MTP0014 & J1911$-$2020 & 59177.610266 & CB & 9.8 & 10 & 1210(47) & 650 & 31834(267) & 471 & L, UHF \\
MTP0016 &  & 59208.082669 & CB & 8.4 & 4 & 4100(50) & $-$ & 0 & 16 & L \\
MTP0017 & J0402$-$6542 & 59214.773203 & CB & 9.4 & 2 & 63846(178) & 133 & 345590(446) & 1446 & UHF \\
MTP0018 & & 59215.933480 & IB & 8.5 & 1 & 410414(790) & $-$ & 0 & 1446 & UHF \\
MTP0020 & J1930$-$1856 & 59324.147363 & CB & 8.5 & 16 & 40648(108) & 235 & 34290(101) & 26 & L, UHF \\
MTP0021 & J0219$-$06 & 59330.512419 & IB, CB & 7.0 & 6 & 22175(106) & $-$ & 0 & 13 & L \\
MTP0022 & & 59344.674851 & CB & 9.2 & 4 & 2860(44) & 0 & 2942(69) & 6 & L \\
MTP0023 & J1319$-$4536 & 59369.833079 & IB, CB & 18.2 & 39 & 42982(272) & 52 & 7284(100) & 799 & L, UHF \\
MTP0024 & & 59372.073761 & IB & 9.9 & 2 & 21254(38) & $-$ & 0 & 6 & L \\
MTP0026 & & 59390.962435 & CB & 7.9 & 1 & 565(28) & $-$ & 0 & 1 & L \\
MTP0029 & & 59446.907329 & CB & 13.9 & 3 & 560(13) & $-$ & 0 & 1 & L \\
MTP0031 & J0917$-$4245 & 59448.261286 & IB, CB & 22.9 & 30 & 9064(67) & $-$ & 0 & 25 & L, UHF \\
MTP0032 & & 59499.647428 & CB & 7.9 & 1 & 569(13) & $-$ & 0 & 1 & L \\
MTP0034 & J1108$-$5946 & 59513.466697 & IB, CB & 8.5 & 135 & 3095(40) & $-$ & 0 & 4 & L \\
MTP0035 & J1308$-$61 & 59513.543102 & IB & 12.3 & 2 & 574(13) & $-$ & 0 & 1 & L \\
MTP0036 & & 59517.055729 & CB & 8.9 & 1 & 201(18) & $-$ & 0 & 1 & L \\
MTP0038 & & 59557.258616 & CB & 16.7 & 1 & 56973(149) & $-$ & 0 & 17 & L \\
MTP0039 & J1533$-$5609 & 59559.185447 & CB & 8.5 & 71 & 54032(137) & 0 & 1291(53) & 33 & L \\
MTP0040 & & 59564.389209 & CB & 9.3 & 1 & 548(13) & 0 & 94(13) & 2 & L \\
MTP0042 & J1641$-$5109 & 59495.789294 & CB & 10.0 & 2 & 545(13) & $-$ & 0 & 1 & L \\
MTP0044 & & 59581.544680 & CB & 15.5 & 9 & 6586(44) & $-$ & 0 & 6 & UHF \\
MTP0045 & & 59568.125229 & CB & 8.8 &28 & 83633(185) & 1 & 3808(71) & 39 & L \\
MTP0046 & & 59594.430739 & CB & 9.0 & 1 & 535(13) & $-$ & 0 & 1 & L \\
MTP0047 & & 59595.470964 & CB & 8.3 & 2 & 181(18) & 6 & 1041(45) & 12 & L \\
MTP0048 & & 59613.248703 & CB & 9.2 & 2 & 557(13) & $-$ & 0 & 1 & L \\
MTP0049 & & 59616.254504 & CB & 10.0 & 1 & 1013(18) & $-$ & 0 & 2 & L \\
\hline
\end{tabular}
\end{table*}

\section{Results}\label{results}
\subsection{Sources published elsewhere}
\begin{figure*}
    \centering
    \subfigure{\includegraphics[width=0.33\textwidth]{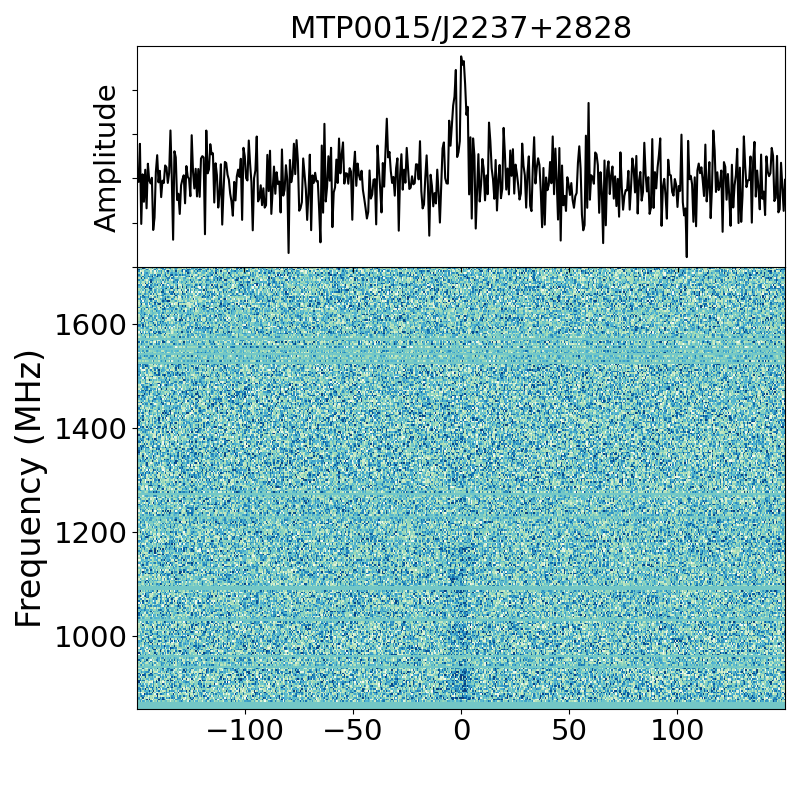}}
    \subfigure{\includegraphics[width=0.33\textwidth]{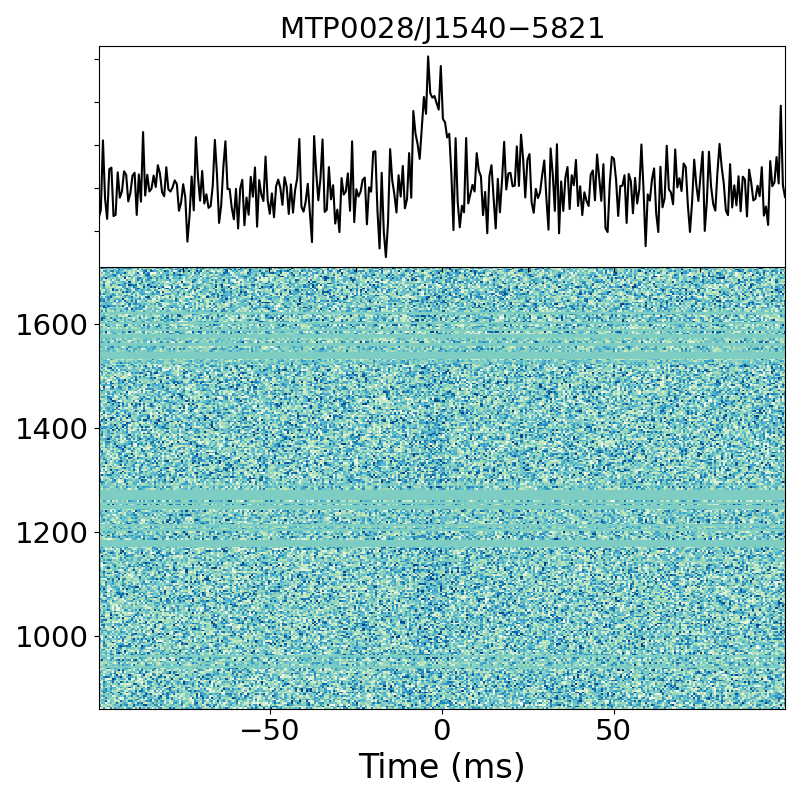}}
    \subfigure{\includegraphics[width=0.33\textwidth]{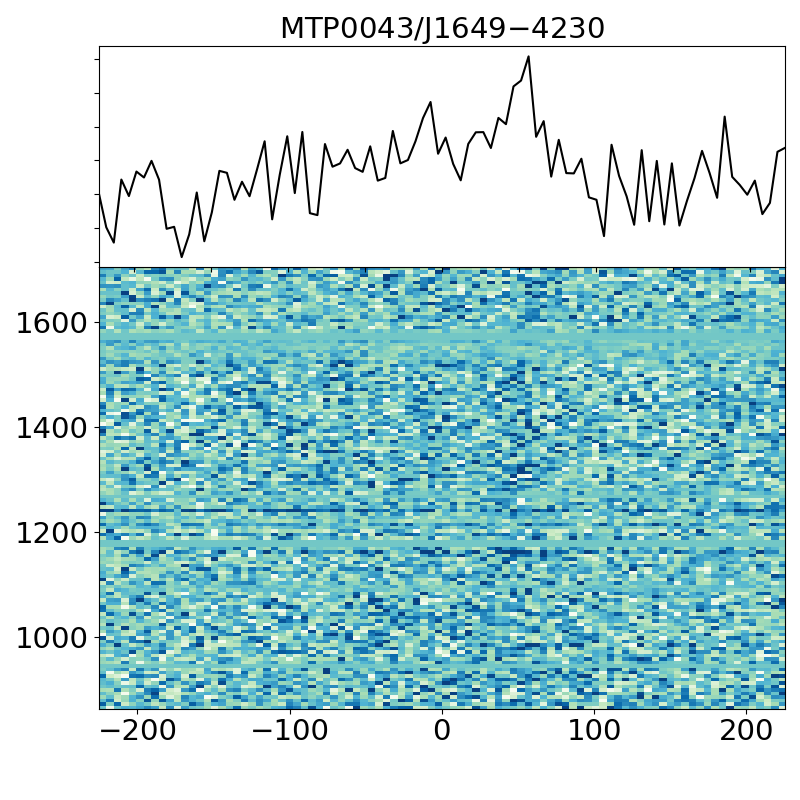}}
    \caption{Dynamic spectra (bottom) of the first pulses seen by MeerTRAP for the RRAT \fifteen{} and the two pulsars \twentyeight{} and \fortythree{}, and their frequency-averaged pulse profiles (top). They have been dedispersed at the best DM which is given in \autoref{tab:MTPinfo}. The blank horizontal lines are masked channels that were affected by RFI.}
    \label{fig:waterfallsA}
\end{figure*}

We provide here information about three sources that were new at the time of detection, but were concurrently seen in other surveys. The properties of the detections and pulses are given in \autoref{tab:detections}. The discovery S/N is that of the first pulse as seen by the real-time MeerTRAP pipeline. The definitions of the targeted and non-targeted detection rates are described in Section \ref{timing}. The position, DM, S/N determined by \spyden{} and spin period values are provided in \autoref{tab:MTPinfo}. DM-derived distances from the \textsc{ne2001} \citep{Cordes2001} and \textsc{ymw16} \citep{Yao2016} Galactic electron density models are obtained using the \texttt{pygedm}\footnote{\href{https://github.com/FRBs/pygedm}{https://github.com/FRBs/pygedm}} tool \citep{Price2021b}. \\
MTP0015/\mbox{PSR J2237+2828} was discovered by MeerTRAP on MJD 59177 in a CB. The pulse profile and the dynamic spectrum are shown in the leftmost plot of \autoref{fig:waterfallsA}. It was also detected by the CHIME telescope on MJD 59193 by \citet{Dong2023}. They found a spin period of 1.077\,s and DM of 38.1\dm. \mbox{PSR J2237+2828} is located only 14\arcmin{} from the L--band gain calibrator\footnote{A list of calibrator sources that are recommended for users of MeerKAT is \href{https://skaafrica.atlassian.net/wiki/spaces/ESDKB/pages/1452310549/Calibration}{provided here} by SARAO.} J2236+2828. Since our first detection, two more pulses have been seen in approximately 98\,minutes of targeted observing time, giving a targeted detection rate of 1.2\,hr$^{-1}$, which is below the rate reported across CHIME observations in \citet{Dong2023}. If we include the discovery pulse, albeit non-targeted, this gives a rate of 1.8\,hr$^{-1}$ which is more closely aligned to the value from CHIME observations. Regardless, both are consistent if a Poissonian pulse rate and uncertainty are assumed. Interestingly, we have not detected a population of fainter pulses when targeting \mbox{PSR J2237+2828}, only very bright pulses of S/N $>$ 50. The source scintillates across the band and has some microstructure within a main peak. The brightness of the pulses and microstructure allowed us to find a structurally-optimised DM using \dmphase{} of 38.1$\pm$0.1\dm, which matches the DM measured by \citet{Dong2023}. \\
MeerTRAP commensally observed with the L--band portion of the Max-Planck-Institut f\"{u}r Radioastronomie (MPIfR)--MeerKAT Galactic Plane Survey \citep[MMGPS--L;][]{Padmanabh2023}. Two of the pulsars discovered up to that point by the survey\footnote{MMGPS discoveries are listed at \href{https://trapum.org/discoveries/}{trapum.org/discoveries}.} have single pulses bright enough to be detected by the MeerTRAP pipeline. These are MTP0028/\twentyeight{} and MTP0043/\fortythree{} and we provide here some information about their properties supplementary to those given by \citet{Padmanabh2023}.
Of all the 78 pulsars discovered by MMGPS--L, \twentyeight{} has the longest period, 3.47472(7)~s and a DM of 427$\pm$3\dm. MeerTRAP detected three pulses in the same CB at much higher DMs between 439--441\dm{}. We checked the data in \mtcutils{}/\texttt{candreport} and visual inspection of a range of DMs showed that MMGPS--L DM value is too low to fully dedisperse the pulse. 
With \scatfit{} we obtained a refined DM value of 440.3$\pm$0.2\dm, which dedisperses the data to produce the dynamic spectrum in \autoref{fig:waterfallsA}, and is consistent to 5-sigma with the MMGPSL--L value. The S/N range of the MeerTRAP pulses is 5-11 when dedispersed at the 427\dm{} and 13-23 at 440.3\dm. We do not have a robust explanation as to why the MMGPS--L DM is so low, though we note there is some RFI still present in the diagnotics plots that may have affected their measurement. We measured TOAs for the three pulses and found they are fully consistent with the separation predicted by the period reported by \citet{Padmanabh2023}. \\ 
\fortythree{} was detected in MMGPS--L at a period of 0.676\,s and a DM of 374$\pm$6\dm. Two faint pulses from \fortythree{}, the first of which is shown in \autoref{fig:waterfallsA} were concurrently detected by MeerTRAP. The TOAs we measured are consistent with their period to 4-sigma. 
They are at least 60\,ms in width and exhibit a tenously visibile precursor component to the main peak. We relay the best DM reported by \citet{Padmanabh2023} of 374$\pm$6\dm{} as our methods did not provide an improved measurement. Due to how TUSE pads the candidate data, a further three spin period lengths were kept. We inspected the DM-time planes at integer periods after both pulses and saw very faint signals where the next pulses should be. Using the MMGPS--L DM we dedispersed individual windows of data 0.2\,s wide around each of the three potential pulses, but did not see anything above a S/N of 7. We therefore claim detections for two pulses only, though it suggests the pulse energy distribution contains some pulses just below our detection threshold.

\subsection{New sources}

\begin{figure*}
    \centering
    \subfigure{\includegraphics[width=0.235\textwidth]{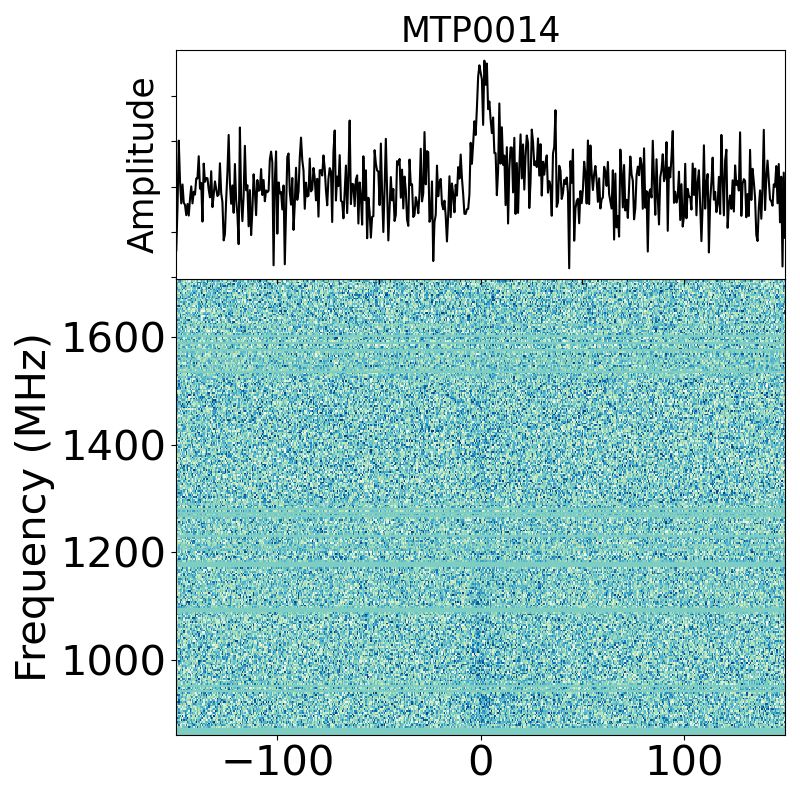}}
    \subfigure{\includegraphics[width=0.235\textwidth]{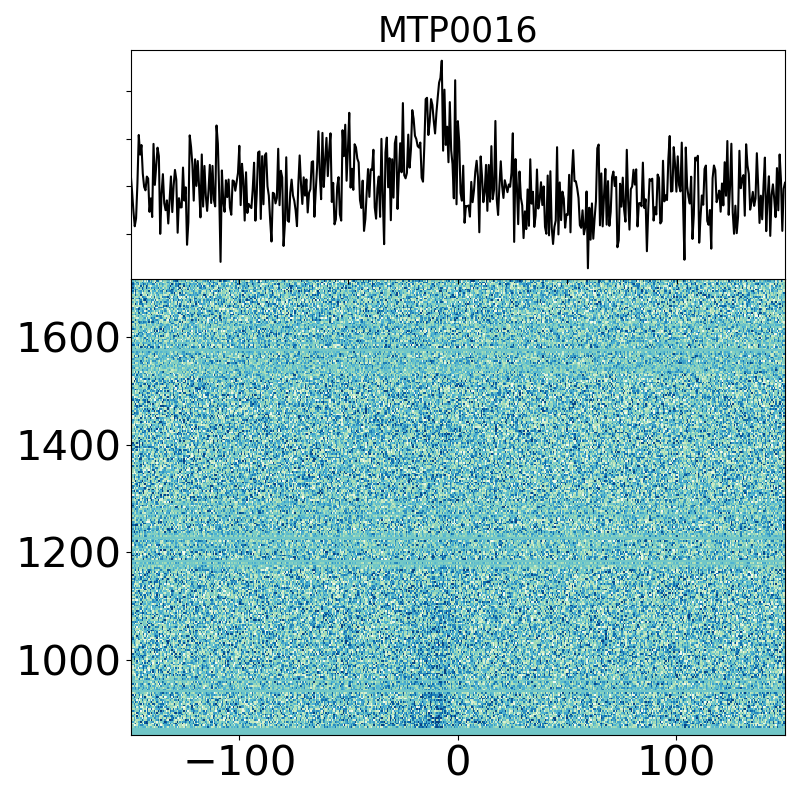}}
    \subfigure{\includegraphics[width=0.235\textwidth]{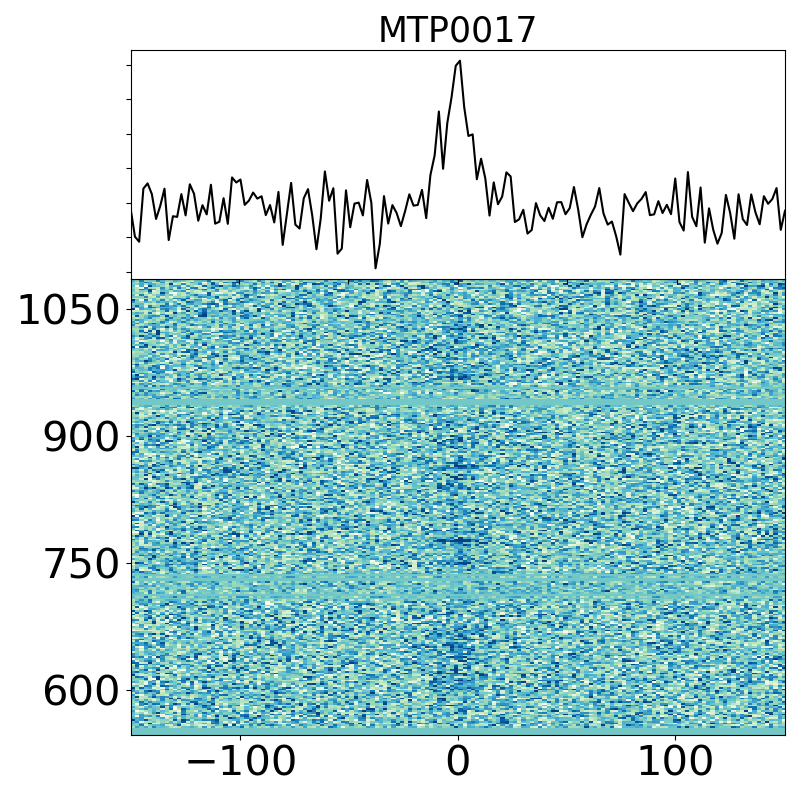}} 
    \subfigure{\includegraphics[width=0.235\textwidth]{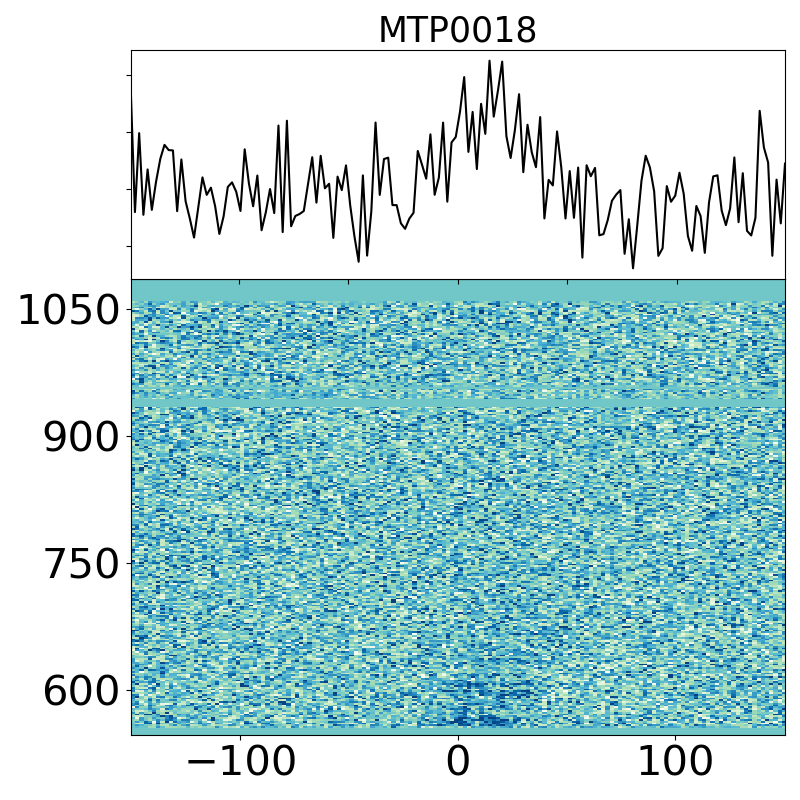}} 
    \subfigure{\includegraphics[width=0.235\textwidth]{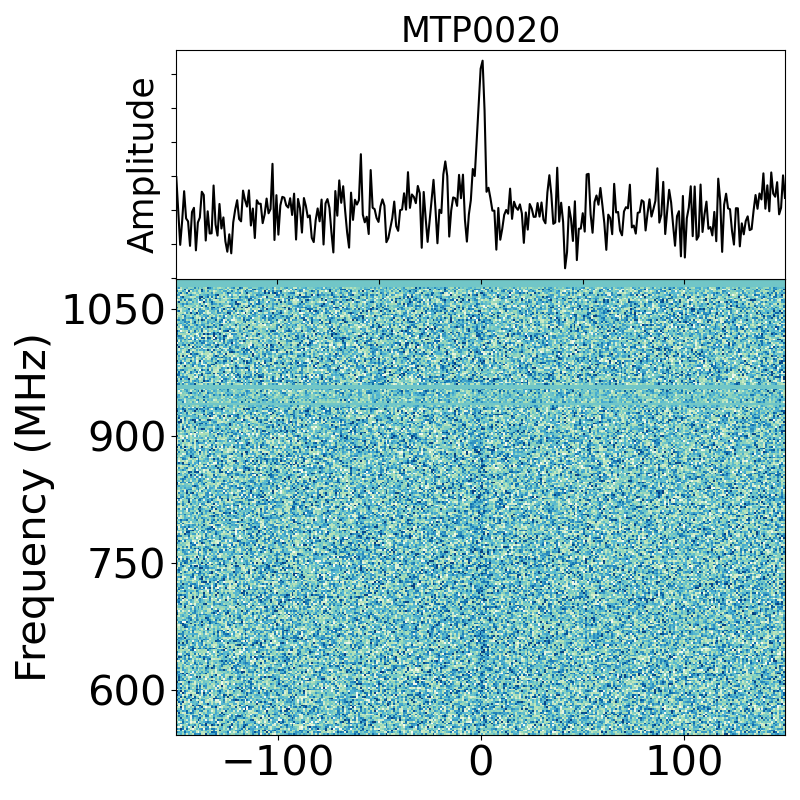}} 
    \subfigure{\includegraphics[width=0.235\textwidth]{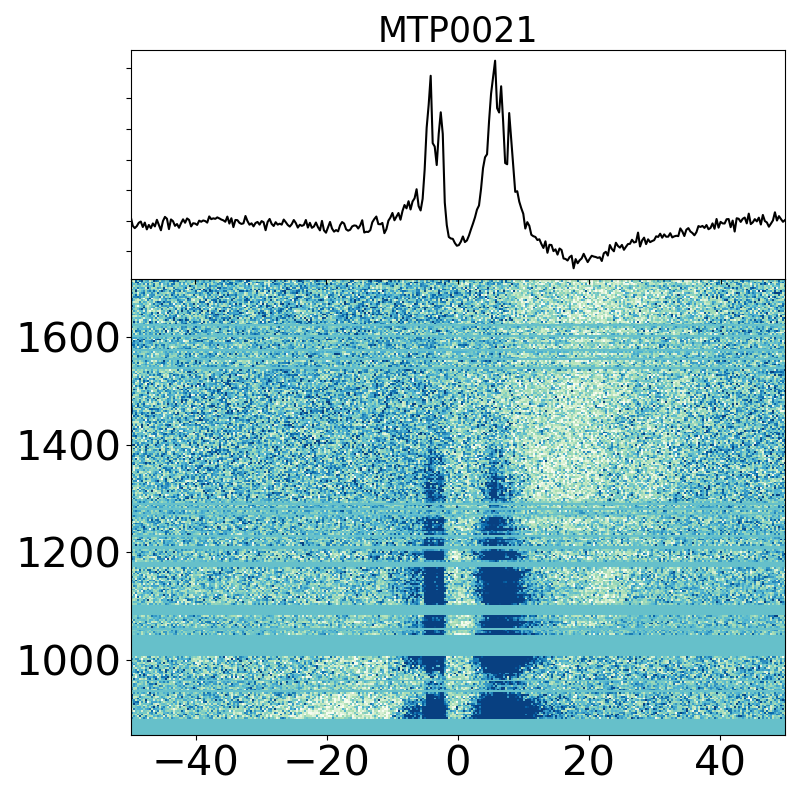}} 
    \subfigure{\includegraphics[width=0.235\textwidth]{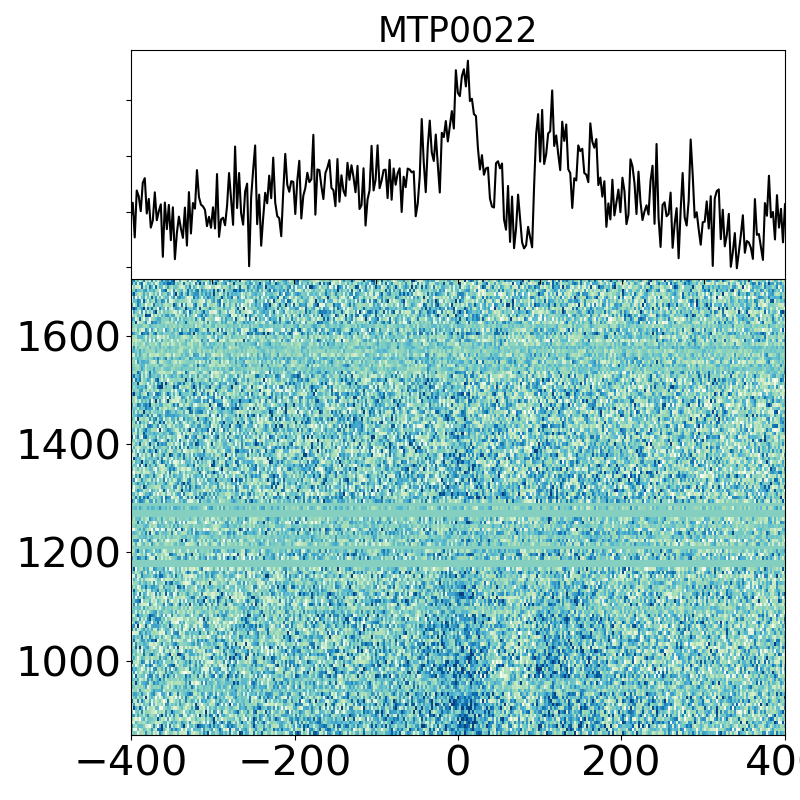}} 
    \subfigure{\includegraphics[width=0.235\textwidth]{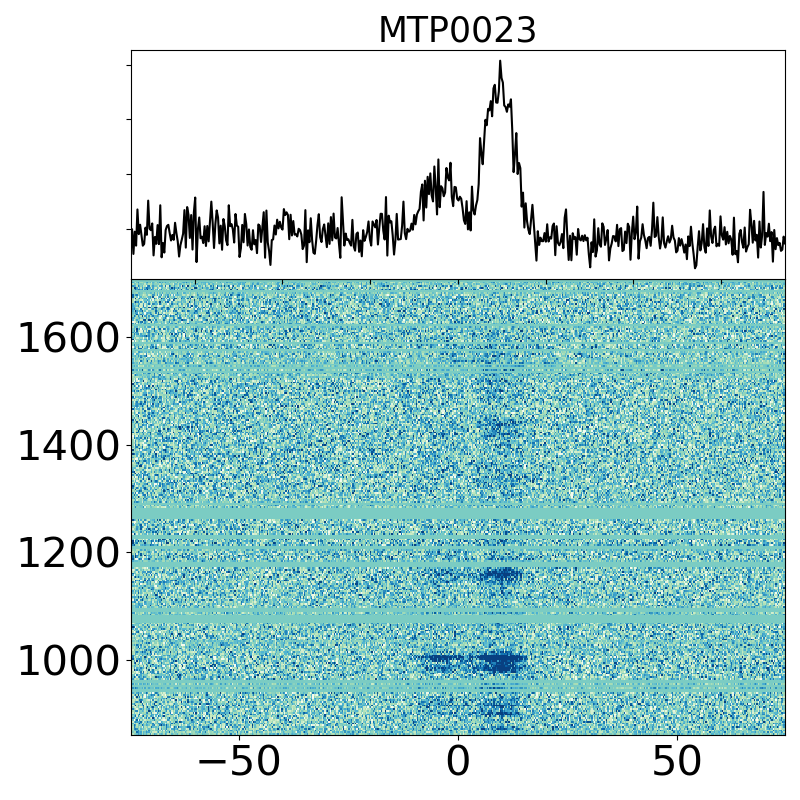}} 
    \subfigure{\includegraphics[width=0.235\textwidth]{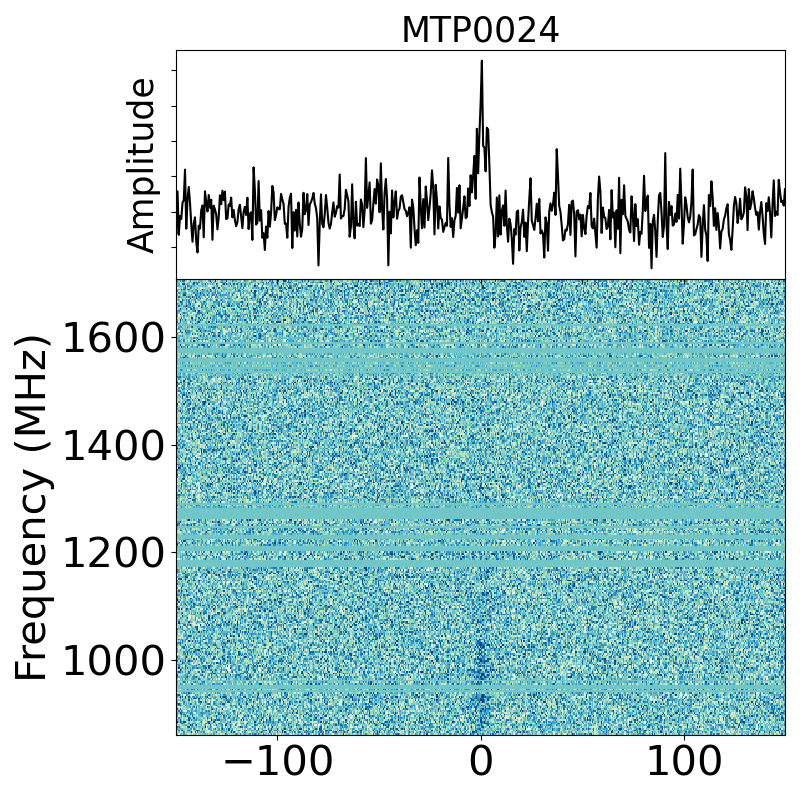}} 
    \subfigure{\includegraphics[width=0.235\textwidth]{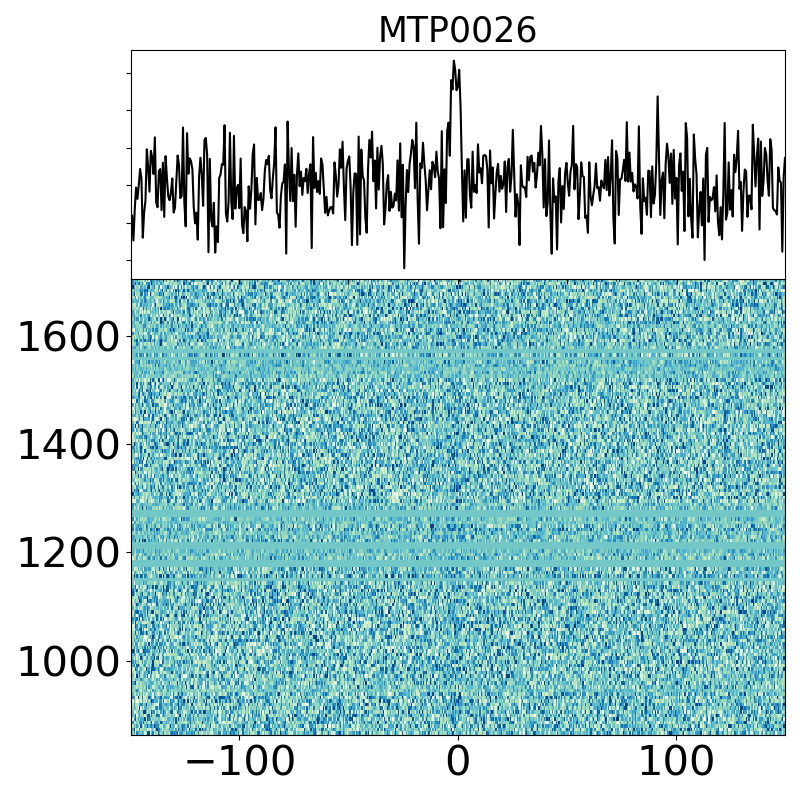}}
    \subfigure{\includegraphics[width=0.235\textwidth]{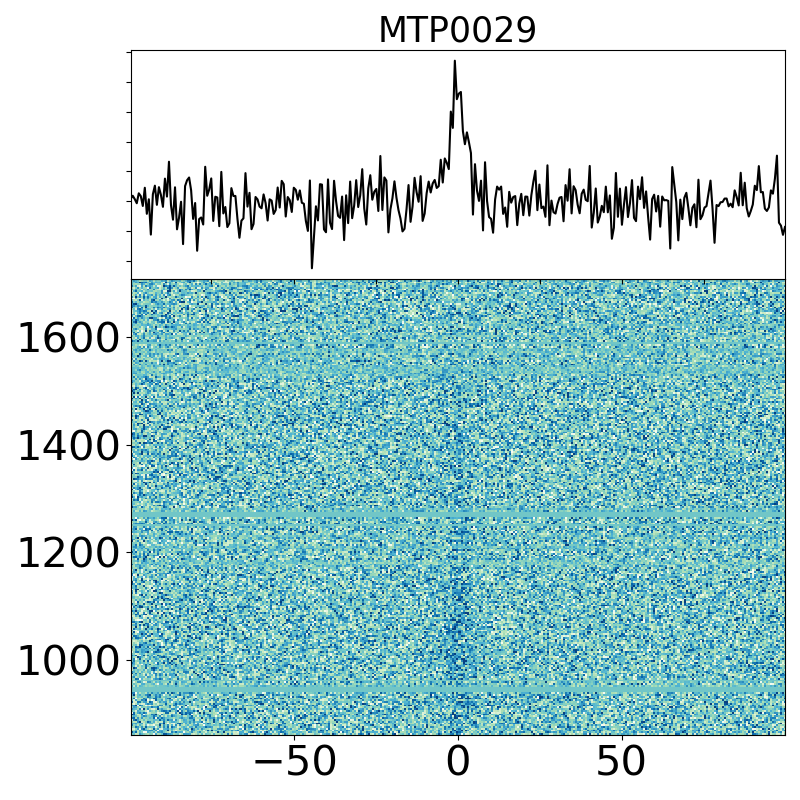}} 
    \subfigure{\includegraphics[width=0.235\textwidth]{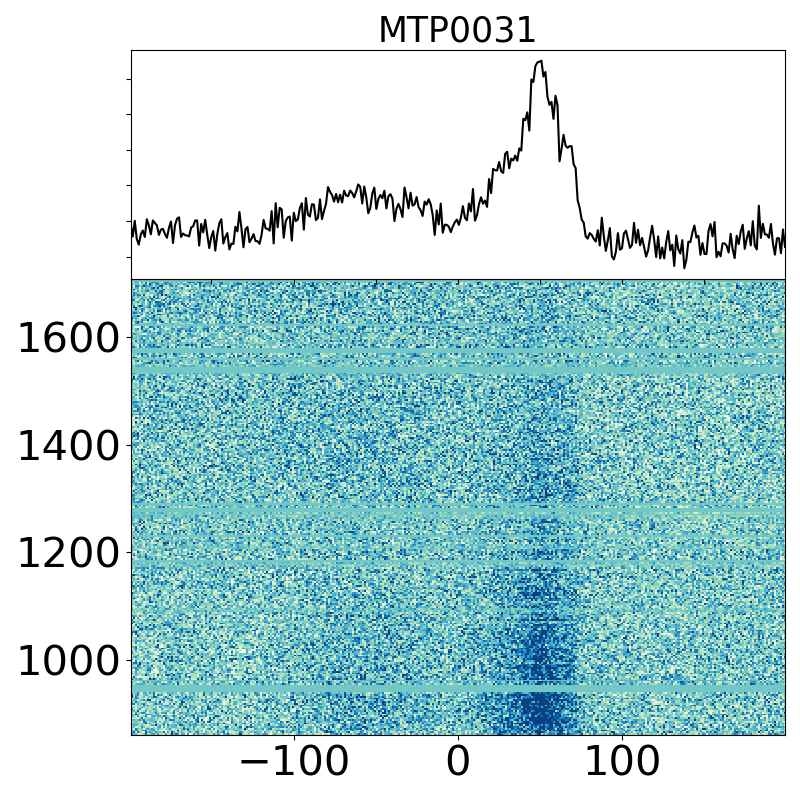}} 
    \subfigure{\includegraphics[width=0.235\textwidth]{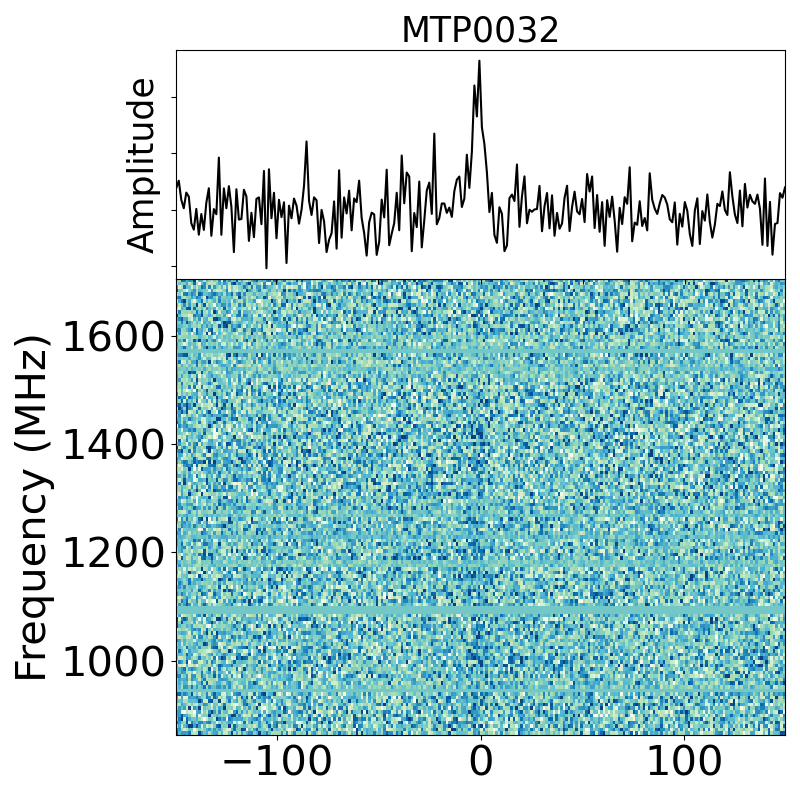}}
    \subfigure{\includegraphics[width=0.235\textwidth]{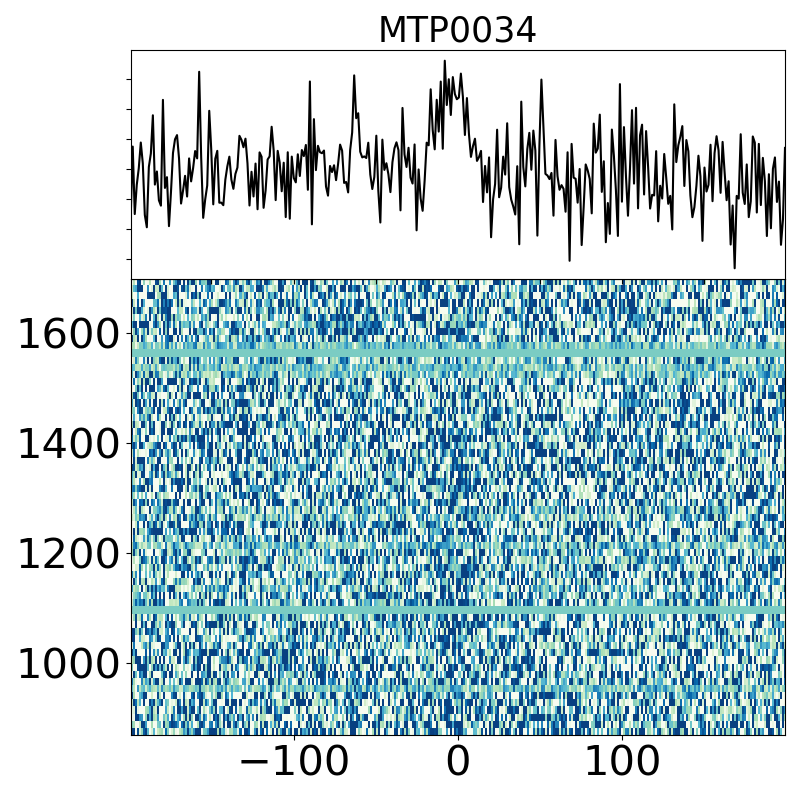}} 
    \subfigure{\includegraphics[width=0.235\textwidth]{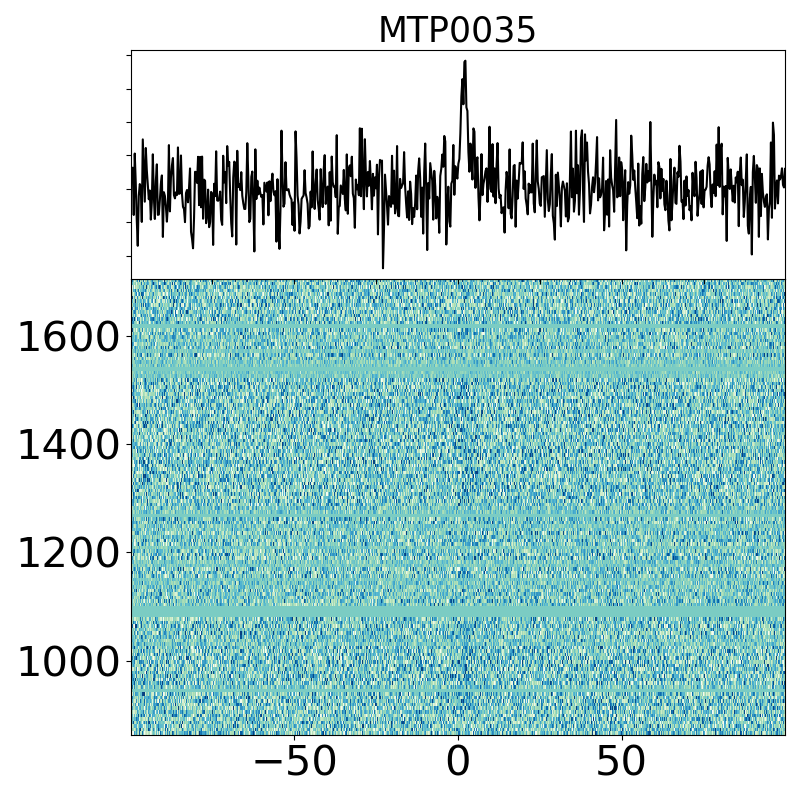}} 
    \subfigure{\includegraphics[width=0.235\textwidth]{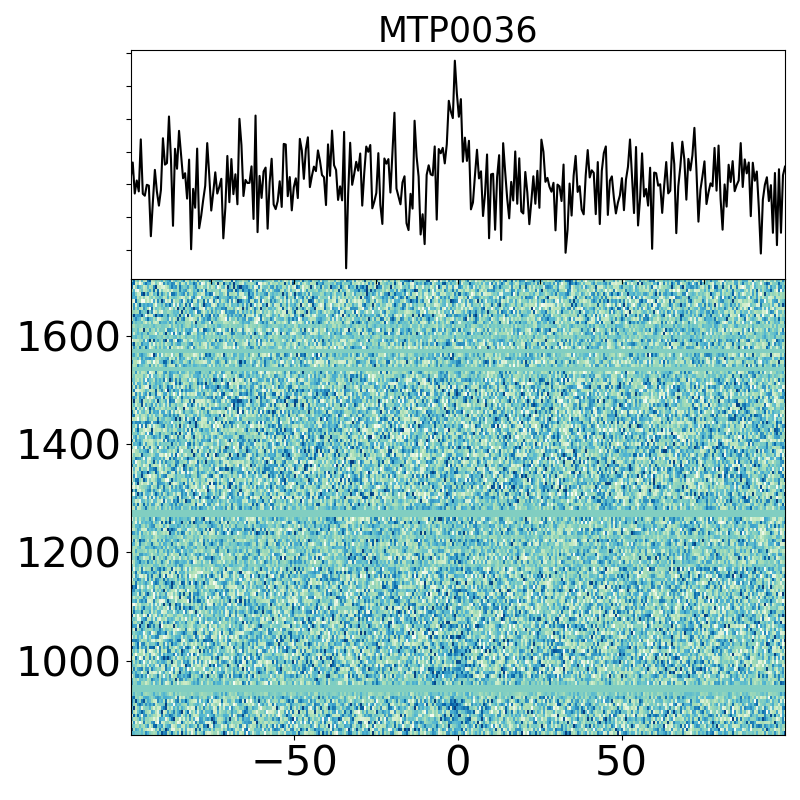}} 
    \subfigure{\includegraphics[width=0.235\textwidth]{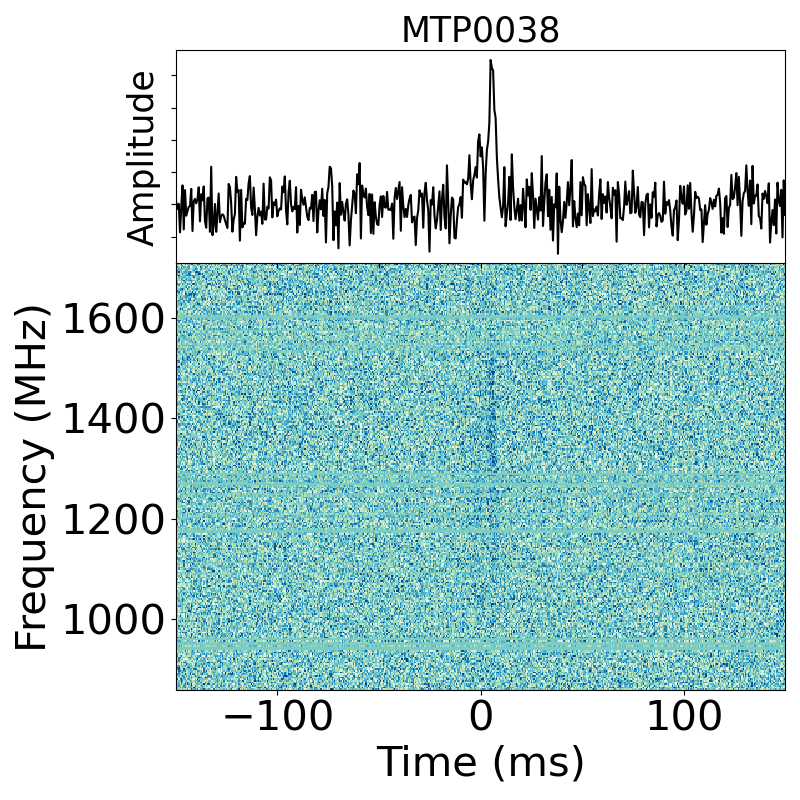}} 
    \subfigure{\includegraphics[width=0.235\textwidth]{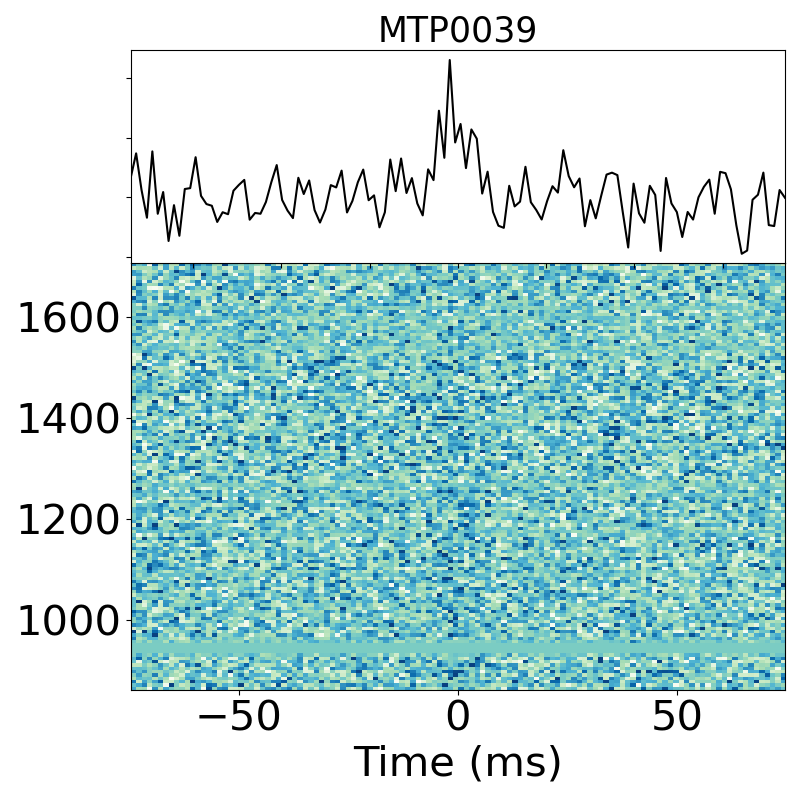}} 
    \subfigure{\includegraphics[width=0.235\textwidth]{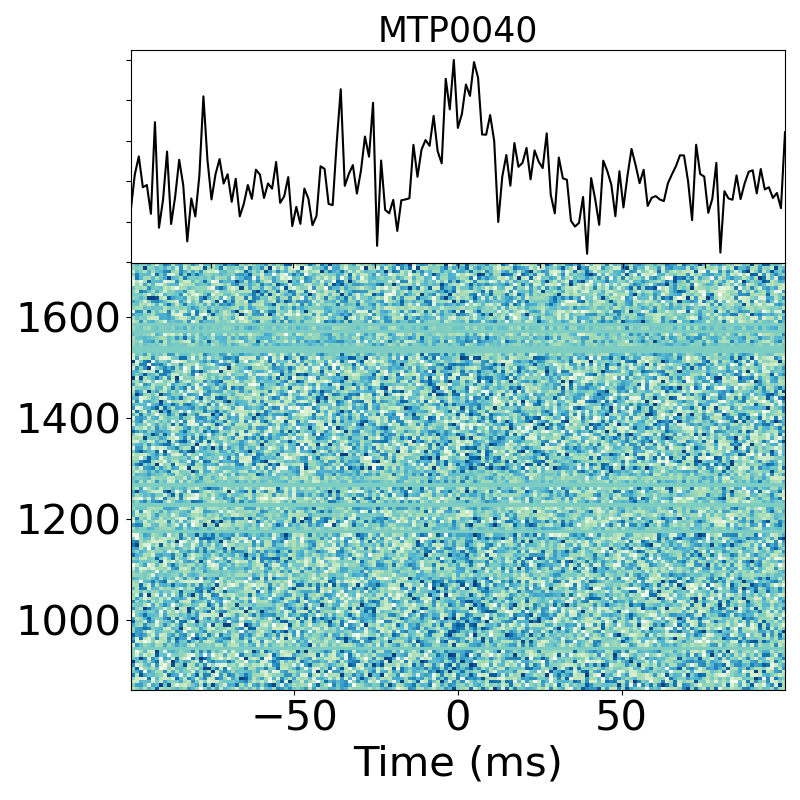}}
    \subfigure{\includegraphics[width=0.235\textwidth]{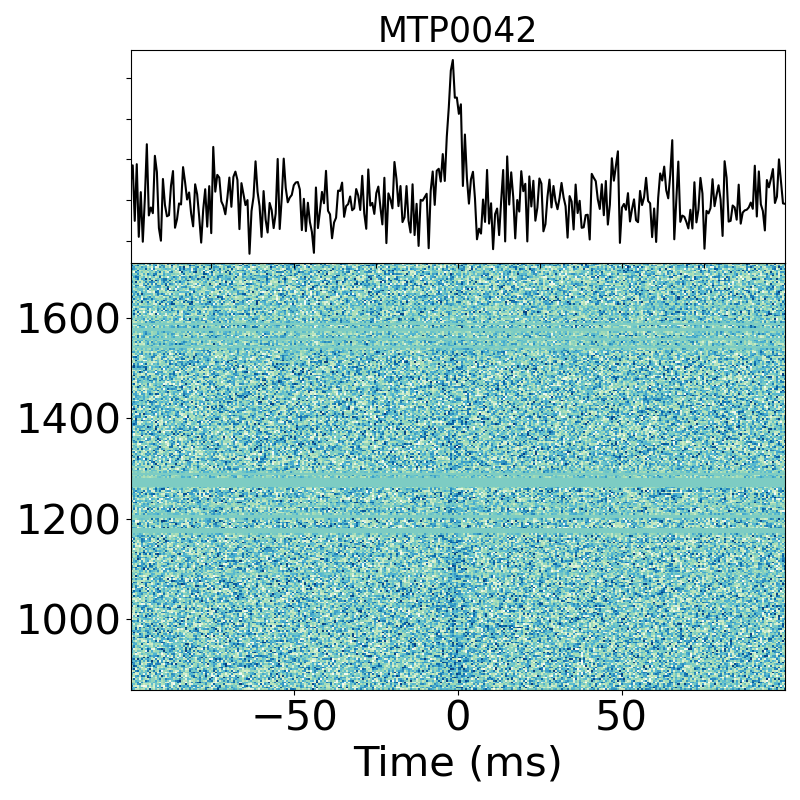}} 
    \caption{Dynamic spectra (bottom) and frequency-averaged pulse profiles (top) for the new Galactic transients. Each pulse shown is the first that MeerTRAP detected. These plots have been produced in the same way as in \autoref{fig:waterfallsA}. Some data have been downsampled in time to improve the visibility of the pulse. Blank horizontal lines are masked channels that were affected by RFI. The pulse of MTP0021 is visibly affected by flux over-subtraction due to zero-DM RFI removal.}
    \label{fig:waterfallsB}
\end{figure*}
\begin{figure*}\ContinuedFloat
    \centering
    \subfigure{\includegraphics[width=0.235\textwidth]{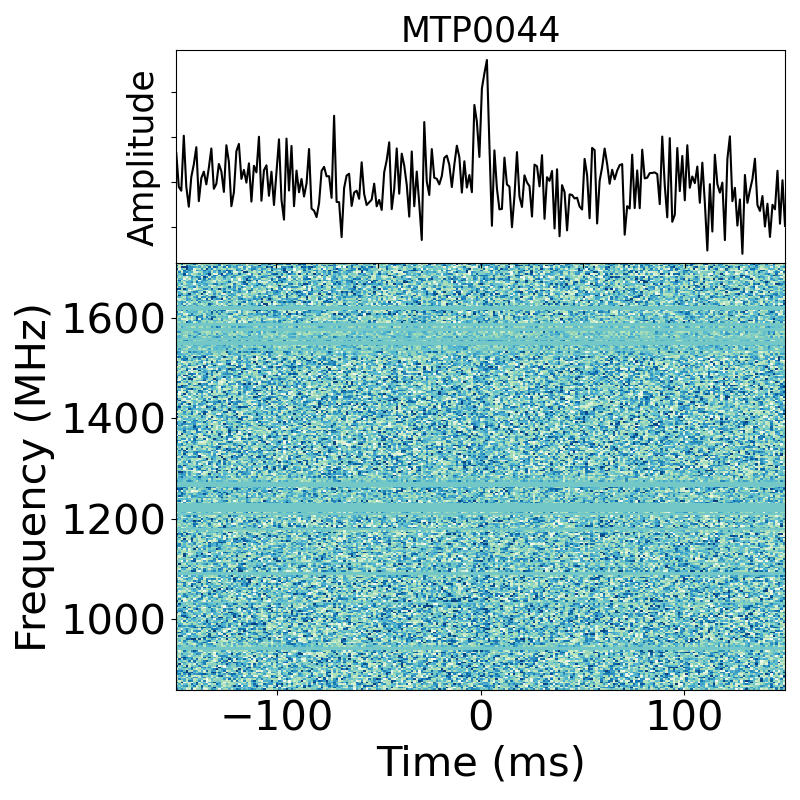}} 
    \subfigure{\includegraphics[width=0.235\textwidth]{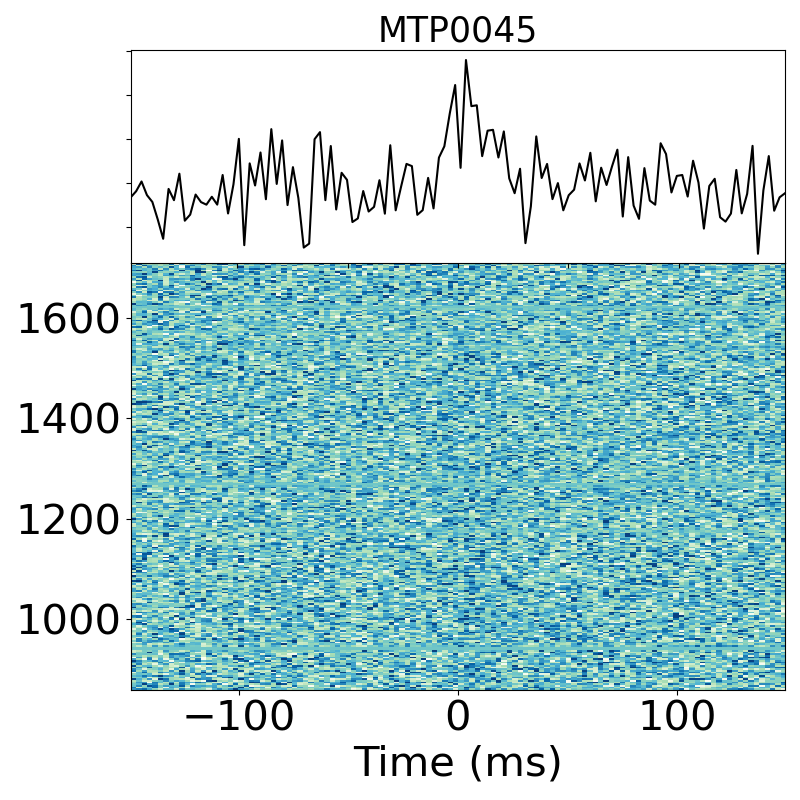}} 
    \subfigure{\includegraphics[width=0.235\textwidth]{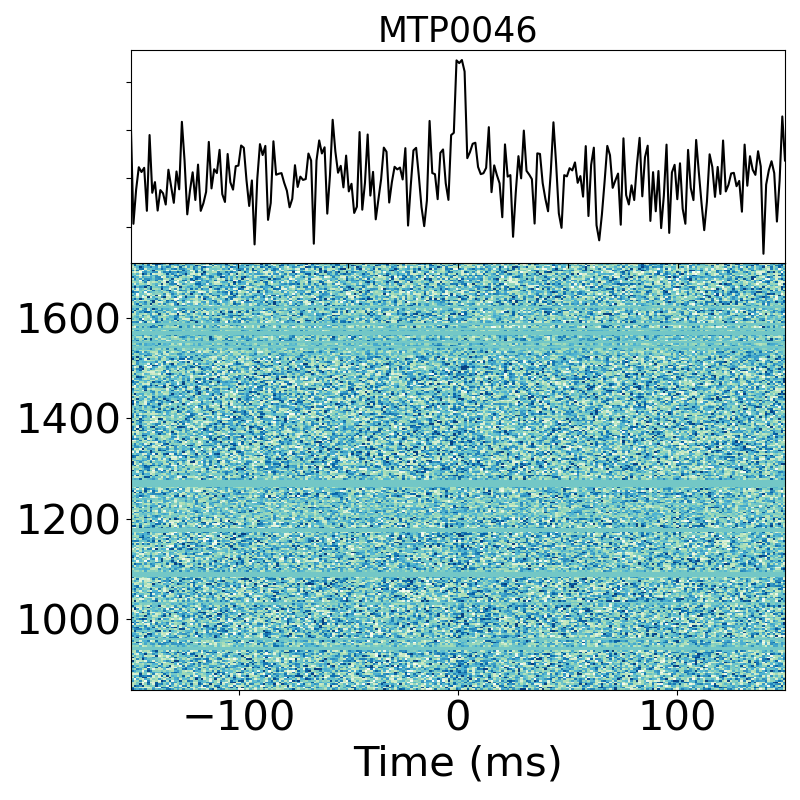}} 
    \subfigure{\includegraphics[width=0.235\textwidth]{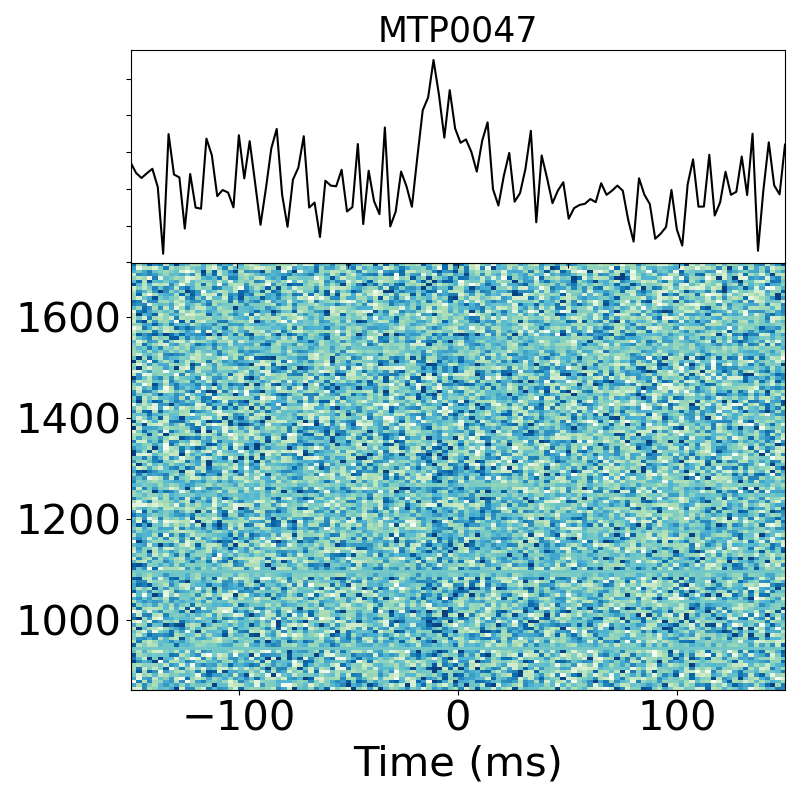}} 
    \subfigure{\includegraphics[width=0.235\textwidth]{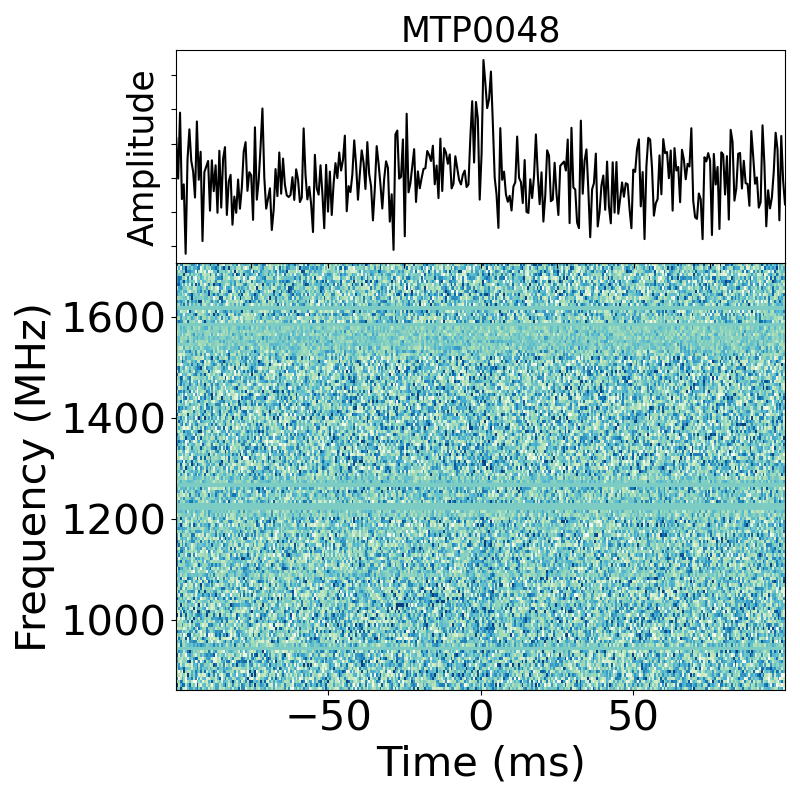}} 
    \subfigure{\includegraphics[width=0.235\textwidth]{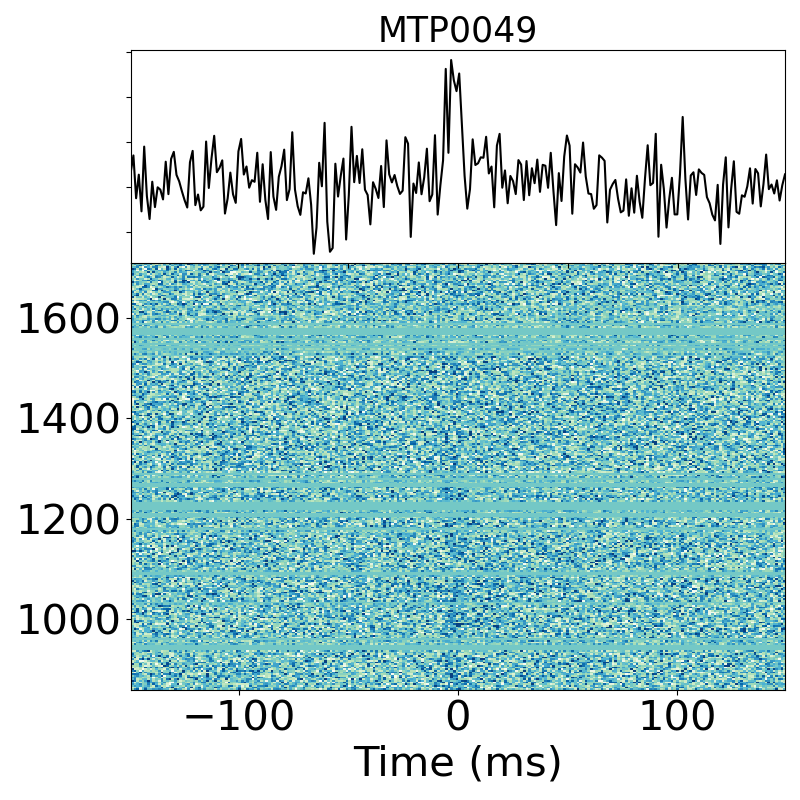}} 
    \caption{(continued)}
\end{figure*}

\begin{table*}
\centering
\caption{Properties of the independently discovered pulsars MeerTRAP detected and the new Galactic radio transients. If a position does not have an uncertainty, this is because a proxy for the positional uncertainty is provided in the text, rather than not at all. For all periods we provide the reference epoch for the midpoint of the TOA range used in the calculation. For five sources, the epoch is provided as part of a complete timing solution in \autoref{tab:timing_solns}. Upper limits on periods correspond to the minimum time between two pulses. References to publications; [1] \citet{Dong2023} and [2] \citet{Padmanabh2023}. }
\label{tab:MTPinfo}
\begin{tabular}{lllccclr}
\hline
MTP name & RA (J2000) & Dec (J2000)  & DM  & Dist. \textsc{ne2001}/\textsc{ymw16} & S/N & Period, $P$ & $P$ Epoch \\
 & (hh:mm:ss) & (dd:mm:ss) & (pc cm$^{-3}$) & (kpc) & & (s) & (MJD) \\
\hline
\multicolumn{8}{l}{Published sources} \\
\hline
MTP0015 & 22:37:29.41 & +28:28:40 & 38.1(1)$^{\dagger}$ & 2.3 / 4 & 12--160 & 1.0773950915(3) [1] & $-$ \\
MTP0028 & 15:40:10.6 & $-$58:21:56 & 440.3(2) & 8.6 / 11.8 & 14--23 & 3.47472(8) [2] & $-$  \\
MTP0043  & 16:49:47 & $-$42:30:21 & 374(6) [2] & 5 / 7.1 & 9--11 & 0.67641(4) [2] & $-$ \\
\hline
\multicolumn{7}{l}{New sources} \\
\hline
MTP0014 & 19:11:16.05(8) & $-$20:20:02(9) & 71.30(3) & 2.1 / 4.2 & 8--54 & 4.46792112045(2) & see \autoref{tab:timing_solns} \\
MTP0016 & 15:25:06.4 & $-$23:22:19 & 41.2(3) & 1.7 / 3.0 & 13--22 & 5.5719(2) & 59208.083669 \\
MTP0017 & 04:02:52.25(3) & $-$65:42:43.46(16) & 31.5(2) & 1.5 / 2.7 & 8--19 & 3.0335229848(1) & see \autoref{tab:timing_solns} \\
MTP0018 & 04:08:00 & $-$65:45:00 & 38.8(3) & 2.3 / $>$25 & 10 & $-$ & $-$ \\
MTP0020 & 19:30:41.880(92) & $-$18:56:28.49(12) & 63.143(9) & 2.1 / 4.3 & 8--75 & 1.76083292608(5) & see \autoref{tab:timing_solns} \\
MTP0021 & 02:19:00 & $-$06:10:00 & 8.46(7)$^{\dagger}$ & 0.4 / 0.4 & 22--132 & 1.87868(4) & 59330.517875 \\
MTP0022 & 12:43:41.07 & $-$64:23:07.2 & 342(2)$^{\dagger}$ & 7.0 / 10.2 & 14--29 & $-$ & $-$ \\
MTP0023 & 13:19:48.24(4) & $-$45:36:04.0(4) & 40.41(8)$^{\dagger}$ & 1.3 / 1.3 & 8--220 & 1.8709058201(1) & see \autoref{tab:timing_solns} \\
MTP0024 & 18:08:27.65 & $-$36:58:43.9 & 41.0(5) & 1.1 / 1.1 & 12--13 & $-$ & $-$ \\
MTP0026 & 15:39:40 & $-$61:10:15 & 206.8(3) & 4.5 / 7.4 & 10 & $-$ & $-$ \\
MTP0029 & 16:48:36.46 & $-$51:42:46.5 & 201.4(2) & 4.1 / 6.1 & 10--17 & $-$ & $-$ \\
MTP0031 & 09:17:28.3(1) & $-$42:45:55(1) & 97.7(3)$^{\dagger}$ & 0.6 / 0.4 & 8--68 & 2.5519(1) & 59699.588283 \\
MTP0032 & 14:52:34.91 & $-$62:35:06.0 & 271.5(5) & 6.1 / 6.1 & 11 & $-$ & $-$ \\
MTP0034 & 11:07:58.56(23) & $-$59:47:01.1(1) & 92.7(4) & 2.1 / 1.5 & 8--28 & 1.516531550(3) & see \autoref{tab:timing_solns} \\
MTP0035 & 13:08:59 & $-$61:17:00 & 224.5(2) & 4.4 / 5.6 & 11--12 & $\leq$3.9548 & 59513.543146 \\
MTP0036 & 10:00:58.60 & $-$51:45:52.9 & 128.9(7) & 3.1 / 0.4 & 9 & $-$ & $-$ \\
MTP0038 & 17:27:25.11 & $-$29:42:20.6 & 126.7(4) & 2.4 / 3.0 & 16 & $-$ & $-$ \\
MTP0039 & 15:33:50.46 & $-$56:09:29.7 & 95.31(9) & 1.8 / 2.5 & 8--39 & 1.06167(1) & 59568.370608 \\
MTP0040 & 13:57:49.48 & $-$65:07:40.7 & 264(1) & 5.6 / 6.4 & 9 & $-$ & $-$ \\
MTP0042 & 16:41:39.43 & $-$51:09:00.8 & 250.4(7) & 4.6 / 5.8 & 13 & $\leq$5.5143 & 59495.789312 \\
MTP0044 & 22:18:23.3(19) & +29:02:56(33) & 55.8(4) & 4.4 / >25 & 8--16 & 17.49487(3) & 59581.575764 \\
MTP0045 & 15:31:08.0(2) & $-$55:57:29(2) & 56.6(6) & 1.3 / 1.2 & 8--17 & 2.919875(2) & 59568.259968 \\
MTP0046 & 15:58:46.25 & $-$48:38:48.4 & 254(1) & 8.1 / 6.8 & 9 & $-$ & $-$ \\
MTP0047 & 18:07:10.36 & $-$11:51:08.2 & 152.4(4) & 3.4 / 3.5 & 8--19 & $-$ & $-$ \\
MTP0048 & 14:29:17.29 & $-$64:01:15.2 & 151.6(5) & 3.5 / 4 & 8--9 & $-$ & $-$ \\
MTP0049 & 14:05:37.10 & $-$65:05:25.3 & 346.5(7) & 8.1 / 9.2 & 9 & $-$ & $-$ \\
\hline
\end{tabular}
\begin{tablenotes}
    \item $^{\dagger}$DM measured using \dmphase{}
\end{tablenotes}
\end{table*}

\begin{table*}
\centering
\caption{Timing solutions for the five RRATs; MTP0014, MTP0017, MTP0020, MTP0023 and MTP0034.}
\label{tab:timing_solns}
\begin{tabular}{llllll}
\hline\hline
\multicolumn{6}{c}{Fit and data-set} \\
\hline
PSR name\dotfill & J1911$-$2020 & J0402$-$6542 & J1930$-$1856 & J1319$-$4536 & J1108$-$5946 \\
MeerTRAP name\dotfill & MTP0014 & MTP0017 & MTP0020 & MTP0023 & MTP0034 \\
MJD range\dotfill & 59301.3---60385.1 & 59214.8---60506.1 & 59324.1---59818.8 & 59369.8---60506.9 & 59513.5---60008.2 \\ 
Data span (yr)\dotfill & 2.97 & 3.54 & 1.35 & 3.11 & 1.35 \\ 
Number of TOAs\dotfill & 648 & 132 & 231 & 91 & 134 \\
Rms timing residual (ms)\dotfill & 11 & 2.5 & 3.8 & 5.2 & 7.4 \\
Weighted fit\dotfill &  Y & Y & Y & Y & Y \\ 
Reduced-$\chi^{2}$\dotfill &  444 & 5.4 & 380 & 1400 & 17 \\ 
\hline
\multicolumn{6}{c}{Measured Quantities} \\ 
\hline
Right ascension, $\alpha$ (hh:mm:ss)\dotfill &  19:11:16.05(8) & 04:02:52.27(3) & 19:30:41.88(9)$^{\dagger}$ & 13:19:48.31(6) & 11:07:58.56(23)$^{\dagger}$ \\ 
Declination, $\delta$ (dd:mm:ss)\dotfill & $-$20:20:02(9) & $-$65:42:43.41(16) & $-$18:56:28.5(12) & $-$45:36:03.0(8) & $-$59:47:01.1(12) \\ 
Spin period, $P$ (s)\dotfill & 4.4679211203(2) & 3.03352298461(7) & 1.76083292621(3) & 1.8709058202(2) & 1.516531549(3) \\ 
First derivative of $P$, $\dot{P}$ ($\times 10^{-15}$\,ss$^{-1}$)\dotfill & 6.726(8) & 5.601(2) & 0.593(7) & 6.975(3) & 0.22(17) \\
Epoch of period determination (MJD)\dotfill & 60098 & 59581.5 & 59581.5 & 59369 & 60001.2 \\ 
Epoch of position determination (MJD)\dotfill & 60098 & 59581.5 & 59581.5 & 59369 & 60001.2 \\ 
Epoch of DM determination (MJD)\dotfill & 60098 & 59581.5 & 59581.5 & 59369 & 60001.2 \\ 
Dispersion measure, DM (cm$^{-3}$pc)\dotfill & 71.30(3) & 31.5(2) & 63.143(9) & 40.41(8) & 92.7(4) \\ 
\hline
\multicolumn{6}{c}{Derived Quantities} \\
\hline
Characteristic age (Myr)\dotfill & 10 & 8.6 & 47 & 4.3 & 110 \\
$\log_{10}$(Surface magnetic field strength, G)\dotfill & 12.7 & 12.6 & 12.0 & 12.6 & 11.8 \\
$\log_{10}$(Spin-down luminosity, ergs/s)\dotfill & 30.5 & 30.9 & 30.6 & 31.6 & 30.4 \\
\hline
\multicolumn{6}{c}{Assumptions} \\
\hline
Clock correction procedure\dotfill & TT(TAI) & TT(TAI) & TT(TAI) & TT(TAI) & TT(TAI) \\
Solar system ephemeris model\dotfill & DE405 & DE405 & DE405 & DE405 & DE405 \\
Binary model\dotfill & NONE & NONE & NONE & NONE & NONE \\
Model version number\dotfill & 5.00 & 5.00 & 5.00 & 5.00 & 5.00 \\ 
\hline
\end{tabular}
\begin{tablenotes}
    \item $^{\dagger}$Position is from an image localisation of transient buffer data, not from fitting in \tempo{}
\end{tablenotes}
\end{table*}

\begin{figure*}
    \centering
    \includegraphics[width=0.8\textwidth]{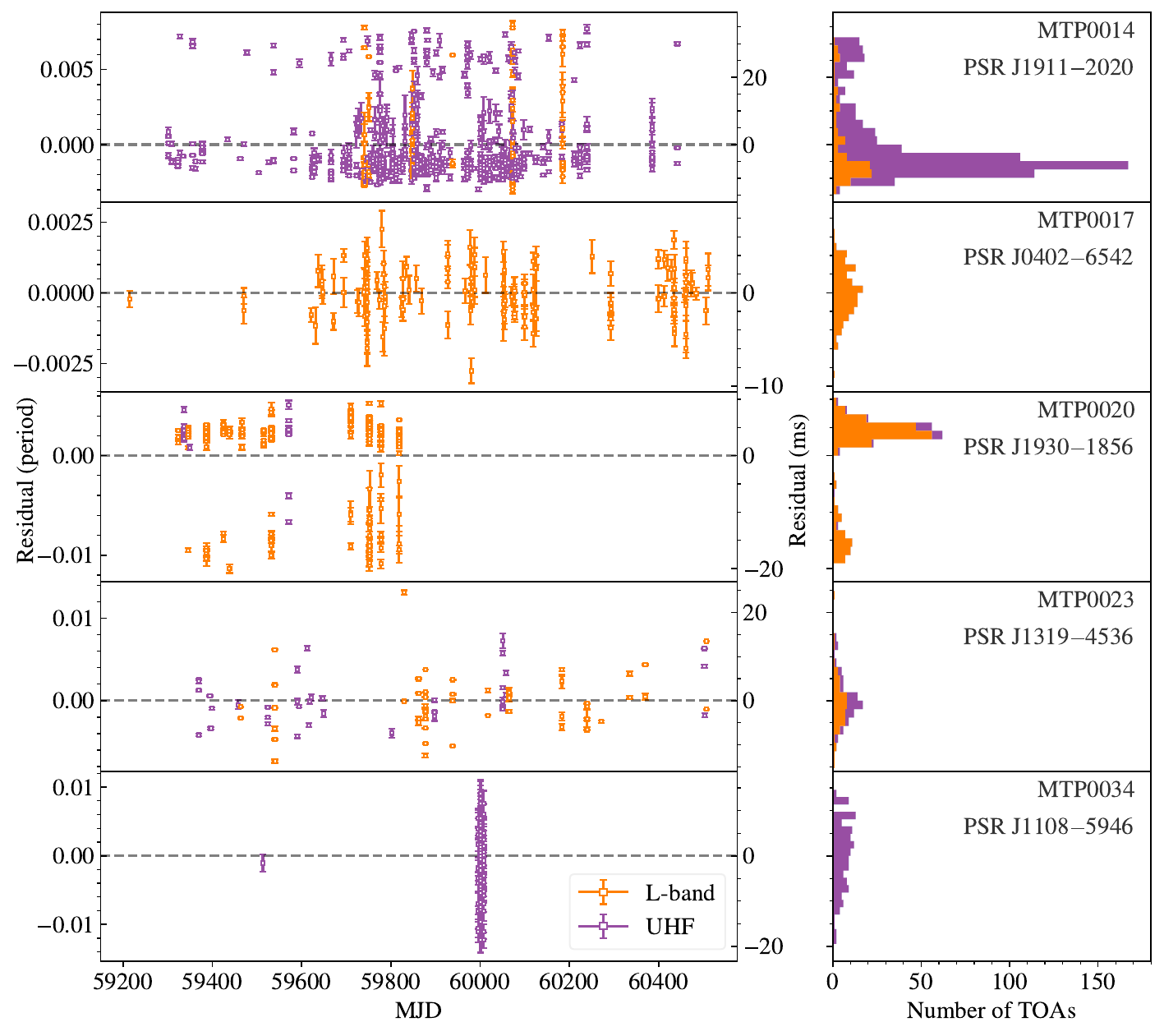}
	\caption{Residuals (\textit{left panels}) and stacked histogram of residuals (\textit{right panels}) from the best fit model given in \autoref{tab:timing_solns} for the five sources with timing solutions.}
	\label{fig:residuals}
\end{figure*}

In this section we present the 26 new Galactic transients we have detected. Information about their discovery and observations is shown in \autoref{tab:detections}, and their measured properties are listed in \autoref{tab:MTPinfo}. For five of these sources we were able to find timing solutions, which are given in \autoref{tab:timing_solns}. We proceed with detailed descriptions of the sources that have been detected on multiple epochs. The other sources are summarised at the end of this section.

\subsubsection*{MTP0014/\fourteen}
\begin{figure*}
    \centering
    \includegraphics[width=0.9\textwidth]{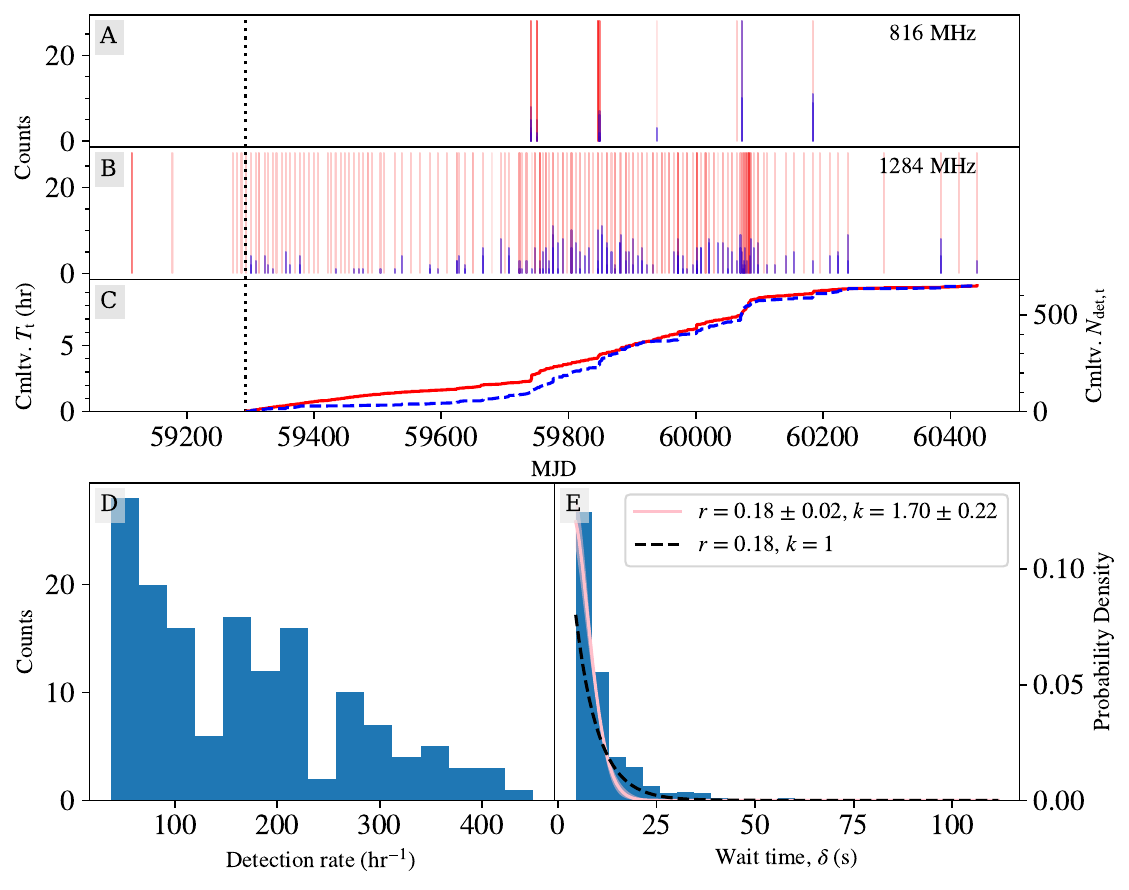}
    \caption{\textit{Upper:} Plots tallying the integration time and number of detections of MTP0014/\fourteen{} at UHF (A) and L--band (B). Red shaded boxes are extent of the observations of the primary target J1911$-$2006. The dashed vertical line marks the date that targeted observations began. Darker shades of red are an effect of plotting narrow shaded regions over a long baseline, so only signify that the calibrator was more frequently visited. Panel C shows the cumulative targeted observation time, $T_{\text{obs}}$ as a solid line and the cumulative number of detections, $N_{\text{pulse}}$ as the dashed line. \textit{Lower:} Panel D: histogram of pulse rates calculated for all targeted observations. Panel E: histogram of the wait times, $\delta$ between consecutive pulses in targeted observations. The best-fit for $\mathcal{W}$($\delta$\,|\,$k$, $r$) and phase-space encapsulating the 3-sigma uncertainty of $k$ and $r$ are overlaid in pink, and the same function for $k$\,=\,1 is shown as the dashed line.}
    \label{fig:MTP0014_act}
\end{figure*}

MTP0014/\fourteen{} was discovered on 2020 November 24 when a cluster of pulses were detected at a DM of $\sim$\,72\dm. The dynamic spectrum and the frequency-integrated pulse profile is shown for the first of these pulses in the top left of \autoref{fig:waterfallsB}. We have since constrained the DM to 71.30$\pm$0.03\dm{} with \scatfit{} using the brightest pulse seen so far, which was on 2023 March 8. A spin period of 4.467\,s was initially found using \rratsolve{} on the discovery pulses. \fourteen{} is located very close to the gain calibrator J1911$-$2006 used by many observers, so of the MeerTRAP sources it has been seen on the most days and has the largest number of detections.  \\ 
On 2023 January 28, \fourteen{} triggered the TB. We constrained the position using the resulting image from the data. We measured TOAs for all the pulses up to 2024 July 25 and, using the image localisation and initial period, were able to find a phase-connected timing solution spanning the three years of data. The resulting position fit of RA 19:11:16.05(8) and Dec $-$20:20:02(9) is more constraining than the image localisation. In \autoref{fig:residuals} we show the timing residuals produced by subtracting the model-predicted arrival times from the measured TOAs. These are plotted for each observing frequency and their distribution is shown as a frequency-stacked histogram. The residuals reveal that \fourteen{} has an average profile consisting of at least three components. Given our methodology of fitting a single Gaussian to the profile, we would expect a trimodal distribution of TOAs with each peak height reflecting the prevalence of each component. This is similar to the magnetar-like \mbox{PSR J1819$-$1458} \citep{Lyne2009, Bhattacharyya2018} where component flux variation produced three bands in the residual-space. Both \mbox{PSR J1819$-$1458} and \fourteen{} have a similar period but the period derivative of \fourteen{} is $\sim$\,100 times smaller, thus has a greater characteristic age by approximately 2 orders of magnitude greater. These calculations assume a dipolar magnetic field configuration. The leading component of \fourteen{} is the dominant or only pulse visible for 76 per cent of the observed pulses, while the middle and trailing component each dominate 12 per cent. On 2022 January 3 at UTC 09:06:42 we detected the only three component profile seen so far. This pulse closely resembles the residual distribution in that the leading component is more prominent and the three components are evenly spaced apart. Due to the dominance of the main component and the lack of structure evident in its residual distribution such as timing noise, the timing would not be improved by subtracting the offset between the residual bands as was the method of \citet{Bhattacharyya2018}. Futhermore, timing noise is much less prominent in older pulsars \citep{Hobbs2010}. \\
We have used the large number of detections to investigate the activity of \fourteen{} further. Our calculations and analysis only use detections made after 2021 March 20 when targeted observations began. The average detection rate across these observations is 74\,hr$^{-1}$, and split by observing band it is 68\,hr$^{-1}$ for L--band 127\,hr$^{-1}$ for UHF. The rate at UHF is expected to be higher due to the pulsar spectral index. If we take a value of $-$1.6 \citep[e.g.][]{Jankowski2018}, and assume that the difference in contribution due to the sky and reciever temperatures is small, we would expect a ratio between the rates of (816/1284)$^{-1.6}\times\sqrt{544/856}$ = 1.65. Instead we observe a factor 1.88, which could be due to a slightly steeper spectral index, or observational effects such as the UHF band being generally cleaner than L--band, or due to the larger size of the UHF CB. We analyse the pulse detections in more detail in \autoref{fig:MTP0014_act}. Panel A and B tally all observations of the primary target J1911$-$2006 and detections of \fourteen. In panel C, the cumulative targeted observation time, $T_{\text{t}}$ against the cumulative number of detections, $N_{\text{det,t}}$ shows that the detectability of \fourteen{} increased after around MJD 59700. This does not appear to be solely the result of UHF detections increasing the mean rate, so we could be observing an increased detection rate from intrinsic changes at this epoch. In panel D, we show the histogram of detection rates. As alluded to in Section \ref{timing}, we see the detection rate distribution is affected by the fact that J1911$-$2006 is a calibrator. Most observations are 120\,s long, corresponding to 95\,s of MeerTRAP time. The rate appears to be boosted at integer values of $N_{\text{det,t}}$ in this time frame. To investigate the nature of the wait times distribution in panel E, we use the same method as \citet{Bezuidenhout2022} to fit a modified Poissonian distribution. If the wait time, $\delta$ between detections that have a underlying pulsation rate, $r$\,=\,1/$P$, was determined by a random process, then their probability distribution would be
\begin{equation}
    \mathcal{P}(\delta\,|\,r) = r\,e^{-\delta r}.
\end{equation}
An example of Poisson-distributed wait times is the giant pulses of the Crab pulsar \citep[e.g.][]{Karuppusamy2010}. Other transient sources often exhibit non-random burst rates, instead being much more clustered in time. This includes some RRATs \citep[e.g.][]{Shapiro-Albert2018, Bezuidenhout2022} but also repeating FRBs, for example FRB 20121102A \citep{Scholz2016} and FRB 20201124A \citep{Lanman2022}. As \citet{Oppermann2018} shows, this can be parameterised by the Weibull function, which includes an extra shape parameter, $k$ such that
\begin{equation}
    \mathcal{W}(\delta\,|\,k, r) = k\,\delta^{-1} [\delta\,  r\, \Gamma(1+1/k) ]^{k}\, e^{-[\delta\,  r\, \Gamma(1+1/k) ]^{k}},
\end{equation}
where the gamma function, $\Gamma$ is given by
\begin{equation}
    \Gamma(x) = \int_{0}^{\infty} t^{x-1} e^{-t} \,dt.
\end{equation}
For $k$\,=\,1, the Weibull function returns to the Poission-distributed exponential. We use the non-linear least-squares method of the \textsc{python} module \texttt{scipy.optimise.curve\_fit} to fit the wait times, and show it overlaid on the distribution in \autoref{fig:MTP0014_act}. The fit returns the rotational frequency, $r$\,=\,1/$P$ and $k$\,>\,1 showing that the pulses are evidently more clustered than would be the case for randomness. The clustering is not a manifestation of a short observation bias, as we exclude durations longer than 120\,s before the fit, as wait times longer than this are not sampled as often.
\subsubsection*{MTP0016/\sixteen}
MTP0016/\sixteen{} was detected during a MeerKAT commissioning observation on 2020 December 25, with the first pulse arriving at UTC 01:59:02. It was seen in one CB centred on RA 15:25:06.41 and Dec $-$23:22:19.1 and located on the edge of the area filled with CBs, but was not seen in the IB. Three more pulses were seen in the same beam over the subsequent 173~s. The pulses had S/N values of 8-14 and DM values between 38-42\dm. Using \scatfit{} we measured the best DM to be 41.2$\pm$0.3\dm, which would place MTP0016 at 1.7~kpc (\textsc{ne2001}) or 3~kpc (\textsc{ywm16}). The nearest pulsar of a similar DM is PSR J1530$-$2114 at 2.5\degr away with DM 37.95\dm{} \citep{Fiore2023}. 
The first pulse, shown in \autoref{fig:waterfallsB}, has a hint of a precursor component, but the rest are single-peaked. With these four arrival times, we used \rratsolve{} to calculate a spin period of 5.5719(2)~s. \sixteen{} was detected in only the first of 16 separate 600-s observations of the primary target, which were spaced almost periodically across approximately 9 hours. The reason \sixteen{} was not detected again when the target was revisited could be explained by its position at the edge of the first CBs. The PSF of the beams will change shape significantly due to the changing availability of baselines and also their projected length in the direction parallel to the line of sight to the source as its elevation changes. It is possible that the elliptical beams rotated such that \sixteen{} fell into a less sensitive part of the CB, and therefore could not be detected again. However, it could also be that the activity or flux density of the source dropped. The measured average detection rate of \sixteen{} was therefore between 3.5\,hr$^{-1}$--52\,hr$^{-1}$ depending on the length of time \sixteen{} was actually detectable, but this cannot be determined until the source is localised further. \sixteen{} is probably located fairly centrally within the CB, as there were no IB detections. Therefore, we provide the region bounded by the CB as a proxy for the positional uncertainty. Using a record of the antennas used and \mosaic, we know the beam shape was 56\arcsec$\times$27\arcsec{} with an orientation of 290.8\degr{}\footnote{The shape is the semi-major and semi-minor axes, $a$,~$b$ and the angle is anti-clockwise starting from $b$ pointed North in the J2000 world coordinate system. The size corresponds to when the sensitivity drops to 25 per cent of the maximum. Beam shapes and orientations throughout this article are in this format.}. This field has not been revisited by MeerTRAP, so no more pulses have been seen since.
\subsubsection*{MTP0017/\seventeen}
\begin{figure*}
    \centering
    \includegraphics[width=0.9\textwidth]{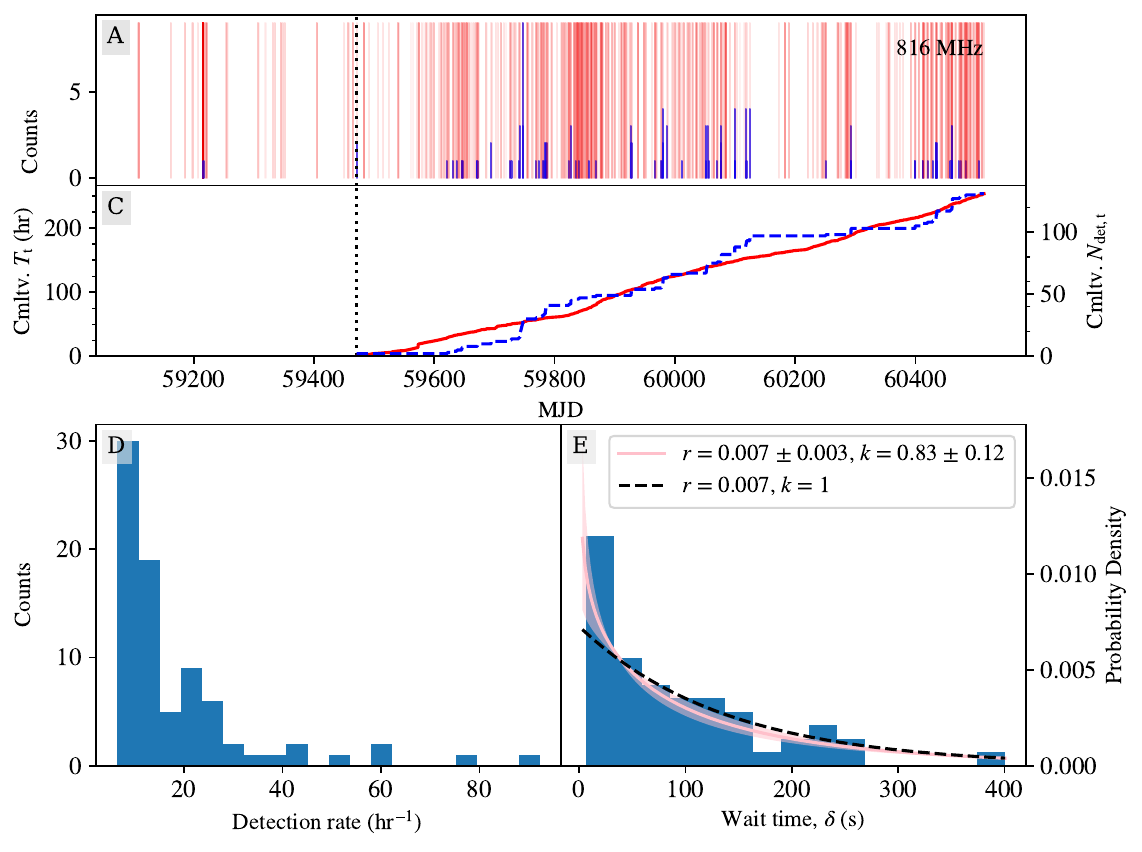}
    \caption{Observations and activity of MTP0017/\seventeen, where the contents of each panel are equivalent to those in \autoref{fig:MTP0014_act}. Panel B is not present as there are no L--band detections of \seventeen{}.}
    \label{fig:MTP0017_act}
\end{figure*}

We first detected MTP0017/\seventeen{} on 2020 December 31 at UHF during a 4\,hour observation of the calibrator J0408$-$6545/B0407$-$65. Only one pulse, which is shown in \autoref{fig:waterfallsB}, was seen. Targeted observations started on 2021 September 13, since then 133 pulses have been seen. The brightest pulse has a S/N of 19, for which we used \scatfit{} to find a DM of 31.5$\pm$0.2\dm{}. The low maximum S/N suggests that \seventeen{} has a pulse energy distribution that is concentrated towards a relatively low luminosity. The height above the Galactic plane predicted by the DM distances are 1.0 (\textsc{ne2001}) and 1.8 (\textsc{ymw16})\,kpc are the highest of the new sources. The mean detection rate across the targeted observations is $\sim$\,1.4\,hr$^{-1}$. This had initially made it difficult to obtain a period using \rratsolve{}, as we could not detect enough pulses during the short calibrator dwells. Eventually we were able to find a period of 3.03\,s. On 2023 August 28, \seventeen{} triggered the TB, which allowed us to get a position accurate to approximately 3\arcsec. Using the period and localisation, we found a long-term phase-connected timing solution using \tempo, which is provided in \autoref{tab:timing_solns}. The characteristic age of about 9\,Myr is consistent with \seventeen{} being an RRAT. The timing position provides a factor of 20 improvement in the localisation precision compared to that provided by imaging. The residuals resulting from the timing model are shown in \autoref{fig:residuals}. The residuals are distributed quasi-randomly about zero, and the reduced-$\chi^{2}$ from the weighted fit is relatively close to unity. \\
We have performed the same calculations of the long-term activity and detection statistics as we did for MTP0014. The calibrator J0408$-$6545/B0407$-$65 is the only primary target during which \seventeen{} has been detected. The tally of observations and detections and the detection rate and wait time statistics are shown in \autoref{fig:MTP0017_act}. The steady increase in the cumulative observation time of the calibrator contrasts with the stepped cumulative detection count. The clusters of detections do not seem to be dependent on how often the calibrator is observed. It therefore appears that \seventeen{} appears to enter epochs of increased activity approximately every 300 days. 
\subsubsection*{MTP0018}
MTP0018 was detected on 2021 January 1 as a single pulse in an incoherent beam centred on RA 04:08:20.38, Dec $-$65:45:09.1. The candidate had a DM of 28.55\dm and S/N 8.5, but it was under-dedispersed due to being affected by short duration zero-DM RFI occurring during the arrival of the dispersed pulse. It was initially suspected that we had made an incoherent mode detection of MTP0017 due to the similar DM and the proximity of the IB to the positions of CBs where MTP0017 has been detected on both days either side. However, after robust cleaning of the data, we constrained the DM to 38.8$\pm$0.3\dm, far from that of MTP0017. The pulse has a width of approximately 33\,ms, wider than any pulse we have detected from MTP0017. The pulse flux is highly concentrated at the bottom of the band, and such a steep spectrum suggests a position far from the IB centre where the lower frequencies remain somewhat sensitive. Interestingly, MTP0018 has not been seen again since, despite approximately 1000 observations of the calibrator in the 4.5 years since this detection. This could be due to being positioned in a very low sensitivity part of the IB. Alternativaly, MTP0018 could be a very infrequently emitting source, or has an unusual burst energy distribution. We note that the DM slightly exceeds the maximum line of sight value predicted by the \textsc{ymw16} model of 37\dm, and close to the \textsc{ne2001} value of 43\dm. This raises the possibility that MTP0018 is located in the Galactic halo but likely not beyond given a predicted total halo contribution to the line of sight of $\sim$\,40-50\dm \citep{Yamasaki2020}, making it unlikely MTP0018 is a one-off FRB. Such an event could however explain the low detection rate. MTP0018 is the only source to have exclusively been seen in a UHF IB.

\subsubsection*{MTP0020/\twenty}
\begin{figure}
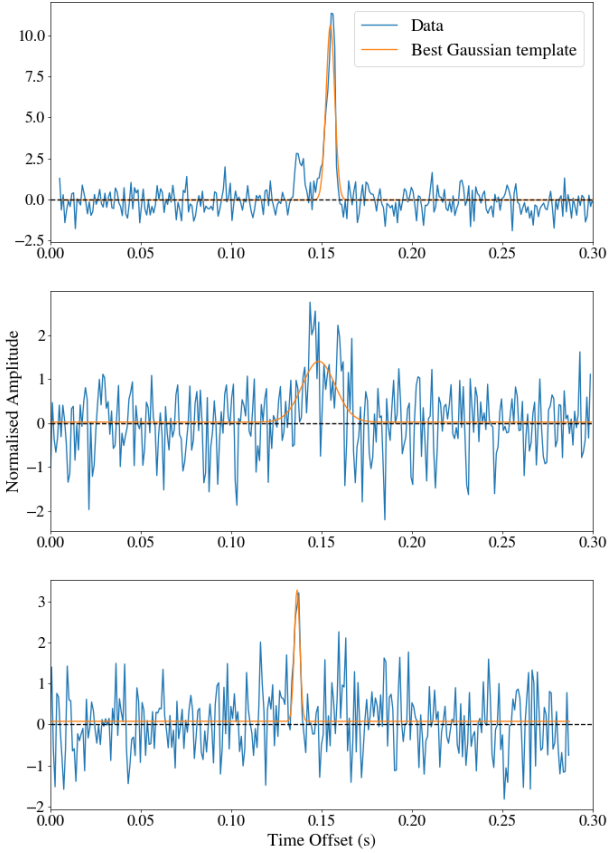

    \centering
    \subfigure{\includegraphics[width=0.95\columnwidth]{Figs/A\_2022\_07\_17\_19-26-47\_MTP0020.png}}
    \subfigure{\includegraphics[width=0.95\columnwidth]{Figs/B\_2022\_07\_17\_19-26-52\_MTP0020.png}}
    \subfigure{\includegraphics[width=0.95\columnwidth]{Figs/C\_2022\_07\_17\_19-26-54\_MTP0020.png}}
	\caption{Plots showing three successive detections of pulses from MTP0020/\twenty{}, overlaid with the best fit Gaussian from our timing method. From top to bottom: (i) a pulse of S/N 31 detected at MJD 59777.810276813 showing both the leading and trailing components of the average profile, (ii) the next detection three rotations later in time showing a pulse of S/N\,=\,8 and possibly two components, where the best Gaussian fit is a wider profile centred approximately at their midpoint and (iii) the S/N\,=\,8 pulse from next rotation showing the narrower leading component. The trailing component is no longer present.}
	\label{fig:MTP0020_modeswitch}
\end{figure}

\begin{figure*}
    \centering
    \includegraphics[width=0.49\textwidth]{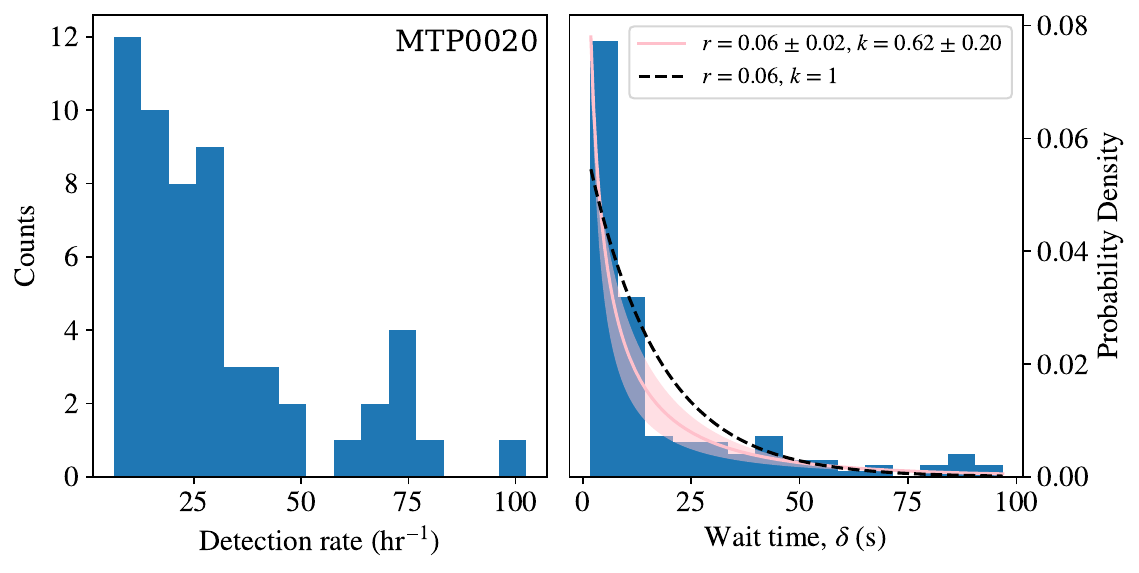}
    \includegraphics[width=0.49\textwidth]{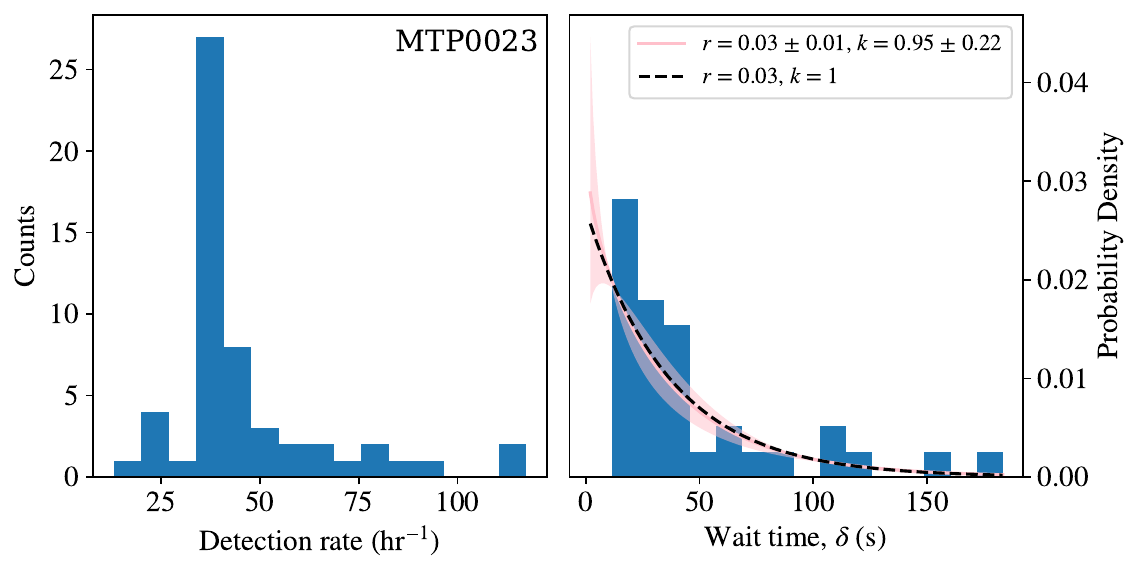}
    \caption{Rates and wait time statistics of targeted detections of MTP0020/\twenty{} (L--band and UHF) and MTP0023/\twentythree{} (UHF). These have been calculated in the same way as for \autoref{fig:MTP0014_act}.}
    \label{fig:MTP0020_act}
\end{figure*}

This source was first detected on 2021 April 20 in a CB during an observation of the binary PSR J1930$-$1852 \citep{Swiggum2015} as part of the MeerTIME project's \citep{Bailes2016} relativistic binary programme \citep{Kramer2021a}. A multibeam localisation was performed using a simultaneous L--band detection in three CBs, and \twenty{} has been detected during every targeted observation since. The pulses are characterised as very narrow and predominantly single component, with a small fraction seen with two components. The DM from \scatfit{} of 63.143$\pm$0.009\dm{} is well constrained due to the small width and sustained brightness across the frequency band. The TB was triggered by a pulse on 2022 June 22, which allowed us to localise \twenty{} to an accuracy of approximately 1\arcsec{}. We selected a bright pulse detected in a CB at UHF on 2022 June 22 for imaging and localisation of the \twenty. We created the two images in Figure~\ref{fig:on-off}, "on" and "off", with each covering the same duration.
We identified a transient source in the on-pulse image, as indicated by the magenta circle in Figure~\ref{fig:on-off}. Considering this source appears only at the time of the pulse detection and there is no other transient source in the image, we determine that this is \twenty{}.
Running {\sc pybdsf} on the astrometry corrected image, we found the source position to be RA 19:30:41.88, Dec $-$18:56:28.49.
The total uncertainty after combining all uncertainties in quadrature was 1.3\,arcsec in RA and 1.2\,arcsec in Dec. \\
As with MTP0014 and MTP0017, we found a period using \rratsolve{} on a cluster of pulse arrival times, and used the localisation to fit a timing model in \tempo. The position could not be improved any further, thus was excluded from the fit. The resulting timing solution is given in \autoref{tab:timing_solns}. The $\dot{P}$ of (5.93$\pm$0.07)$\times$10$^{-16}$\,ss$^{-1}$ gives a large characteristic age of 47\,Myr. The residuals that result from the fit, shown in \autoref{fig:residuals}, are distributed across two strips. Based on this and having not detected any pulses with three or more components, we infer that the integrated profile is double-peaked. The trailing component is often narrower, which is reflected in the smaller TOA uncertainties in the upper strip. Some residuals lie in the gap between the strips, which we infer is the result of detecting \twenty{} as the emission switches between the two components. The Gaussian is fitted to both components at the same time and the uncertainties on these TOAs are larger due to the larger Gaussian widths. We demonstrate an example of this effect in \autoref{fig:MTP0020_modeswitch}. Pulse (i) and (ii) are separated by three rotations, whereas pulse (ii) and (iii) are consecutive. The pulses have been aligned by shifting the time series by the timing residual value. We see that the amplitude of one pulse increases while the other decreases, rather than appearing stochastically and independently of each other.  \\ The average detection rate of targeted observations is 25\,hr$^{-1}$:  10\,hr$^{-1}$ at L--band and 26\,hr$^{-1}$ at UHF. The ratio between these of 2.6 is an even more pronounced departure from that expected based on nominal sensitivity that we calculated for MTP0014, but these rates are likely to be very uncertain, especially at L--band where there have been only 2 targeted observations. We repeated the analysis of detection statistics as for MTP0014 and MTP0017, and show the results in \autoref{fig:MTP0020_act}. We do not show the detection tally due to the sparse cadence of targeted observations.
The Weibull fit finds $k$\,=\,0.62$\pm$0.20, which does not strongly suggest the pulses are any more clustered than randomness. There have been no observations of the primary target \mbox{PSR J1930$-$1852}, nor any detections of \twenty{} since 2022 August 27.

\subsubsection*{MTP0021/\twentyone}
\begin{figure}
    \centering
    \includegraphics[width=1.0\columnwidth]{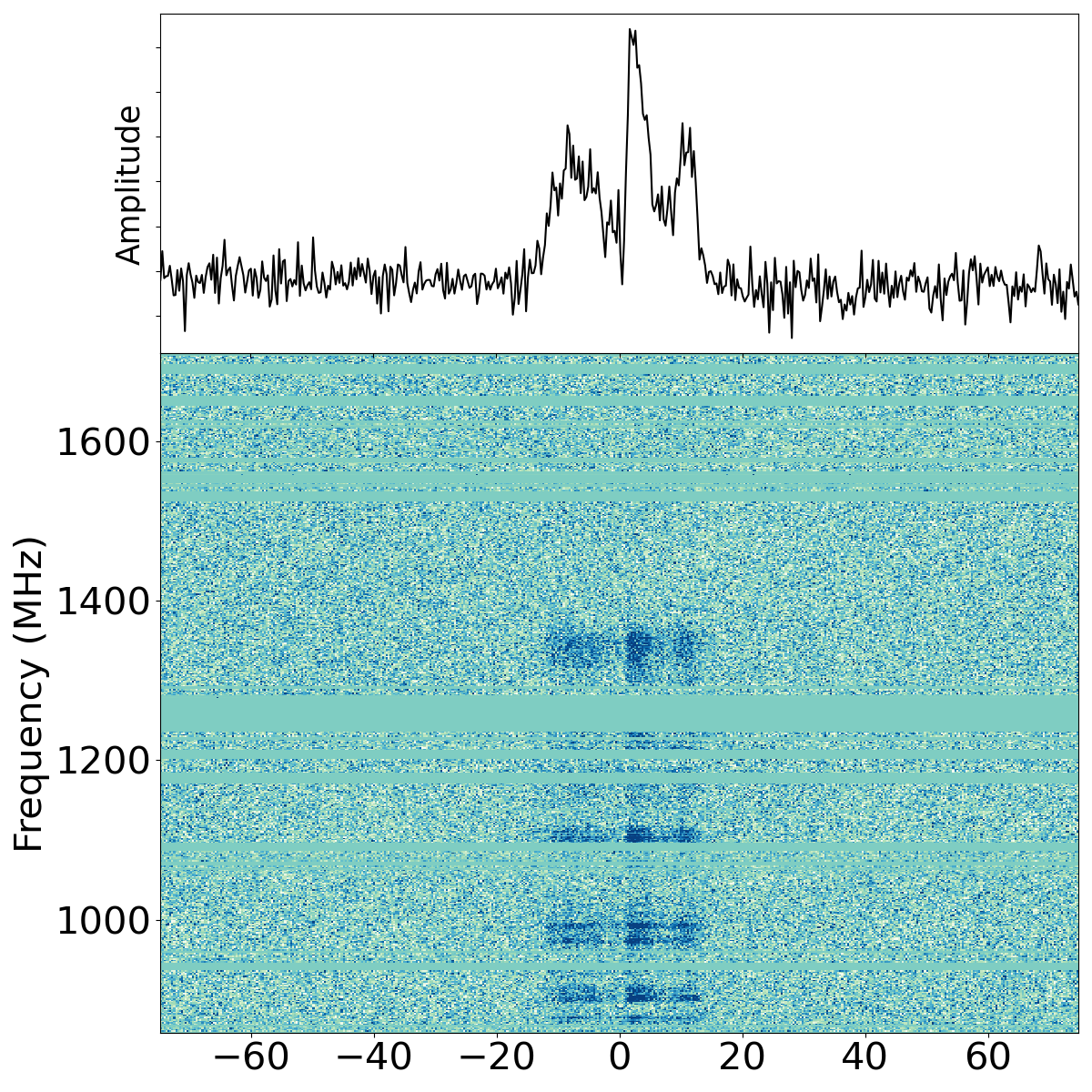}    \includegraphics[width=1.0\columnwidth]{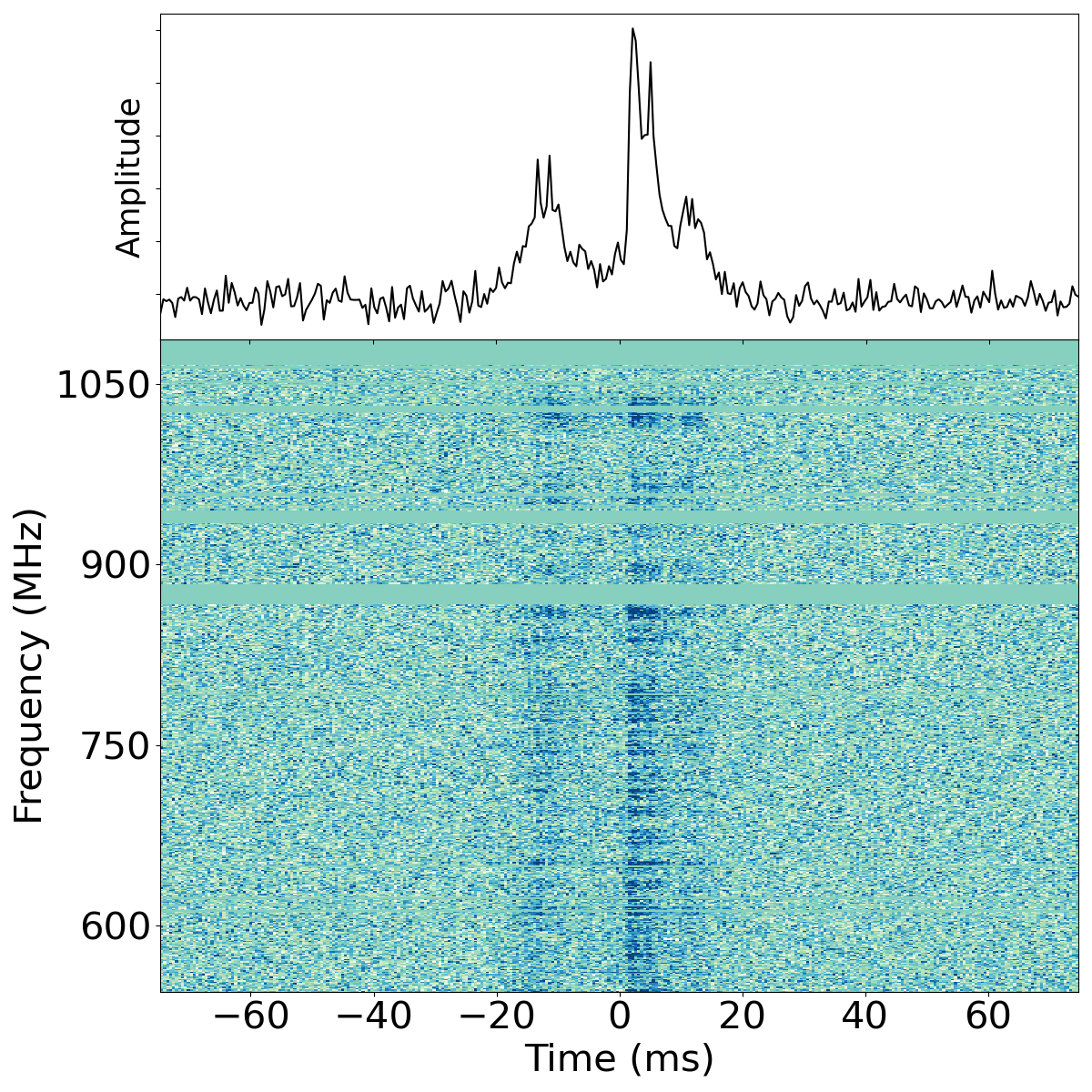}
    \caption{Dynamic spectra (bottom) and frequency-integrated pulse profiles (top) of two pulses from \twentythree: a UHF CB detection on 2022 October 10 (lower plots) and an L--band IB detection on 2024 June 11 (upper plots). The data were dedispersed at a DM of 40.41\dm. The blank horizontal lines are masked channels that were corrupted by RFI.}
    \label{fig:MTP0023-waterfalls}
\end{figure}

\begin{figure}
    \centering
    \includegraphics[width=1.0\columnwidth]{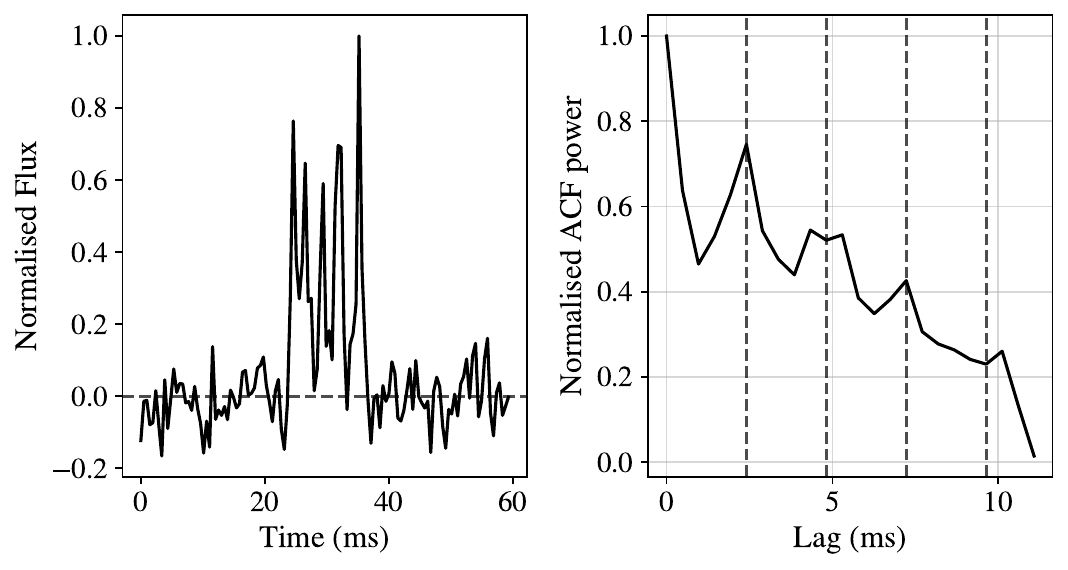}
    \caption{Baseline-subtracted and frequency-integrated pulse profile \textit{(left)} from \twentythree{} on 2022 October 25 and the autocorrelation function of the profile \textit{(right)}. The first dip in the function's power is at the $\approx$1\,ms width of the micropulses and the periodicity of 2.4\,ms, or 5 time samples, is shown as a dashed line which line up with the function peaks.}
    \label{fig:acf}
\end{figure}
\twentyone{} was first detected in several CBs and the IB on 2021 April 26 as a series of 6 very bright pulses between S/N 27 and 134. \twentyone{} has the lowest DM of any MeerTRAP discovery at 8.46$\pm$0.07\dm, which we obtained using \dmphase{}. It was detected at DM values above the 20\dm{} threshold due to its brightness and prominent substructure. The pulses were brightest in the IB (the first pulse is shown in \autoref{fig:waterfallsB}) and there is no spatial cohesion between the five CBs, so those are very likely sidelobe detections. \twentyone{} should therefore be located in the region outside the area filled with CBs. The pulse profile consists of two components with microstructure present. The flux deficit around the pulse is an effect of the zero-DM filter removing mean channel values that have been boosted by the brightness of the pulse. This has caused the frequency-averaged flux to dip below the noise level. The observation, which was part of the MIGHTEE \citep[MeerKAT 
International GHz Tiered Extragalactic Exploration;][]{Taylor2017} survey, lasted approximately 6\,hrs. However, the pulses were all detected within 6 minutes. Using \rratsolve{} we find a period of 1.88\,s, ruling out any possibility that this source is an LPT. Nevertheless, the pulse profile and frequency evolution bear a striking resemblance to those of the 76-second pulsar \citep{Caleb2022}. There are also similarities with the radio pulses of magnetars like XTE J1810$-$197 \citep{Caleb2022b, Bause2024}. The lack of a localisation has precluded multiwavelength follow-up observations thus far. The 20\dm{} threshold explains why we only detected pulses of high S/N, as they had sufficient brightness to be seen at higher DMs. We likely missed a population of fainter pulses, unless \twentyone{} has an unusual burst energy distribution. The exceedingly low average detection rate of 0.6\,hr$^{-1}$ should then serve as a lower limit. Both Galactic electron density models predict a distance of about 400\,pc.

\subsubsection*{MTP0022}
MTP0022 was discovered when two pulses were seen 189\,s apart in a single coherent beam on 2021 May 10. The detections were during a MeerTIME observation of the binary pulsar PSR J1244$-$6359 \citep{Ng2015}. The first pulse seen is shown in \autoref{fig:waterfallsB} and happens to be the brightest. We used \dmphase{} to measure a DM of 342$\pm$2\dm{}. The pulse is very wide at 0.2-0.3\,s towards the bottom of the frequency band, suggests a long rotational period of perhaps several seconds. MTP0022 is probably located within the CB, so the positional uncertainty is defined by the beam shape, which was 55\arcsec{}$\times$37\arcsec{} at an angle of 17.8\degr. MTP0022 has not been seen again in approximately 49\,minutes of targeted observations of PSR J1244$-$6359.

\subsubsection*{MTP0023/\twentythree}
The first pulse detected from MTP0023/\twentythree{}, shown in \autoref{fig:waterfallsB}, was on 2021 June 4 at L--band. It was bright enough to be detected with a S/N of 18 in the IB and the sidelobes of six CBs. Several more pulses were detected in the IB over $\sim$\,3 hours, which allowed us to derive a period of 6.18\,s using \rratsolve{}. We have since detected a population of nearly 100 pulses from this source, exclusively during observations of the gain calibrator \mbox{J1318$-$4620 | 1315$-$46}. We find a diverse set of pulse shapes; some pulses are singly peaked, some show multiple components and some also or exclusively exhibit microstructure. To measure the DM, we ultimately used \dmphase{} with a bright multi-component pulse rather than the outright brightest pulse which was single peaked. The profile and dynamic spectrum of the pulse used is shown in the lower plot of \autoref{fig:MTP0023-waterfalls}, which has three main components and some microstructure, and returns a DM of 40.41$\pm$0.08\dm{}. The average detection rate across targeted observations at UHF is 26\,hr$^{-1}$, similar to that of MTP0020. They also share similar spin periods and both have high pulse-to-pulse variability. \twentythree{} has been detected independently by the CRACO project \citep{Wang2024b}. During observations for the Evolutionary Map of the Universe \citep[EMU;][]{Norris2021} survey at 799.5-1087.5\,MHz, \citet{Wang2024b} detected 20 pulses in a total of 10 hours. The period, DM and position they inferred from their detections are consistent MeerTRAP's measurements. \\
We were able to localise \twentythree{} using \seekat{} for a UHF multibeam detection. The threshold on the angular distance from the primary target for a targeted CB to be deployed is the size of the IB at the middle of the band. \twentythree{} lies within the UHF IB but not the L--band IB, therefore it has only targeted during UHF observations. The detection rate and wait time statistics of these observations are shown in \autoref{fig:MTP0020_act}. The Weibull fit to the wait times has a large uncertainty but the best fit is almost consistent with the Poissonian case of $k$\,=\,1. A UHF CB detection of \twentythree{} on 2024 May 13 triggered the TB, which improved the positional uncertainty down to just 1.3\arcsec. Ultimately we achieved a precision of $<$0.1\arcsec{} by fitting a timing solution, which is provided in \autoref{tab:timing_solns}. From the measured $\dot{P}$ of 6.97$\times$10$^{-15}$ss$^{-1}$ we find \twentythree{} to be the youngest of the five RRATs we have timed at an age of $\sim$\,4\,Myr, and that of the largest spin-down luminosity. \\
\twentythree{} also often exhibits extremely variable emission across the frequency band characteristic of diffractive scintillation by the interstellar medium.
In the upper plot of \autoref{fig:MTP0023-waterfalls}, we show a similar triple-peaked pulse to that in the lower plot that was detected nearly 2 years later in the IB at L--band. The frequency overlap between L--band and UHF makes clear the increasingly wide scintillation bandwidth at higher frequencies. The scintillation bandwidth, $\Delta\nu_{\text{s}}$ is defined as the frequency width of a scintle, and relates to the scattering timescale, $\tau_{\text{s}}$ of the pulse as $2\uppi\Delta\nu_{\text{s}}\tau_{\text{s}}\,\sim\,1$ \citep{Cordes1998}. We were unable to measure $\tau_\text{s}$, so instead we estimate it's value using the empirical dependence on DM derived by \citet{Lewandowski2015}. Using these relations, we find the predicted values to be $\sim$\,12\,MHz at 816\,MHz and $\sim$\,70\,MHz at 1284\,MHz, which appear to agree with the flux variation visible in these pulses. We note the very sharp drop in flux above 1400\,MHz which differs from the L--band IB discovery pulse in \autoref{fig:waterfallsB}, where emission can be seen above this frequency. \twentythree{} is located just beyond the edge of the L--band IB, so the frequency dependence of the IB size results in a drop in sensitivity at higher frequencies. We suspect the presence of the sharp cut-off is a combination of an intrinsically weaker pulse with stronger amplification in scintillation bands below 1400\,MHz. We have also analysed the microstructure and see strong evidence that it has a periodicity. In \autoref{fig:acf} we show the frequency-integrated profile of a pulse consisting solely of micropulses seen on 2022 October 25. We also show the autocorrelation function (ACF) of this profile. The periodicity of 5 bins or 2.4\,ms compared to the spin period agrees very well with the empirical relation found by \citet{Kramer2024} which predicts a micropulse periodicity of $\sim$\,0.001$P$. The same periodic microstructure is also visible in the pulse in the lower plot of \autoref{fig:MTP0023-waterfalls}.

\subsubsection*{MTP0024}
MTP0024 was detected on 2021 June 7 and 8 during a TRAPUM \citep{Stappers2016b} observation of the accreting millisecond pulsar \mbox{SAX J1808.4$-$3658}. We show the first pulse in \autoref{fig:waterfallsB} for which we find a DM of 41.0$\pm$0.5\dm{} with \scatfit{}. The detection was in the IB only which makes it certain that the emission does not originate from \mbox{SAX J1808.4$-$3658}. Two pulses were detected, each during the second and third of three 1\,hr observations a day apart. This observation block was repeated in May 2022 but MeerTRAP did not see any more pulses. Across the six observations, the detection rate for MTP0024 is therefore approximately 0.3\,hr$^{-1}$.
\subsubsection*{MTP0031/\thirtyone}
MTP0031/\thirtyone{} was detected in two CBs and the IB during an L--band observation of the Vela Supercluster on 2021 August 22. It was later detected in multiple CBs simultaneously during a later MMGPS--L observation, which enabled a multibeam localisation using \seekat{}. In addition, there were enough pulses detected to measure a spin period of 2.55\,s using \rratsolve, inferring \thirtyone{} to be an RRAT. The pulses are predominantly composed of a main peak preceded by a weaker component of similar width, with a total on-pulse region of $\sim$\,200\,ms. The first pulse shown in \autoref{fig:waterfallsB} shows this structure. The pulse shape has been seen to evolve; the precursor is occasionally brighter than the main peak, and the brighter pulses show a multipeaked structure within the two components. Using the brightest of these, we measured a DM of 97.7$\pm$0.3\dm{} using \dmphase. We later saw \thirtyone{} in the IB of three other MMGPS--L observations where the IB was overlapping this position. The TB was triggered during the most recent of these on 2022 July 10 in the L--band IB. The position from \seekat{} was extremely useful for confirming \thirtyone{} in the wide-field image for two reasons. Firstly, the time difference between the images required by such a broad pulse was large. Then, even though the image S/N is boosted being an IB detection, the pulse had a low time-domain S/N of 8.3. The astrometrically corrected image position is RA 09:17:28.30, Dec $-$42:45:54.51 with an uncertainty of 1.2\,arcsec in both RA and Dec. The on and off images can be found in Figure~\ref{fig:on-off}. We have also detected four pulses from \thirtyone{} during UHF timing observations of the MMGPS--L discovered pulsar \mbox{PSR J0917$-$4413}. We have not been able to target \thirtyone{} with a dedicated CB as it has always been just beyond our angular offset threshold set by the size of the IB. Thus we calculate a non-targeted detection rate of 12\,hr$^{-1}$, though this is not a totally reliable value due to the different angular offsets from the primary target when detections were made. As of now, the long gaps between detections have prevented us from obtaining a phase-connected timing solution.

\subsubsection*{MTP0034/\thirtyfour}
The new RRAT MTP0034/\thirtyfour{} was discovered when a single pulse was detected in the IB of a MMGPS--L observation on 2021 October 26. The pulse, shown in \autoref{fig:waterfallsB} had been detected at a S/N of 8.5. Then in February and March of 2023, 134 pulses were detected in CBs and the IB on three separate days. These were 15-minute L--band observations of a nearby X-ray binary system HD96670 by ThunderKAT \citep{Fender2016}. 
We used the brightest of these to measure the DM with \scatfit{} to be 92.7$\pm$0.4\dm. Using \rratsolve{} and a cluster of pulses seen at L--band on 2023 February 26, we measured a spin period of 1.52\,s. During this same observation, the TB was triggered twice. We localised \thirtyfour{} to RA 11:07:58.56, Dec $-$59:47:01.10 with an uncertainty of 1.8\,arcsec and 1.2\,arcsec in RA and Dec, respectively, using the voltage data of the brightest detection. See Figure~\ref{fig:on-off} for the on and off images of this source. The remarkable detection rate averaged across these observations is 160\,hr$^{-1}$, the largest of all of the reported transients. \thirtyfour{} was detected during all four observations, but we reiterate the caveat in Section \ref{timing} that any non-targeted detection rates serve only as an upper limit given there may be other nearby targets observed where no pulses were seen. HD96670 is approximately 15 arcmin closer to \thirtyfour{} than the IB of the discovery MMGPS--L observation, which would explain why we observed a lower detection rate in the MMGPS--L IB.
\\ We calculated TOAs for all the pulses and, using the image localisation and the spin period, we were able to fit a more accurate period across the ThunderKAT observations in \tempo. It appears that the large number of TOAs delivered a period accurate enough to phase-connect the first detection 500\,days prior. We included the period derivative in the fit to obtain a 2.2$\times$10$^{-16}$\,ss$^{-1}$ significant to only 1-sigma. Detections spanning a greater length of time will be needed to constrain $\dot{P}$ further. We suspect that the true $\dot{P}$ is below the best fit value, as the inferred characteristic age is perhaps unreasonably high. We checked how much the $\dot{P}$ measurement could be affected by apparent spin period variation due to the Earth's orbital motion and an erroneous position. Assuming the RA uncertainty of 1.7\,arcsec is the same in celestial coordinates, the maximum possible change in spin period is 0.8\,$\upmu$s. This would change $\dot{P}$ by 5.2$\times$10$^{-17}$\,ss$^{-1}$, which is smaller than the uncertainty of our measurement. The size of our uncertainty on $\dot{P}$ ultimately limits how we can characterise \thirtyfour, though we have constrained it to be within the RRAT population.

\subsubsection*{MTP0035}
On 2021 October 26 we discovered MTP0035 during a MMGPS--L observation. Two pulses of S/N 12.3 were detected in quick succession in the IB with DM values of 224-225\dm. The first is shown in \autoref{fig:waterfallsB}, and we used \scatfit{} to measure a DM of 224.5$\pm$0.2\dm. The pulses are separated by 3.9548$\pm$0.0003\,s, thus the period could be any integer factor up to this value. The position is still very uncertain as the source has not been detected in coherent mode nor has it triggered the TB.
\subsubsection*{MTP0039/\thirtynine}
MTP0039/\thirtynine{} was first detected on 2021 December 11 at UTC 04:27:02 during an L--band observation of the galaxy PGC3097177 \citep[HYPERLEDA;][]{Patel2018}. Over the duration of approximately 7 hours a total of 31 pulses were seen in CBs only. The first of these is shown in \autoref{fig:waterfallsB}. \thirtynine{} was then detected 9 days later during a repeat of the observation block. We measured a DM of 95.31$\pm$0.09\dm{} using \scatfit{} for the brightest pulse which had a S/N of 36.5 measured with \spyden{}. The spin period was calculated using \rratsolve{} for clusters of pulses from both observations and a consistent value of 1.06\,s was found. We could not perform a multibeam localisation, as there is no instance of a single pulse of \thirtynine{} being detected in more than one CB. This is peculiar as there have been several pulses of S/N $>$ 15 in beams of varying positions and orientations, so one might expect two beams to have coincidentally overlapped at the true position. This suggests that \thirtynine{} is located in the middle of these CB positions. In any case, \thirtynine{} has not been seen since the TB began operating, precluding an image-based localisation. We therefore provide a crude estimate of the position uncertainty as the region covered by the CB of the brightest pulse. This beam was located at RA 15:33:50.46 and Dec $-$56:09:29.7 and had a shape and orientation of 57\arcsec$\times$40\arcsec{} and 345.8\degr{}. Since these detections, MeerTRAP has targeted this position 14 times, but no detections have been seen in $\sim$\,22 minutes. The average detection rate of the non-targeted observations is about 4.7\,hr$^{-1}$. If we consider the burst rate to be a random process, this gives an uncertainty of 2 detections per hour which is consistent with rates too low to expect a detection in 22 minutes. 
\subsubsection*{MTP0042}
MTP0042 has only been detected during a MMGPS--L observation on 2021 October 08. Two pulses were seen in a single CB. We used the brightest of these to find a DM of 250.4$\pm$0.7\dm{} with \scatfit{}. The closest pulsar on the sky with a similar DM is \mbox{PSR J1621$-$5039} at more than 3\degr{} away. MTP0042 is likely located within the CB, which was positioned at RA 16:41:39.43, Dec $-$51:09:00.8 with a size of 46\arcsec$\times$25\arcsec{} and orientation of 8.5\degr. We calculated the difference in time between the TOAs of the two pulses to be 5.5143(5)\,s, and thus set this as the upper limit on the spin period where, as with MTP0035, any integer factor up to and including this value is also valid.
\subsubsection*{MTP0044/\fortyfour}
MTP0044/\fortyfour{} was discovered on 2022 January 02 during a UHF observation of the radio source J2218+2828, as part of the MeerKAT Absorption Line Survey \citep[MALS;][]{Gupta2016}. 9 pulses were seen during approximately 110\,mins, with DM values of 54-56\dm{}. The first is shown in \autoref{fig:waterfallsB} and is also the brightest. With \scatfit{} we fit a refined DM of 55.8$\pm$0.4\dm. Using \rratsolve{} and the TOAs for all pulses, we find the best solution of the period is 17.49616(4)\,s. \fortyfour{} has therefore the third slowest spin period of a confirmed radio-emitting NS, behind the 23.5\,s \mbox{PSR J0250+5854} \citep{Tan2018} and the 76\,s \mbox{PSR J0901$-$4046} \citep{Caleb2022}. More detections are required to confirm the period. MALS had earlier observed the same source with the L--band receivers on 2020 September 02, but unfortunately MeerTRAP was not commensally observing that day.  \\
We have used eight of the pulses to attempt a localisation. Eight, instead of all nine pulses were chosen because they are of a single component, so their S/N values are more reliable. We used \mosaic{} to simulate the CB PSFs and combined them to form an average PSF weighted by the S/N of the respective pulse. We take the maximum of the average PSF to be the best position to be RA 22:18:23.3(19), Dec +29:02:56(33). This should be treated as an average position. The pulses can be grouped into two types; three have broader single-component profiles and six only have one or more narrow subpulses.
Interestingly, the DM exceeds the maximum line of sight value of the \textsc{ymw16} model of 52.9\dm. The DM of the Galactic pulsar \mbox{PSR J2222+2923}, which is approximately 1\degr away, does not exceed either model \citep{Deneva2024} but is within 3 DM units. \fortyfour{} is probably a Galactic source, potentially in the inner Galactic halo unless the electron content is underestimated in this region. We note that its positional uncertainty overlaps the centre of the galaxy cluster \mbox{RM J221826.6+290308.6} \citep{Rykoff2014}, located approximately 46\,arcsec away. 

\subsubsection*{MTP0045/\fortyfive}
MTP0045/\fortyfive{} was detected at L--band on 2021 December 20, during the same observation of the galaxy PGC3097177 during which MTP0039 was seen for the second time. 28 pulses from MTP0045 were detected, and we used the brightest of these to refine the DM to 56.6$\pm$0.8\dm{} using \scatfit. A period of 2.92\,s was calculated using \rratsolve{}. \fortyfive{} was detected again on 2024 May 07 at L--band during the only targeted observation since discovery. One pulse was seen in the central CB of the targeted area, which is consistent given the mean detection rate from the discovery observation of $\sim$\,1\,hr$^{-1}$. This pulse triggered the TB, so we localised \fortyfive{} using the voltage data to RA 15:31:07.98, Dec $-$55:57:28.53 with an uncertainty of 1.4\,arcsec in RA and 1.5\,arcsec in Dec. For the on and off images of this source see Figure~\ref{fig:on-off}.
\subsubsection*{MTP0047}
MTP0047 was first detected as a faint pulse of S/N\,=\,8.3 on 2022 January 16 at UTC 11:18:11 during a MeerTIME observation of \mbox{PSR J1806$-$1154}. The DM refined by \scatfit{} is 152.4$\pm$0.4\dm. MTP0047 has been seen again during observations of \mbox{PSR J1806$-$1154}; one pulse was seen between two non-targeted observations, and six pulses during the nine subsequent observations where MTP0047 was targeted by MeerTRAP. The pulses are all similar in width and frequency dependence as the discovery pulse in \autoref{fig:waterfallsB}. The targeted detection rate of $\sim$\,21\,hr$^{-1}$ is similar to the other new RRATs MTP0020 and MTP0023. The positional uncertainty we can infer is from the discovery beam, which was centred on RA 18:07:10.36, Dec $-$11:51:08.2 with a shape of 63\arcsec$\times$43\arcsec{} angled at $-$49.3\degr. We noticed a small peak approximately 0.705(2)\,s after the pulse detected on 2022 February 19. The feature has a S/N of 5, thus it is not sufficient to claim this separation to be the upper limit on the spin period. However, we note that the two closest pulses were detected 31.850(2)\,s apart, which would be consistent with 45 rotations to 2-sigma.

\subsubsection*{Nine additional sources of single epoch detections}
Some of the new sources have one or a small number of detections and were seen during just one observation. We are therefore not able to infer much about their properties, but provide a summary of these nine sources here in the order they were seen.
During an MMGPS--L observation on 2021 June 25 at UTC 23:05:54, we detected MTP0026 as a faint pulse with a S/N ratio of 7.9. The S/N measured with \spyden{} rises to 9.8 when the data are dedispersed at the \scatfit-refined DM of 206.8\dm. The two DM-predicted distances disagree by nearly 3\,kpc, but nonetheless both would place this source at hundreds of parsecs above the Galactic plane, thus not dissimilar environments. The dedispersed pulse, shown in \autoref{fig:waterfallsB} is narrow at approximately 4\,ms wide. It is visible across the frequency band, so the source is most likely located within the CB in which it was detected. The shape of the beam was 32\arcsec$\times$24\arcsec{} and orientated at an angle of 366.3\degr. \\
MTP0029 was also discovered during an MMGPS--L observation on 2021 August 20. Three pulses were seen 253\,s apart with DM values between 201-202\dm{}. The first pulse was the brightest at S/N 13.9 and was detected at UTC 21:46:33 and is shown in \autoref{fig:waterfallsB}. We refined the DM to 201.4\dm{} using \scatfit{} on this pulse. The CB was centred on RA 16:48:36.46 and Dec $-$51:42:46.5, which had a shape of 71\arcsec{}$\times$38\arcsec{} at an angle of 5.5\degr. The first pulse was also seen at S/N 10.9 in an adjacent beam. A minimum of three beams are required for a robust multibeam localisation. In this observation, \citet{Padmanabh2023} discovered the pulsar PSR J1650$-$5025 at a similar DM of 213\dm. MTP0029 was seen in a CB 78\,arcmin from their pulsar position, and a topocentric period of 59.6\,ms does not connect the TOAs we measure for the MeerTRAP pulses. We therefore exclude the possibility that these pulses originate from \mbox{PSR J1650$-$5025}. \\
MTP0032 was detected as a single pulse during an MMGPS--L observation on 2021 October 12. Similar to MTP0026, the S/N was 7.9. The S/N is boosted to 11.0 after better RFI cleaning and when dedispersed at the \scatfit{}-refined DM of 271.5\dm. The beam was 38\arcsec{}$\times$24\arcsec{} in size, oriented at 336.7\degr and centred on the coordinates RA 14:52:34.91 and Dec $-$62:35:06.0. As can be seen in \autoref{fig:waterfallsB}, the pulse is consistently bright across the band. Even though we do not know the intrinsic pulse spectrum, it remains more likely that MTP0032 is located within the defined beam shape, rather than in a sidelobe. \\
We detected MTP0036 on 2021 October 30 at UTC 01:20:14 during an Open Time observation of the Vela Supercluster at L--band. The pulse was seen in a CB with RA 10:00:58.60, Dec $-$51:45:52.9, a size of 44\arcsec$\times$26\arcsec{} and an orientation of 355.4\degr. Using \scatfit{}, we find measure the DM to be 128.9\dm. The pulse, shown in \autoref{fig:waterfallsB} has a S/N of 9.4, a width of $\sim$\,5\,ms and broadens at lower frequencies. The distances of 3.1\,kpc (\textsc{ne2001}) and 0.4\,kpc (\textsc{ymw16}) predicted by the DM are very discrepant. If we assume MTP0036 is an old pulsar/RRAT then it could be far above the Galactic plane, which would suggest the \textsc{ymw16} distance of 3.1\,kpc is more realistic. The non-targeted detection rate of $\sim$\,18\,hr$^{-1}$ is high amongst the new sources, but the time on source of only 200\,s is too low to conclude if this is indeed a representative value of the intrinsic activity.\\ 
MTP0038 was detected during a 1 hour MeerTIME observation of the relativistic binary pulsar PSR J1727$-$2946 on 2021 December 9 at UTC 06:12:24. The CB position was RA 17:27:25.11, Dec $-$29:42:20.6, the shape was 55\arcsec$\times$26\arcsec{} and the orientation angle was 339.4\degr. The pulse, shown in \autoref{fig:waterfallsB} is dedispersed at the DM of 126.7 which we obtained with \scatfit{}. The pulse has an interesting frequency dependence. The flux is concentrated at the centre of the band. 
This is unlikely to be due to  MTP0038 being located in a sidelobe of the CB because we would expect an IB detection and much more rapid frequency evolution of the flux. Instead, we suggest the frequency dependence could be explained by scintillation. We note the similarity of the frequency evolution to those of bursts seen from repeating FRBs, for example FRB 20240114A \citep{Tian2024, Kumar2024}. We attempted to use the sharpness of the trailing edge to fit for DM using \dmphase, but the S/N was not sufficient, even when the data were downsampled in time. We did not detect MTP0038 in the other four L--band observations of PSR J1727$-$2946 totalling $\sim$\,4\,hours, or in $\sim$\,10.8 hours of observations at UHF. The average detection rate then is 0.6\,hr$^{-1}$, though more observations would be required to confirm if this is due to a constant and intrinsically low activity or due to a highly variable burst rate. \\
MTP0040 was detected in a CB with S/N\,=\,9 during a MMGPS--L observation on 2021 December 16. Using \scatfit{} the DM was refined to 264$\pm$1\dm. The position, shape and orientation of the beam was RA, 13:57:49.48 and Dec $-$65:07:40.7, 38\arcsec$\times$28\arcsec{} and 327.4\degr{} respectively. Since its discovery, MeerTRAP was able to target this position briefly on 2024 February 17, but no pulses were detected in 94$\pm$13\,s.  \\
MTP0046 was discovered on 2022 January 15 at UTC 10:20:15. This was detected during a MMGPS--L observation, in a CB centred on RA 15:58:46.25, Dec $-$48:38:48.4 and of size and orientation of 34\arcsec$\times$24\arcsec and 344.2\degr. We have refined the DM of the pulse using \scatfit{} to 254$\pm$1\dm. The pulsar PSR B1557$-$50 has a very similar DM, but is 2.1\degr away. The pulsar PSR J1554$-$4854 was discovered by \citep{Padmanabh2023} during this observation, and has a DM of 255.6$\pm$0.3 fully consistent with that of MTP0046. However, the position of the pulsar is 46\,arcmin away from the beam MTP0046 was detected in. The frequency evolution of the pulse does not suggest that MTP0046 is a sidelobe detection of either PSR J1554$-$4854 or PSR B1557$-$50, therefore we conclude that this is a new source. \\
MTP0048 was seen as two pulses detected on 2022 February 03 in a single CB. The position of the beam was RA 14:29:17.29, Dec $-$64:01:15.2, the shape was 30\arcsec$\times$24\arcsec{} and the orientation was 311.3\degr. We used \scatfit{} to find a DM of 151.6$\pm$0.5\dm, which is consistent with PSR J1446$-$6405 which was discovered by \citet{Padmanabh2023} in the same observation. However, they localised the pulsar to a position 48\,arcmin away and the spin period of 9\,ms is shorter than the on-pulse region of the MeerTRAP pulses. Therefore we are confident that these are different sources. \\
Finally, MTP0049 was discovered when a single pulse was seen on 2022 February 06 in a CB with the shape 39\arcsec$\times$25\arcsec{} and orientation of 327.5\degr. Using \scatfit{} we measured the DM to be 346.7$\pm$0.7\dm, which is the largest of all the new sources presented here. The observing block was repeated about 4 months later, but no detections of MTP0049 were made.

\section{Discussion}\label{disc}
\begin{figure}
    \centering
    \subfigure{\includegraphics[width=0.95\columnwidth]{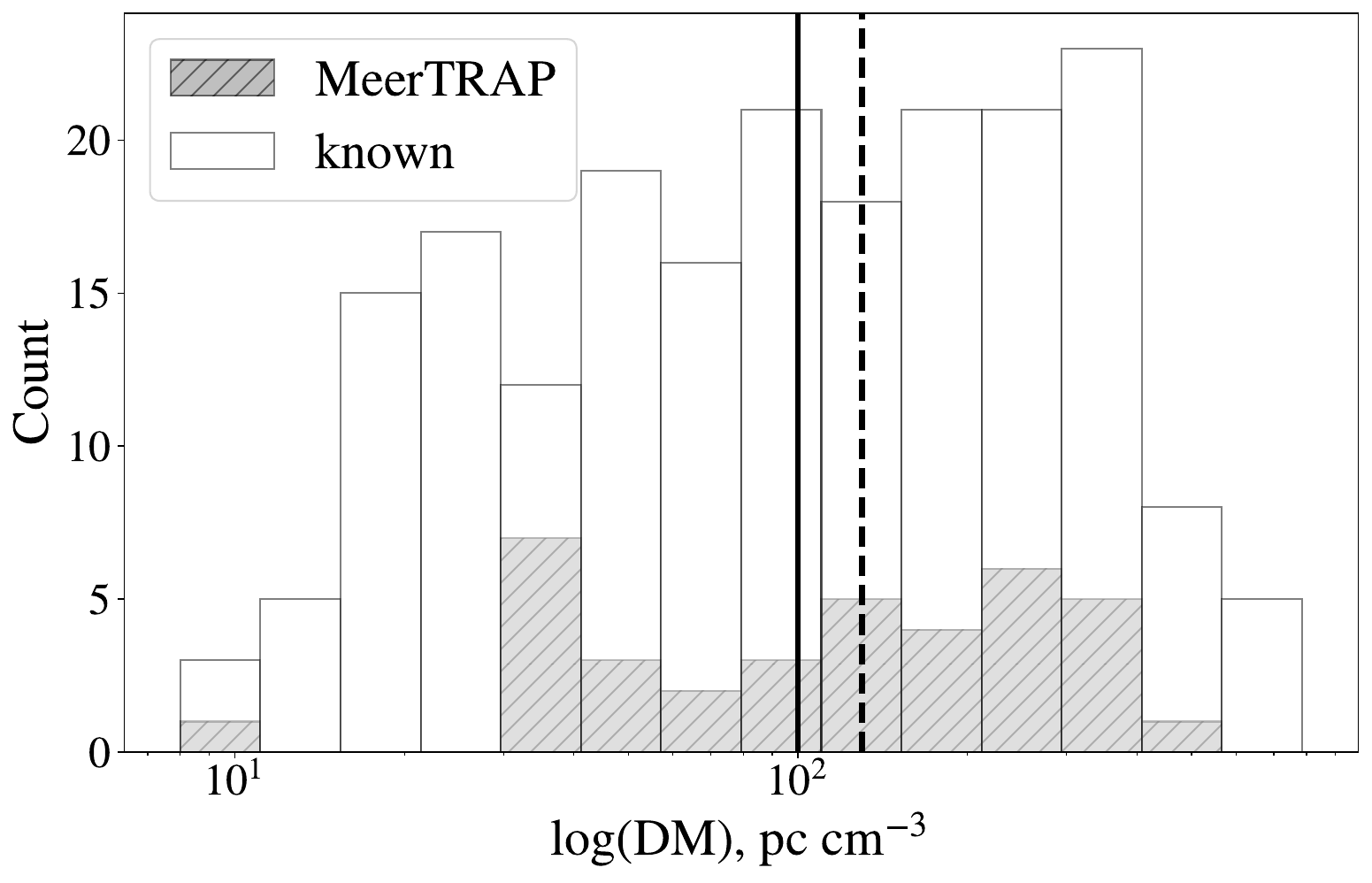}}
    \subfigure{\includegraphics[width=0.95\columnwidth]{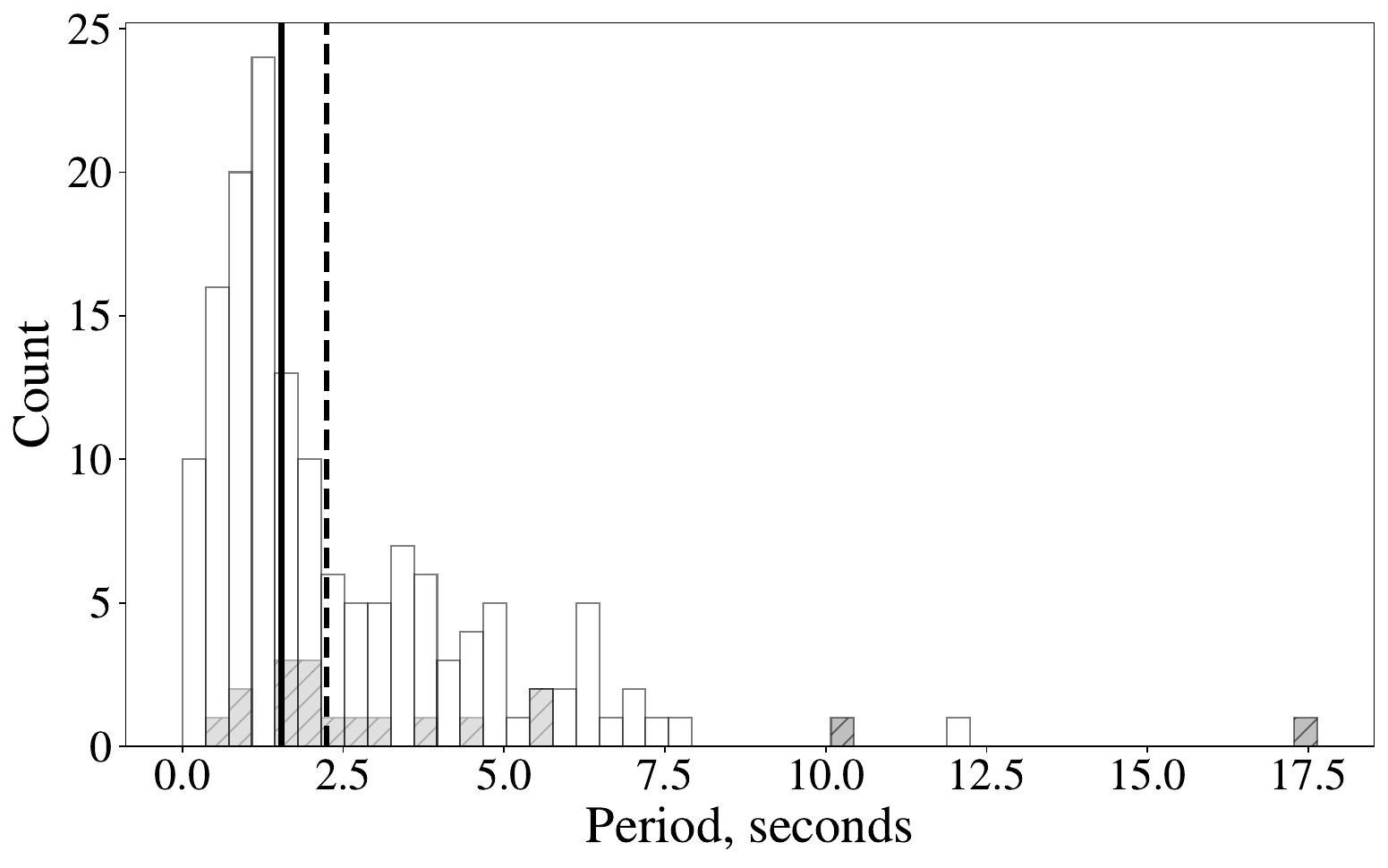}}
	\caption{\textit{Upper}: Histogram of the DM values of the Galactic transients discovered by MeerTRAP (shaded), compared to those of the rest of the RRAT population (un-shaded). The dashed line is the median value of the MeerTRAP DMs and the solid line is that of the known RRATs. \textit{Lower}: Same for the period values of known and the new RRATs. Data for the non-MeerTRAP RRATs is from v2.2.0 of the ATNF Catalogue \citep{Manchester2005}.}
	\label{fig:DMhist}
\end{figure}

\begin{figure}
    \centering
    \includegraphics[width=1.0\columnwidth]{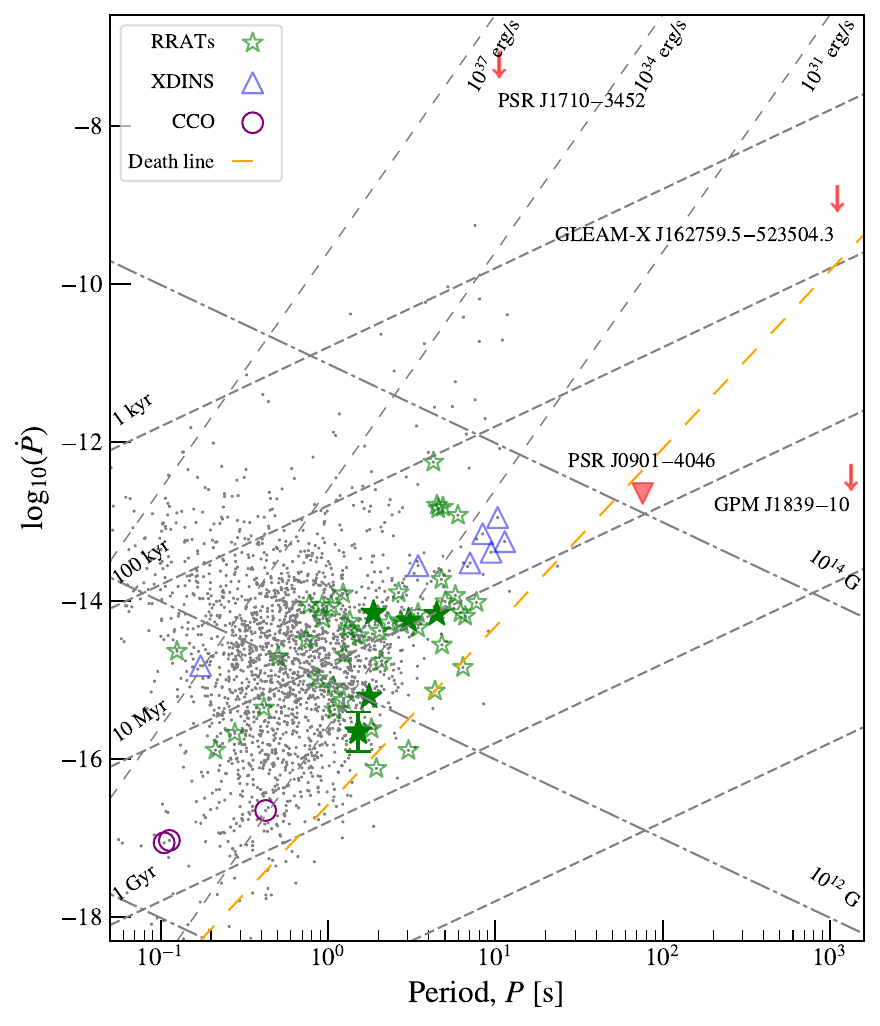}
    \caption{$P$-$\dot{P}$ diagram showing the five RRATs with timing solutions reported in this paper as filled in stars. All other RRATs are unfilled stars. The death line is that of Equation (4) in \citet{Zhang2000}. The long period sources that could be NSs; \mbox{GLEAM-X J162759.5$-$523504.3} \citep{Hurley-Walker2022} and \mbox{GPM J1839$-$10} \citep{Hurley-Walker2023}, or are definitively NSs; \mbox{PSR J0901$-$4046} \citep{Caleb2022}) are shown in red where a down arrow signifies the value is an upper limit on $\dot{P}$. The NS classes of thermal X-ray isolated emitters (XDINS) and the central compact objects (CCOs) are also shown.}\label{fig:ppdot}
\end{figure}
\subsection{The MeerTRAP population}

MeerTRAP has discovered these new sources on a better than monthly cadence, highlighting the importance of such commensal observations. The ability to piggy-back on a diverse set of MeerKAT observations has provided discoveries with a diverse set of properties. In the past, there has been a bias towards discovering RRATs of higher DM due to many previous single pulse surveys focusing on searching the Galactic plane. The sky coverage away from the plane where the line of sight electron content is low has increased the likelihood of discovering low DM sources such as MTP0020, MTP0021 and MTP0044. Long integration times provided by the repeated coverage of primary targets such as calibrators, or of deep field imaging, allow more intermittent sources to be detected. Such sources are often missed by more uniform yet shallower single pass surveys. Nevertheless, a plurality of 12 sources were seen during observations of the MMGPS--L survey. Due to it's focus on the Galactic plane, MMGPS--L has provided the opportunity to sample more distant RRATs than previous surveys, further helped by the high instantaneous sensitivity of MeerKAT. Of these 12 sources, two are IB detections so would not have been visible in their survey at the time we saw them. Of the 10 CB discoveries, eight were not detected by the pulsar searching pipeline. This demonstrates and reinforces the distinguishing property of RRATs and the prediction by \citet{Cordes2003} that some periodic sources will only be detectable in single pulse searches. \\
Interestingly, the fraction of new RRATs discovered by commensal searches on the Galactic Plane Pulsar Snapshot \citep[GPPS;][]{Zhou2023} survey with the Five-hundred meter Aperture Spherical Telescope \citep[FAST;][]{Li2016} compared to pulsars was $76/566$\,=\,0.12 at the time they were reported\footnote{\href{http://zmtt.bao.ac.cn/GPPS/GPPSv2.11.0.html}{http://zmtt.bao.ac.cn/GPPS/} v2.11.0}. This is similar to the equivalent fraction for our detections and MMGPS--L pulsar discoveries of $10/78$\,=\,0.13. This is clearly an important consideration for future pulsar searches that commensally searching for single pulses boosts the discovery yield by over 10 per cent. Interestingly, despite advances in survey sensitivities, RRATs continue to be found at a slower rate than normal pulsars, despite their larger predicted population \citep[e.g.][]{Keane2008}. \\
The discovery of the 17.5-second source MTP0044 further demonstrates the important role of single pulse searches. In this period regime, conventional periodicity searches would only integrate over $\sim$50 rotations, thus failing to obtain a signal of sufficient S/N, especially for sources as intermittent as MTP0044. We choose to refrain from labelling MTP0044 as an RRAT as no periodicity searches of the discovery observation data was possible. MTP0044 cannot yet be placed in $P$-$\dot{P}$ space, but is similar in spin period to sources beyond the pulsar death line, for which explanations of their coherent radio emission have been difficult to pin down. The 421-second transient discovered by CHIME \citep{Dong2024} highlights the success of single pulse searches in the period range $10<P<1000$\,s. Above this period range, the on-pulse region becomes too wide for pulsar-like duty cycles. Despite this, these searches may yet be sensitive to microstructure within pulses of even slower NSs, such as the LPTs identified by image-domain searches with the SKA percursors ASKAP \citep[VAST;][]{Murphy2021} and the MWA \citep[GLEAM-X;][]{Hurley-Walker2022b}. \\
MeerTRAP sources show some exceptionality compared to the bulk of the known RRAT population. A histogram of DMs of all Galactic transients from MeerTRAP compared to the rest of the RRAT population is shown in \autoref{fig:DMhist}. The general trend suggested by \citet{Bezuidenhout2022} that MeerTRAP is sampling a population of larger DMs remains true. The MeerTRAP sources have a median value of 130\dm{}, higher than that of 100\dm{} for known RRATs. Further to this, both MTP0018 and MTP0044 exceed the Galactic DM contribution predicted by the {\textsc ymw16} model. Unless the Galactic electron density is being underestimated along these lines of sight, MeerTRAP could be sensitive to a sample of Galactic halo transients. Curiously, despite DM values exceeding 400\dm{}, we were not able to measure a scattering timescale for any of the brightest pulses from these sources. \\
The five sources for which we have found timing solutions are shown on the $P$-$\dot{P}$ plot in \autoref{fig:ppdot}. The distribution of these sources and the known RRAT population when compared with the characteristic age, magnetic field strength and spin-down luminosity contours suggests that MeerTRAP is sampling RRATs with a lower spin-down luminosity, but not any particular extremity in magnetic field strength or age. It is not clear on this plot what the distribution in $P$ of each population is, so we show same comparison as before but for spin period in the lower plot of \autoref{fig:DMhist}. The distributions tentatively show that MeerTRAP may be sampling the more slowly rotating NS population, as the median of the MeerTRAP sources is 2.4\,s compared to the value of 1.5\,s for the known RRATs. \\

\begin{figure*}
    \centering
    \includegraphics[width=0.99\textwidth]{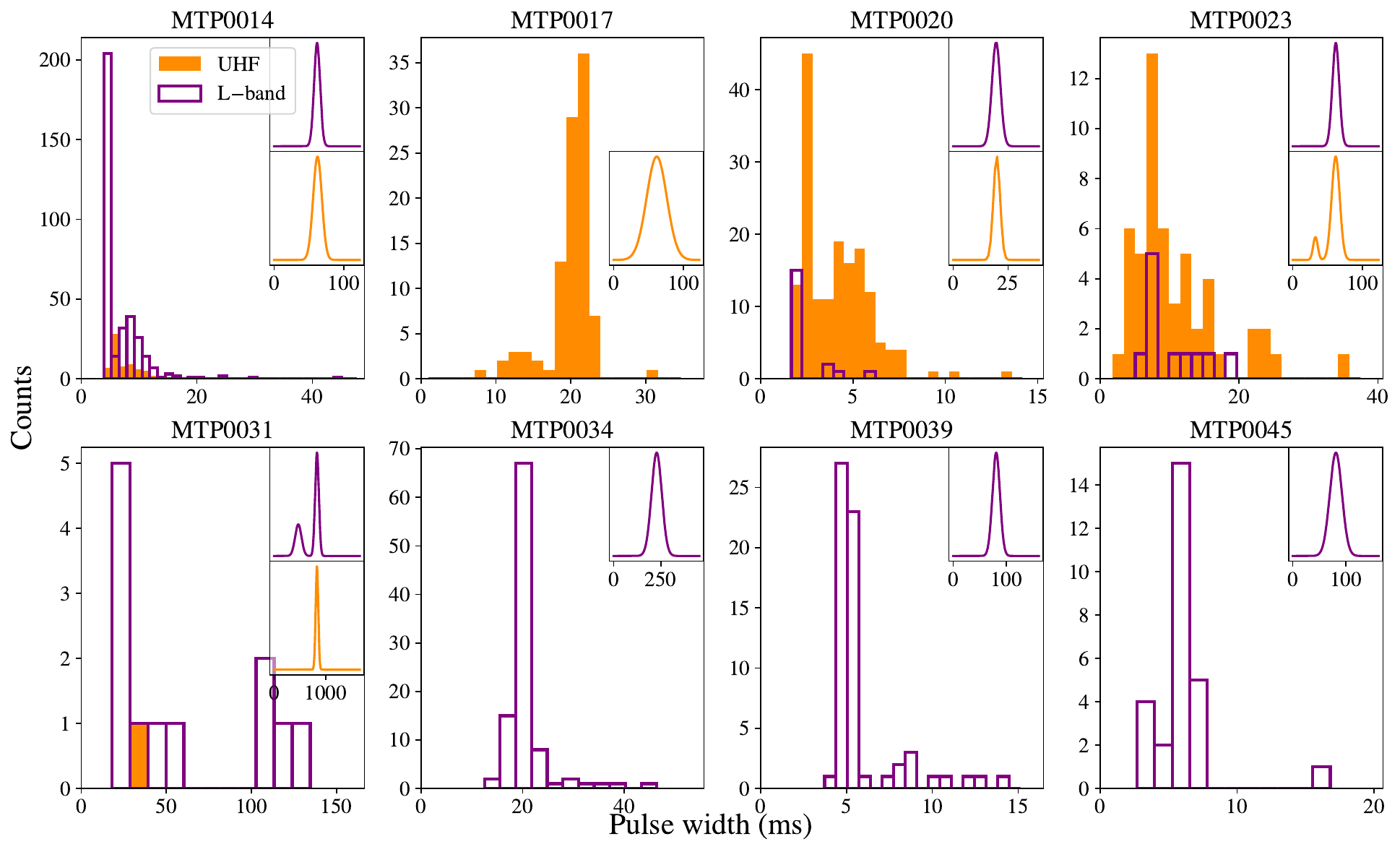}
    \caption{Histograms of $w_{50}$ in milliseconds for eight of the new RRATs. The templates for the respective band they have been fitted to are shown on the right flank of each panel, also in units of milliseconds. The widths have not been scaled to a common frequency.}
    \label{fig:widths}
\end{figure*}

\subsection{RRAT timing and pulse widths}
Our method for measuring TOAs produced accurate and precise arrival times. Once a spin period accurate to a few hundred microseconds was found, we found that sources with many detections such as MTP0014 and MTP0020 did not require a position more accurate than that provided by a CB detection to phase connect pulses several weeks apart. This is despite pronounced pulse jitter of $\sim$\,30\,ms due to pulse-to-pulse profile variation or the variable compatibility of the Gaussian with the pulse shapes. However, localisations of near arcsecond precision provided by \seekat{} or imaging were essential for phase-connecting the TOAs of sources with fewer and less frequent detections, e.g. MTP0023 and MTP0034. We did not use our timing solutions to make integrated pulse profiles of the five timed RRATs. Instead, we propose that the histogram of residuals in \autoref{fig:residuals} can potentially serve as a proxy for this. An obvious exception is MTP0023; the residual distribution of MTP0023 is unimodal, thus does not reflect the presence of multiple components that we repeatedly observe. We suspect that, unlike for the other four timed RRATs, the overlap between components is too great to allow the TOAs measured in this way to align into residual bands. To check how robust the other residual distributions are at showing the average pulse profile, we calculated the widths of the pulses for these five RRATs. We also do this for three other new RRATs with a sufficiently large population of pulses: MTP0031, MTP0039 and MTP0045. \\ 
To measure the pulse width, specifically the width at 50 per cent of the peak flux, $w_{50}$, we use the \texttt{fitvonMises} function in the pulsar data analysis package \salsa\footnote{\href{https://github.com/weltevrede/psrsalsa}{https://github.com/weltevrede/psrsalsa} by Patrick Weltevrede} \citep{Weltevrede2016}. The data are reduced using tools from the pulsar data analysis package \psrchive\footnote{\url{https://psrchive.sourceforge.net/index.shtml}}~\citep{Hotan2004,VanStraten2011}. We start by using \psrchive/\texttt{paas} to make a noise-free template of the brightest pulse from the source at each frequency band it has been detected at. Then, for each pulse, we produce its dedispersed time series by using \psrchive/\dspsr{} and \psrchive/\texttt{pam} to convert and downsample the data. Then, we interactively select the off-pulse region before fitting the template to the pulse with the \salsa/\texttt{fitvonMises} function, which fits the pulse shape by refining the concentration parameter of the von-Mises distribution. The model has noise added to it using the noise statistics of the off-pulse region we had selected. Finally, we find $w_{50}$ and its uncertainty with \texttt{fitvonMises}, bootstrapped with 100 iterative fits of the model to the pulse. For each pulse, we make 10 such measurements of $w_{50}$ and calculate the mean weighted by their uncertainties to obtain our final value for $w_{50}$. Pulses for which the data are too badly affected by RFI or where the pulse is too faint for the template to be fitted are rejected. \\

We found that fitting >$N$ von-Mises functions to a pulse profile with $N$ components became problematic as the spare fit component(s) would find power within noise features and disrupt the fit. As a result, we are mostly measuring widths of individual components, rather than the average profile. For sources which have only shown single component pulses like MTP0017, MTP0034, MTP0039 and MTP0045, we assumed that the average pulse profile is dominated by, or entirely consists of, a single component. The residual distributions for MTP0017 and MTP0034 suggests this is true for these sources and supports this assumption. We also base the assumption on having not actually seen a multicomponent pulse profile from these four RRATs. For MTP0014, we inspected all the detections by eye and found that approximately 10 per cent have a two components, thus fitting a single component template is appropriate for the majority of pulses. Similarly for MTP0020, we know from its timing residuals and component switching that it is rare for more than one component to be present. MTP0023 and MTP0031 have decidedly more complicated pulse-to-pulse variation and a two-component template was required for one of the frequency bands. For MTP0023, a majority of UHF pulses had two components, similar in form to the discovery pulse in \autoref{fig:waterfallsB}, whereas the L--band pules were mainly singly-peaked. This is probably due to the increased sensitivity of the UHF targeted observations, not due to frequency-dependent profile changes. A significant proportion of pulses from MTP0031 show the main pulse and precursor, so a two-component template was used.  \\
The $w_{50}$ distributions and the shape of the templates used are shown in \autoref{fig:widths}. The templates are not intended to show frequency evolution of the pulse shape, instead they only show the noise-free model of the brightest pulse of that frequency. The width distribution for MTP0014 at L--band is the most populous and appears to agree with the residual distributions of \autoref{fig:residuals}. We see two distinct peaks; a small range of narrow pulses followed by a broad range of wider pulses. This is to be expected given the relative share of TOAs we detect from each component, and suggests that the leading component is wider than the trailing component(s). The same is seen for MTP0020, though the reverse is true for widths; the trailing component dominates and is the wider of the two. Interestingly, the widths of MTP0017 are bimodal, peaking at approximately 13\,ms and 22\,ms. This is not predicted by the residuals, as their distribution is not clearly bimodal, and there is no preference for smaller error TOAs in residual space. This could mean that there are indeed two components are never concurrent or are closely overlapping such that they are resolved out by the pulse jitter. \\ 
In most of the pulses between MTP0023 and MTP0031, the combined width of two components was not measured as the precursor did not contain enough power. Four pulses from MTP0031 had precursors strong enough to be included in the width measurement, for which we found widths between 103$\pm$6\,ms and 124$\pm$1\,ms. These correspond to a duty cycle of between 4-5 per cent. These width distributions have allowed a better understanding of the residual distributions in \autoref{fig:residuals} and how they are affected by the pulse-to-pulse variability unique to each RRAT. We note that the ability to measure widths of only a few milliseconds demonstrates the importance of retaining sub-millisecond sampling times for single pulse data. 

\section{Conclusions}\label{conc}
We have presented the latest 26 Galactic transients to be discovered by MeerTRAP, following the 14 that have already been published \citep{Bezuidenhout2022, Caleb2022, Surnis2023}. The majority are identified as members of the RRAT class of neutron stars. We also presented detections of three sources independently discovered elsewhere, including two pulsars that were detected whilst piggy-backing on the MMGPS--L observations that were later identified in periodicity searches. We calculated arrival times and found phase-connected timing solutions for five RRATs and estimate spin periods for an additional eight. For MTP0044/\fortyfour{}, we found a spin period of 17.5\,s, adding to a growing population of slowly rotating neutron stars that challenge magnetospheric radio emission models. This continues the trend of single pulse searches uncovering longer period transients. The complex structure, brightness and variability of pulses from MTP0021/\twentyone{} and MTP0023/\twentythree{} recommend them for further study. MeerTRAP is sampling a population of higher DM, longer period Galactic transients. The DMs of MTP0018 and MTP0044/\fortyfour{} infer their locations could be in the Galactic halo depending on the predicted electron density along their line of sight. The large fraction of sources to have been seen only further hints at existence of a significant population of these sources within the Milky Way Galaxy. The importance of commensal single pulse searches operating as close to full-time as possible continues to be demonstrated.

\section*{Acknowledgements}
The MeerKAT telescope is operated by the South African Radio Astronomy Observatory (SARAO), which is a facility of the National Research Foundation, itself an agency of the Department of Science and Innovation. All the authors thank the MeerKAT LSP teams for allowing commensal observing and the staff at SARAO for scheduling MeerKAT observations. MeerTRAP observations use the FBFUSE and TUSE computing clusters for data acquisition and storage. These instruments were designed, funded and installed by the Max-Planck Institut f{\"u}r Radioastronomie (MPIfR) and the Max-Planck-Gesellschaft.
This project has received funding from the European Research Council (ERC) under the European Union’s Horizon 2020 research and innovation programme (grant agreement no. 694745). JDT acknowledges funding from the United Kingdom's Research and Innovation Science and Technology Facilities Council (STFC) Doctoral Training Partnership, project code 2659479. M.C. acknowledges support of an Australian Research Council Discovery Early Career Research Award (project number DE220100819) funded by the Australian Government.
IPM further acknowledges funding from an NWO Rubicon Fellowship, project number 019.221EN.019.
For the purpose of open access, the author has applied a Creative Commons Attribution (CC BY) licence to any Author Accepted Manuscript version arising.
J.D.T. thanks Dr. O. G. Dodge for his help with scripting.
The authors would like to thank the reviewer for their helpful comments and recommendations that improved this publication.
This research used version 2.2.0 of the ATNF Pulsar Catalogue.
This research has made use of the SIMBAD data base, operated at CDS, Strasbourg, France \citep{Wenger2000}. This research has made use of NASA’s Astrophysics Data System Bibliographic Services.
\section*{Data Availability}
The data underlying this article are available in Zenodo, at \href{https://doi.org/10.5281/zenodo.14646142}{10.5281/zenodo.14646142}. Other data will be made available to those with a reasonable request to the authors.

\bibliographystyle{mnras}
\bibliography{main} 

\begin{thebibliography}{}
\makeatletter
\relax
\def\mn@urlcharsother{\let\do\@makeother \do\$\do\&\do\#\do\^\do\_\do\%\do\~}
\def\mn@doi{\begingroup\mn@urlcharsother \@ifnextchar [ {\mn@doi@} {\mn@doi@[]}}
\def\mn@doi@[#1]#2{\def\@tempa{#1}\ifx\@tempa\@empty \href {http://dx.doi.org/#2} {doi:#2}\else \href {http://dx.doi.org/#2} {#1}\fi \endgroup}
\def\mn@eprint#1#2{\mn@eprint@#1:#2::\@nil}
\def\mn@eprint@arXiv#1{\href {http://arxiv.org/abs/#1} {{\tt arXiv:#1}}}
\def\mn@eprint@dblp#1{\href {http://dblp.uni-trier.de/rec/bibtex/#1.xml} {dblp:#1}}
\def\mn@eprint@#1:#2:#3:#4\@nil{\def\@tempa {#1}\def\@tempb {#2}\def\@tempc {#3}\ifx \@tempc \@empty \let \@tempc \@tempb \let \@tempb \@tempa \fi \ifx \@tempb \@empty \def\@tempb {arXiv}\fi \@ifundefined {mn@eprint@\@tempb}{\@tempb:\@tempc}{\expandafter \expandafter \csname mn@eprint@\@tempb\endcsname \expandafter{\@tempc}}}

\bibitem[\protect\citeauthoryear{{Bailes} et~al.,}{{Bailes} et~al.}{2016}]{Bailes2016}
{Bailes} M.,  et~al., 2016, in MeerKAT Science: On the Pathway to the SKA. p.~11 (\mn@eprint {arXiv} {1803.07424}), \mn@doi{10.22323/1.277.0011}

\bibitem[\protect\citeauthoryear{{Barr}}{{Barr}}{2018}]{Barr2018}
{Barr} E.~D.,  2018, in {Weltevrede} P.,  {Perera} B.~B.~P.,  {Preston} L.~L.,   {Sanidas} S.,  eds, ~ Vol. 337, Pulsar Astrophysics the Next Fifty Years. pp 175--178, \mn@doi{10.1017/S1743921317009036}

\bibitem[\protect\citeauthoryear{{Bause}, {Herrmann}  \& {Spitler}}{{Bause} et~al.}{2024}]{Bause2024}
{Bause} M.~L.,  {Herrmann} W.,   {Spitler} L.~G.,  2024, \mn@doi [\aap] {10.1051/0004-6361/202348878}, \href {https://ui.adsabs.harvard.edu/abs/2024A&A...686A.144B} {686, A144}

\bibitem[\protect\citeauthoryear{{Beniamini}, {Wadiasingh}, {Hare}, {Rajwade}, {Younes}  \& {van der Horst}}{{Beniamini} et~al.}{2023}]{Beniamini2023}
{Beniamini} P.,  {Wadiasingh} Z.,  {Hare} J.,  {Rajwade} K.~M.,  {Younes} G.,   {van der Horst} A.~J.,  2023, \mn@doi [\mnras] {10.1093/mnras/stad208}, \href {https://ui.adsabs.harvard.edu/abs/2023MNRAS.520.1872B} {520, 1872}

\bibitem[\protect\citeauthoryear{Bezuidenhout et~al.,}{Bezuidenhout et~al.}{2022}]{Bezuidenhout2022}
Bezuidenhout M.~C.,  et~al., 2022, \mn@doi [\mnras] {10.1093/mnras/stac579}, 512

\bibitem[\protect\citeauthoryear{Bezuidenhout et~al.,}{Bezuidenhout et~al.}{2023}]{Bezuidenhout2023}
Bezuidenhout M.~C.,  et~al., 2023, \mn@doi [RAS Techniques and Instruments] {10.1093/rasti/rzad007}, 2, 114

\bibitem[\protect\citeauthoryear{{Bhattacharyya} et~al.,}{{Bhattacharyya} et~al.}{2018}]{Bhattacharyya2018}
{Bhattacharyya} B.,  et~al., 2018, \mn@doi [\mnras] {10.1093/mnras/sty923}, \href {https://ui.adsabs.harvard.edu/abs/2018MNRAS.477.4090B} {477, 4090}

\bibitem[\protect\citeauthoryear{{Burke-Spolaor} \& {Bailes}}{{Burke-Spolaor} \& {Bailes}}{2010}]{Burke-Spolaor2010}
{Burke-Spolaor} S.,  {Bailes} M.,  2010, \mn@doi [\mnras] {10.1111/j.1365-2966.2009.15965.x}, \href {https://ui.adsabs.harvard.edu/abs/2010MNRAS.402..855B} {402, 855}

\bibitem[\protect\citeauthoryear{{Burke-Spolaor} et~al.,}{{Burke-Spolaor} et~al.}{2011}]{Burke-Spolaor2011}
{Burke-Spolaor} S.,  et~al., 2011, \mn@doi [\mnras] {10.1111/j.1365-2966.2011.18521.x}, \href {https://ui.adsabs.harvard.edu/abs/2011MNRAS.416.2465B} {416, 2465}

\bibitem[\protect\citeauthoryear{{CHIME/FRB Collaboration} et~al.,}{{CHIME/FRB Collaboration} et~al.}{2018}]{CHIME2018}
{CHIME/FRB Collaboration} et~al., 2018, \mn@doi [\apj] {10.3847/1538-4357/aad188}, \href {https://ui.adsabs.harvard.edu/abs/2018ApJ...863...48C} {863, 48}

\bibitem[\protect\citeauthoryear{Caleb et~al.,}{Caleb et~al.}{2022a}]{Caleb2022}
Caleb M.,  et~al., 2022a, \mn@doi [Nature Astronomy] {10.1038/s41550-022-01688-x}, 6, 1

\bibitem[\protect\citeauthoryear{{Caleb} et~al.,}{{Caleb} et~al.}{2022b}]{Caleb2022b}
{Caleb} M.,  et~al., 2022b, \mn@doi [\mnras] {10.1093/mnras/stab3223}, \href {https://ui.adsabs.harvard.edu/abs/2022MNRAS.510.1996C} {510, 1996}

\bibitem[\protect\citeauthoryear{{Caleb} et~al.,}{{Caleb} et~al.}{2024}]{Caleb2024}
{Caleb} M.,  et~al., 2024, \mn@doi [Nature Astronomy] {10.1038/s41550-024-02277-w}, \href {https://ui.adsabs.harvard.edu/abs/2024NatAs.tmp..107C} {}

\bibitem[\protect\citeauthoryear{Camilo et~al.,}{Camilo et~al.}{2018}]{Camilo2018a}
Camilo F.,  et~al., 2018, \mn@doi [\apj] {10.3847/1538-4357/aab35a}, 856, 180

\bibitem[\protect\citeauthoryear{Chen, Barr, Karuppusamy, Kramer  \& Stappers}{Chen et~al.}{2021}]{Chen2021}
Chen W.,  Barr E.,  Karuppusamy R.,  Kramer M.,   Stappers B.,  2021, \mn@doi [Journal of Astronomical Instrumentation] {10.1142/S2251171721500136}, 10, 2150013

\bibitem[\protect\citeauthoryear{{Cordes} \& {Chatterjee}}{{Cordes} \& {Chatterjee}}{2019}]{Cordes2019}
{Cordes} J.~M.,  {Chatterjee} S.,  2019, \mn@doi [\araa] {10.1146/annurev-astro-091918-104501}, \href {https://ui.adsabs.harvard.edu/abs/2019ARA&A..57..417C} {57, 417}

\bibitem[\protect\citeauthoryear{Cordes \& Lazio}{Cordes \& Lazio}{2002}]{Cordes2001}
Cordes J.~M.,  Lazio T.~J.~W.,  2002, \mn@doi [arXiv e-prints] {10.48550/arXiv.astro-ph/0207156}, pp astro--ph/0207156

\bibitem[\protect\citeauthoryear{{Cordes} \& {McLaughlin}}{{Cordes} \& {McLaughlin}}{2003}]{Cordes2003}
{Cordes} J.~M.,  {McLaughlin} M.~A.,  2003, \mn@doi [\apj] {10.1086/378231}, \href {https://ui.adsabs.harvard.edu/abs/2003ApJ...596.1142C} {596, 1142}

\bibitem[\protect\citeauthoryear{{Cordes} \& {Rickett}}{{Cordes} \& {Rickett}}{1998}]{Cordes1998}
{Cordes} J.~M.,  {Rickett} B.~J.,  1998, \mn@doi [\apj] {10.1086/306358}, \href {https://ui.adsabs.harvard.edu/abs/1998ApJ...507..846C} {507, 846}

\bibitem[\protect\citeauthoryear{{Deneva} et~al.,}{{Deneva} et~al.}{2016}]{Deneva2016}
{Deneva} J.~S.,  et~al., 2016, \mn@doi [\apj] {10.3847/0004-637X/821/1/10}, \href {https://ui.adsabs.harvard.edu/abs/2016ApJ...821...10D} {821, 10}

\bibitem[\protect\citeauthoryear{{Deneva}, {McLaughlin}, {Olszanski}, {Lewis}, {Pang}, {Freire}, {Bagchi}  \& {Stovall}}{{Deneva} et~al.}{2024}]{Deneva2024}
{Deneva} J.~S.,  {McLaughlin} M.,  {Olszanski} T.~E.~E.,  {Lewis} E.~F.,  {Pang} D.,  {Freire} P.~C.~C.,  {Bagchi} M.,   {Stovall} K.,  2024, \mn@doi [\apjs] {10.3847/1538-4365/ad19da}, \href {https://ui.adsabs.harvard.edu/abs/2024ApJS..271...23D} {271, 23}

\bibitem[\protect\citeauthoryear{{Dong} et~al.,}{{Dong} et~al.}{2023}]{Dong2023}
{Dong} F.~A.,  et~al., 2023, \mn@doi [\mnras] {10.1093/mnras/stad2012}, \href {https://ui.adsabs.harvard.edu/abs/2023MNRAS.524.5132D} {524, 5132}

\bibitem[\protect\citeauthoryear{{Dong} et~al.,}{{Dong} et~al.}{2024}]{Dong2024}
{Dong} F.~A.,  et~al., 2024, \mn@doi [arXiv e-prints] {10.48550/arXiv.2407.07480}, \href {https://ui.adsabs.harvard.edu/abs/2024arXiv240707480D} {p. arXiv:2407.07480}

\bibitem[\protect\citeauthoryear{{Driessen} et~al.,}{{Driessen} et~al.}{2022}]{Driessen2022}
{Driessen} L.~N.,  et~al., 2022, \mn@doi [\mnras] {10.1093/mnras/stac756}, \href {https://ui.adsabs.harvard.edu/abs/2022MNRAS.512.5037D} {512, 5037}

\bibitem[\protect\citeauthoryear{{Driessen} et~al.,}{{Driessen} et~al.}{2024}]{Driessen2024}
{Driessen} L.~N.,  et~al., 2024, \mn@doi [\mnras] {10.1093/mnras/stad3329}, \href {https://ui.adsabs.harvard.edu/abs/2024MNRAS.527.3659D} {527, 3659}

\bibitem[\protect\citeauthoryear{{Duchesne} et~al.,}{{Duchesne} et~al.}{2024}]{Duchesne2024}
{Duchesne} S.~W.,  et~al., 2024, \mn@doi [\pasa] {10.1017/pasa.2023.60}, \href {https://ui.adsabs.harvard.edu/abs/2024PASA...41....3D} {41, e003}

\bibitem[\protect\citeauthoryear{Eatough, Keane  \& Lyne}{Eatough et~al.}{2009}]{Eatough2009}
Eatough R.~P.,  Keane E.~F.,   Lyne A.~G.,  2009, \mn@doi [Monthly Notices of the Royal Astronomical Society] {10.1111/j.1365-2966.2009.14524.x}, 395, 410

\bibitem[\protect\citeauthoryear{{Fender} et~al.,}{{Fender} et~al.}{2016}]{Fender2016}
{Fender} R.,  et~al., 2016, in MeerKAT Science: On the Pathway to the SKA. p.~13 (\mn@eprint {arXiv} {1711.04132}), \mn@doi{10.22323/1.277.0013}

\bibitem[\protect\citeauthoryear{{Fiore} et~al.,}{{Fiore} et~al.}{2023}]{Fiore2023}
{Fiore} W.,  et~al., 2023, \mn@doi [\apj] {10.3847/1538-4357/aceef7}, \href {https://ui.adsabs.harvard.edu/abs/2023ApJ...956...40F} {956, 40}

\bibitem[\protect\citeauthoryear{{Gupta} et~al.,}{{Gupta} et~al.}{2016}]{Gupta2016}
{Gupta} N.,  et~al., 2016, in MeerKAT Science: On the Pathway to the SKA. p.~14 (\mn@eprint {arXiv} {1708.07371}), \mn@doi{10.22323/1.277.0014}

\bibitem[\protect\citeauthoryear{{Hale} et~al.,}{{Hale} et~al.}{2021}]{Hale2021}
{Hale} C.~L.,  et~al., 2021, \mn@doi [\pasa] {10.1017/pasa.2021.47}, \href {https://ui.adsabs.harvard.edu/abs/2021PASA...38...58H} {38, e058}

\bibitem[\protect\citeauthoryear{{Hessels} et~al.,}{{Hessels} et~al.}{2019}]{Hessels2019}
{Hessels} J.~W.~T.,  et~al., 2019, \mn@doi [\apjl] {10.3847/2041-8213/ab13ae}, \href {https://ui.adsabs.harvard.edu/abs/2019ApJ...876L..23H} {876, L23}

\bibitem[\protect\citeauthoryear{Hobbs, Edwards  \& Manchester}{Hobbs et~al.}{2006}]{Hobbs2006}
Hobbs G.~B.,  Edwards R.~T.,   Manchester R.~N.,  2006, {TEMPO2, a new pulsar-timing package - I. An overview}, \mn@doi{10.1111/j.1365-2966.2006.10302.x}

\bibitem[\protect\citeauthoryear{{Hobbs}, {Lyne}  \& {Kramer}}{{Hobbs} et~al.}{2010}]{Hobbs2010}
{Hobbs} G.,  {Lyne} A.~G.,   {Kramer} M.,  2010, \mn@doi [\mnras] {10.1111/j.1365-2966.2009.15938.x}, \href {https://ui.adsabs.harvard.edu/abs/2010MNRAS.402.1027H} {402, 1027}

\bibitem[\protect\citeauthoryear{Hosenie}{Hosenie}{2021}]{Hosenie2021}
Hosenie Z.,  2021, PhD thesis, The University of Manchester, \url {https://research.manchester.ac.uk/en/studentTheses/feature-detection-and-classification-in-streaming-and-non-streami}

\bibitem[\protect\citeauthoryear{Hotan, van Straten  \& Manchester}{Hotan et~al.}{2004}]{Hotan2004}
Hotan A.~W.,  van Straten W.,   Manchester R.~N.,  2004, \mn@doi [\pasa] {10.1071/AS04022}, 21, 302

\bibitem[\protect\citeauthoryear{{Hurley-Walker} et~al.,}{{Hurley-Walker} et~al.}{2022a}]{Hurley-Walker2022b}
{Hurley-Walker} N.,  et~al., 2022a, \mn@doi [\pasa] {10.1017/pasa.2022.17}, \href {https://ui.adsabs.harvard.edu/abs/2022PASA...39...35H} {39, e035}

\bibitem[\protect\citeauthoryear{{Hurley-Walker} et~al.,}{{Hurley-Walker} et~al.}{2022b}]{Hurley-Walker2022}
{Hurley-Walker} N.,  et~al., 2022b, \mn@doi [\nat] {10.1038/s41586-021-04272-x}, \href {https://ui.adsabs.harvard.edu/abs/2022Natur.601..526H} {601, 526}

\bibitem[\protect\citeauthoryear{{Hurley-Walker} et~al.,}{{Hurley-Walker} et~al.}{2023}]{Hurley-Walker2023}
{Hurley-Walker} N.,  et~al., 2023, \mn@doi [\nat] {10.1038/s41586-023-06202-5}, \href {https://ui.adsabs.harvard.edu/abs/2023Natur.619..487H} {619, 487}

\bibitem[\protect\citeauthoryear{{Jankowski}}{{Jankowski}}{2022}]{Jankowski2022b}
{Jankowski} F.,  2022, {Scatfit: Scattering fits of time domain radio signals (Fast Radio Bursts or pulsars)}, Astrophysics Source Code Library, record ascl:2208.003

\bibitem[\protect\citeauthoryear{Jankowski, van Straten, Keane, Bailes, Barr, Johnston  \& Kerr}{Jankowski et~al.}{2018}]{Jankowski2018}
Jankowski F.,  van Straten W.,  Keane E.~F.,  Bailes M.,  Barr E.~D.,  Johnston S.,   Kerr M.,  2018, \mn@doi [\mnras] {10.1093/mnras/stx2476}, 473, 4436

\bibitem[\protect\citeauthoryear{{Jankowski} et~al.,}{{Jankowski} et~al.}{2022}]{Jankowski2022}
{Jankowski} F.,  et~al., 2022, in {Ruiz} J.~E.,  {Pierfedereci} F.,   {Teuben} P.,  eds,  Astronomical Society of the Pacific Conference Series Vol. 532, Astronomical Society of the Pacific Conference Series. p.~273 (\mn@eprint {arXiv} {2012.05173}), \mn@doi{10.48550/arXiv.2012.05173}

\bibitem[\protect\citeauthoryear{Jankowski et~al.,}{Jankowski et~al.}{2023}]{Jankowski2023}
Jankowski F.,  et~al., 2023, \mn@doi [\mnras] {10.1093/mnras/stad2041}, 524, 4275

\bibitem[\protect\citeauthoryear{Jonas}{Jonas}{2018}]{Jonas2018}
Jonas J.,  2018, in Proceedings of MeerKAT Science: On the Pathway to the SKA. Proceedings of Science, p.~1, \mn@doi{10.22323/1.277.0001}

\bibitem[\protect\citeauthoryear{{Karako-Argaman} et~al.,}{{Karako-Argaman} et~al.}{2015}]{Karako-Argaman2015}
{Karako-Argaman} C.,  et~al., 2015, \mn@doi [\apj] {10.1088/0004-637X/809/1/67}, \href {https://ui.adsabs.harvard.edu/abs/2015ApJ...809...67K} {809, 67}

\bibitem[\protect\citeauthoryear{{Karuppusamy}, {Stappers}  \& {van Straten}}{{Karuppusamy} et~al.}{2010}]{Karuppusamy2010}
{Karuppusamy} R.,  {Stappers} B.~W.,   {van Straten} W.,  2010, \mn@doi [\aap] {10.1051/0004-6361/200913729}, \href {https://ui.adsabs.harvard.edu/abs/2010A&A...515A..36K} {515, A36}

\bibitem[\protect\citeauthoryear{Keane \& Kramer}{Keane \& Kramer}{2008}]{Keane2008}
Keane E.~F.,  Kramer M.,  2008, \mn@doi [\mnras] {10.1111/j.1365-2966.2008.14045.x}, 391, 2009

\bibitem[\protect\citeauthoryear{{Keane} \& {McLaughlin}}{{Keane} \& {McLaughlin}}{2011}]{Keane2011b}
{Keane} E.~F.,  {McLaughlin} M.~A.,  2011, \mn@doi [Bulletin of the Astronomical Society of India] {10.48550/arXiv.1109.6896}, \href {https://ui.adsabs.harvard.edu/abs/2011BASI...39..333K} {39, 333}

\bibitem[\protect\citeauthoryear{{Keane}, {Ludovici}, {Eatough}, {Kramer}, {Lyne}, {McLaughlin}  \& {Stappers}}{{Keane} et~al.}{2010}]{Keane2010}
{Keane} E.~F.,  {Ludovici} D.~A.,  {Eatough} R.~P.,  {Kramer} M.,  {Lyne} A.~G.,  {McLaughlin} M.~A.,   {Stappers} B.~W.,  2010, \mn@doi [\mnras] {10.1111/j.1365-2966.2009.15693.x}, \href {https://ui.adsabs.harvard.edu/abs/2010MNRAS.401.1057K} {401, 1057}

\bibitem[\protect\citeauthoryear{{Keane} et~al.,}{{Keane} et~al.}{2018}]{Keane2018}
{Keane} E.~F.,  et~al., 2018, \mn@doi [\mnras] {10.1093/mnras/stx2126}, \href {https://ui.adsabs.harvard.edu/abs/2018MNRAS.473..116K} {473, 116}

\bibitem[\protect\citeauthoryear{{Kramer} et~al.,}{{Kramer} et~al.}{2021}]{Kramer2021a}
{Kramer} M.,  et~al., 2021, \mn@doi [\mnras] {10.1093/mnras/stab375}, \href {https://ui.adsabs.harvard.edu/abs/2021MNRAS.504.2094K} {504, 2094}

\bibitem[\protect\citeauthoryear{{Kramer}, {Liu}, {Desvignes}, {Karuppusamy}  \& {Stappers}}{{Kramer} et~al.}{2024}]{Kramer2024}
{Kramer} M.,  {Liu} K.,  {Desvignes} G.,  {Karuppusamy} R.,   {Stappers} B.~W.,  2024, \mn@doi [Nature Astronomy] {10.1038/s41550-023-02125-3}, \href {https://ui.adsabs.harvard.edu/abs/2024NatAs...8..230K} {8, 230}

\bibitem[\protect\citeauthoryear{{Kumar}, {Maan}  \& {Bhusare}}{{Kumar} et~al.}{2024}]{Kumar2024}
{Kumar} A.,  {Maan} Y.,   {Bhusare} Y.,  2024, \mn@doi [arXiv e-prints] {10.48550/arXiv.2406.12804}, \href {https://ui.adsabs.harvard.edu/abs/2024arXiv240612804K} {p. arXiv:2406.12804}

\bibitem[\protect\citeauthoryear{{Lanman} et~al.,}{{Lanman} et~al.}{2022}]{Lanman2022}
{Lanman} A.~E.,  et~al., 2022, \mn@doi [\apj] {10.3847/1538-4357/ac4bc7}, \href {https://ui.adsabs.harvard.edu/abs/2022ApJ...927...59L} {927, 59}

\bibitem[\protect\citeauthoryear{Lehmensiek \& Theron}{Lehmensiek \& Theron}{2012}]{Lehmensiek2012}
Lehmensiek R.,  Theron I.~P.,  2012, in Proceedings of the 2012 International Conference on Electromagnetics in Advanced Applications, ICEAA'12. pp 321--324, \mn@doi{10.1109/ICEAA.2012.6328642}

\bibitem[\protect\citeauthoryear{Lehmensiek \& Theron}{Lehmensiek \& Theron}{2014}]{Lehmensiek2014a}
Lehmensiek R.,  Theron I.~P.,  2014, in The 8th European Conference on Antennas and Propagation (EuCAP 2014). pp 880--884, \mn@doi{10.1109/EuCAP.2014.6901903}

\bibitem[\protect\citeauthoryear{Lewandowski, Kowali{\'{n}}ska  \& Kijak}{Lewandowski et~al.}{2015}]{Lewandowski2015}
Lewandowski W.,  Kowali{\'{n}}ska M.,   Kijak J.,  2015, \mn@doi [\mnras] {10.1093/mnras/stv385}, 449, 1570

\bibitem[\protect\citeauthoryear{{Li} \& {Pan}}{{Li} \& {Pan}}{2016}]{Li2016}
{Li} D.,  {Pan} Z.,  2016, \mn@doi [Radio Science] {10.1002/2015RS005877}, \href {https://ui.adsabs.harvard.edu/abs/2016RaSc...51.1060L} {51, 1060}

\bibitem[\protect\citeauthoryear{{Lyne}, {McLaughlin}, {Keane}, {Kramer}, {Espinoza}, {Stappers}, {Palliyaguru}  \& {Miller}}{{Lyne} et~al.}{2009}]{Lyne2009}
{Lyne} A.~G.,  {McLaughlin} M.~A.,  {Keane} E.~F.,  {Kramer} M.,  {Espinoza} C.~M.,  {Stappers} B.~W.,  {Palliyaguru} N.~T.,   {Miller} J.,  2009, \mn@doi [\mnras] {10.1111/j.1365-2966.2009.15668.x}, \href {https://ui.adsabs.harvard.edu/abs/2009MNRAS.400.1439L} {400, 1439}

\bibitem[\protect\citeauthoryear{{Malenta} et~al.,}{{Malenta} et~al.}{2020}]{Malenta2020}
{Malenta} M.,  et~al., 2020, in {Pizzo} R.,  {Deul} E.~R.,  {Mol} J.~D.,  {de Plaa} J.,   {Verkouter} H.,  eds,  Astronomical Society of the Pacific Conference Series Vol. 527, Astronomical Data Analysis Software and Systems XXIX. p.~457

\bibitem[\protect\citeauthoryear{Manchester, Hobbs, Teoh  \& Hobbs}{Manchester et~al.}{2005}]{Manchester2005}
Manchester R.~N.,  Hobbs G.~B.,  Teoh A.,   Hobbs M.,  2005, \mn@doi [\aj] {10.1086/428488}, 129, 1993

\bibitem[\protect\citeauthoryear{{McLaughlin} et~al.,}{{McLaughlin} et~al.}{2006}]{McLaughlin2006}
{McLaughlin} M.~A.,  et~al., 2006, \mn@doi [\nat] {10.1038/nature04440}, \href {https://ui.adsabs.harvard.edu/abs/2006Natur.439..817M} {439, 817}

\bibitem[\protect\citeauthoryear{{McMullin}, {Waters}, {Schiebel}, {Young}  \& {Golap}}{{McMullin} et~al.}{2007}]{McMullin2007}
{McMullin} J.~P.,  {Waters} B.,  {Schiebel} D.,  {Young} W.,   {Golap} K.,  2007, in {Shaw} R.~A.,  {Hill} F.,   {Bell} D.~J.,  eds,  Astronomical Society of the Pacific Conference Series Vol. 376, Astronomical Data Analysis Software and Systems XVI. p.~127

\bibitem[\protect\citeauthoryear{Men et~al.,}{Men et~al.}{2019}]{Men2019}
Men Y.~P.,  et~al., 2019, \mn@doi [\mnras] {10.1093/mnras/stz1931}, 488, 3957

\bibitem[\protect\citeauthoryear{{Michilli} et~al.,}{{Michilli} et~al.}{2018}]{Michilli2018}
{Michilli} D.,  et~al., 2018, \mn@doi [\mnras] {10.1093/mnras/sty2072}, \href {https://ui.adsabs.harvard.edu/abs/2018MNRAS.480.3457M} {480, 3457}

\bibitem[\protect\citeauthoryear{{Morello}, {Rajwade}  \& {Stappers}}{{Morello} et~al.}{2022}]{Morello2022}
{Morello} V.,  {Rajwade} K.~M.,   {Stappers} B.~W.,  2022, \mn@doi [\mnras] {10.1093/mnras/stab3493}, \href {https://ui.adsabs.harvard.edu/abs/2022MNRAS.510.1393M} {510, 1393}

\bibitem[\protect\citeauthoryear{{Morello}, {Rajwade}  \& {Stappers}}{{Morello} et~al.}{2023}]{Morello2023}
{Morello} V.,  {Rajwade} K.~M.,   {Stappers} B.~W.,  2023, {IQRM: IQRM interference flagging algorithm for radio pulsar and transient searches}, Astrophysics Source Code Library, record ascl:2311.008

\bibitem[\protect\citeauthoryear{{Murphy} et~al.,}{{Murphy} et~al.}{2021}]{Murphy2021}
{Murphy} T.,  et~al., 2021, \mn@doi [\pasa] {10.1017/pasa.2021.44}, \href {https://ui.adsabs.harvard.edu/abs/2021PASA...38...54M} {38, e054}

\bibitem[\protect\citeauthoryear{{Ng} et~al.,}{{Ng} et~al.}{2015}]{Ng2015}
{Ng} C.,  et~al., 2015, \mn@doi [\mnras] {10.1093/mnras/stv753}, \href {https://ui.adsabs.harvard.edu/abs/2015MNRAS.450.2922N} {450, 2922}

\bibitem[\protect\citeauthoryear{{Norris} et~al.,}{{Norris} et~al.}{2021}]{Norris2021}
{Norris} R.~P.,  et~al., 2021, \mn@doi [\pasa] {10.1017/pasa.2021.42}, \href {https://ui.adsabs.harvard.edu/abs/2021PASA...38...46N} {38, e046}

\bibitem[\protect\citeauthoryear{{Offringa} et~al.,}{{Offringa} et~al.}{2014}]{Offringa2014}
{Offringa} A.~R.,  et~al., 2014, \mn@doi [\mnras] {10.1093/mnras/stu1368}, \href {https://ui.adsabs.harvard.edu/abs/2014MNRAS.444..606O} {444, 606}

\bibitem[\protect\citeauthoryear{{Oppermann}, {Yu}  \& {Pen}}{{Oppermann} et~al.}{2018}]{Oppermann2018}
{Oppermann} N.,  {Yu} H.-R.,   {Pen} U.-L.,  2018, \mn@doi [\mnras] {10.1093/mnras/sty004}, \href {https://ui.adsabs.harvard.edu/abs/2018MNRAS.475.5109O} {475, 5109}

\bibitem[\protect\citeauthoryear{Padmanabh et~al.,}{Padmanabh et~al.}{2023}]{Padmanabh2023}
Padmanabh P.~V.,  et~al., 2023, \mn@doi [\mnras] {10.1093/mnras/stad1900}, 524, 1291

\bibitem[\protect\citeauthoryear{{Patel} et~al.,}{{Patel} et~al.}{2018}]{Patel2018}
{Patel} C.,  et~al., 2018, \mn@doi [\apj] {10.3847/1538-4357/aaee65}, \href {https://ui.adsabs.harvard.edu/abs/2018ApJ...869..181P} {869, 181}

\bibitem[\protect\citeauthoryear{{Petroff}, {Hessels}  \& {Lorimer}}{{Petroff} et~al.}{2022}]{Petroff2022}
{Petroff} E.,  {Hessels} J.~W.~T.,   {Lorimer} D.~R.,  2022, \mn@doi [\aapr] {10.1007/s00159-022-00139-w}, \href {https://ui.adsabs.harvard.edu/abs/2022A&ARv..30....2P} {30, 2}

\bibitem[\protect\citeauthoryear{Price, Flynn  \& Deller}{Price et~al.}{2021}]{Price2021b}
Price D.~C.,  Flynn C.,   Deller A.,  2021, \mn@doi [\pasa] {10.1017/pasa.2021.33}, 38, e038

\bibitem[\protect\citeauthoryear{{Rajwade} et~al.,}{{Rajwade} et~al.}{2021}]{Rajwade2021}
{Rajwade} K.,  et~al., 2021, in 43rd COSPAR Scientific Assembly. Held 28 January - 4 February. p.~1194

\bibitem[\protect\citeauthoryear{{Rajwade} et~al.,}{{Rajwade} et~al.}{2022}]{Rajwade2022}
{Rajwade} K.~M.,  et~al., 2022, \mn@doi [\mnras] {10.1093/mnras/stac1450}, \href {https://ui.adsabs.harvard.edu/abs/2022MNRAS.514.1961R} {514, 1961}

\bibitem[\protect\citeauthoryear{{Rajwade} et~al.,}{{Rajwade} et~al.}{2024}]{Rajwade2024}
{Rajwade} K.~M.,  et~al., 2024, \mn@doi [\mnras] {10.1093/mnras/stae1652}, \href {https://ui.adsabs.harvard.edu/abs/2024MNRAS.tmp.1616R} {}

\bibitem[\protect\citeauthoryear{{Rykoff} et~al.,}{{Rykoff} et~al.}{2014}]{Rykoff2014}
{Rykoff} E.~S.,  et~al., 2014, \mn@doi [\apj] {10.1088/0004-637X/785/2/104}, \href {https://ui.adsabs.harvard.edu/abs/2014ApJ...785..104R} {785, 104}

\bibitem[\protect\citeauthoryear{{Sand} et~al.,}{{Sand} et~al.}{2024}]{Sand2024}
{Sand} K.~R.,  et~al., 2024, \mn@doi [arXiv e-prints] {10.48550/arXiv.2408.13215}, \href {https://ui.adsabs.harvard.edu/abs/2024arXiv240813215S} {p. arXiv:2408.13215}

\bibitem[\protect\citeauthoryear{{Sanidas}, {Caleb}, {Driessen}, {Morello}, {Rajwade}  \& {Stappers}}{{Sanidas} et~al.}{2018}]{Sanidas2018}
{Sanidas} S.,  {Caleb} M.,  {Driessen} L.,  {Morello} V.,  {Rajwade} K.,   {Stappers} B.~W.,  2018, in {Weltevrede} P.,  {Perera} B.~B.~P.,  {Preston} L.~L.,   {Sanidas} S.,  eds,  IAU Symposium Vol. 337, Pulsar Astrophysics the Next Fifty Years. pp 406--407, \mn@doi{10.1017/S1743921317009310}

\bibitem[\protect\citeauthoryear{{Scholz} et~al.,}{{Scholz} et~al.}{2016}]{Scholz2016}
{Scholz} P.,  et~al., 2016, \mn@doi [\apj] {10.3847/1538-4357/833/2/177}, \href {https://ui.adsabs.harvard.edu/abs/2016ApJ...833..177S} {833, 177}

\bibitem[\protect\citeauthoryear{{Seymour}, {Michilli}  \& {Pleunis}}{{Seymour} et~al.}{2019}]{Seymour2019}
{Seymour} A.,  {Michilli} D.,   {Pleunis} Z.,  2019, {DM\_phase: Algorithm for correcting dispersion of radio signals}, Astrophysics Source Code Library, record ascl:1910.004

\bibitem[\protect\citeauthoryear{Shapiro-Albert, McLaughlin  \& Keane}{Shapiro-Albert et~al.}{2018}]{Shapiro-Albert2018}
Shapiro-Albert B.~J.,  McLaughlin M.~A.,   Keane E.~F.,  2018, \mn@doi [\apj] {10.3847/1538-4357/aae2b2}, 866

\bibitem[\protect\citeauthoryear{Stappers \& Kramer}{Stappers \& Kramer}{2016}]{Stappers2016b}
Stappers B.~W.,  Kramer M.,  2016, in Proceedings of Science. p.~9, \mn@doi{10.22323/1.277.0009}

\bibitem[\protect\citeauthoryear{{Surnis} et~al.,}{{Surnis} et~al.}{2023}]{Surnis2023}
{Surnis} M.~P.,  et~al., 2023, \mn@doi [\mnras] {10.1093/mnrasl/slad082}, \href {https://ui.adsabs.harvard.edu/abs/2023MNRAS.526L.143S} {526, L143}

\bibitem[\protect\citeauthoryear{{Swiggum} et~al.,}{{Swiggum} et~al.}{2015}]{Swiggum2015}
{Swiggum} J.~K.,  et~al., 2015, \mn@doi [\apj] {10.1088/0004-637X/805/2/156}, \href {https://ui.adsabs.harvard.edu/abs/2015ApJ...805..156S} {805, 156}

\bibitem[\protect\citeauthoryear{{Tan} et~al.,}{{Tan} et~al.}{2018}]{Tan2018}
{Tan} C.~M.,  et~al., 2018, \mn@doi [\apj] {10.3847/1538-4357/aade88}, \href {https://ui.adsabs.harvard.edu/abs/2018ApJ...866...54T} {866, 54}

\bibitem[\protect\citeauthoryear{{Taylor} \& {Jarvis}}{{Taylor} \& {Jarvis}}{2017}]{Taylor2017}
{Taylor} A.~R.,  {Jarvis} M.,  2017, in Materials Science and Engineering Conference Series. IOP, p. 012014, \mn@doi{10.1088/1757-899X/198/1/012014}

\bibitem[\protect\citeauthoryear{{Tian} et~al.,}{{Tian} et~al.}{2024}]{Tian2024}
{Tian} J.,  et~al., 2024, \mn@doi [\mnras] {10.1093/mnras/stae2013}, \href {https://ui.adsabs.harvard.edu/abs/2024MNRAS.tmp.1973T} {}

\bibitem[\protect\citeauthoryear{{Tyul'bashev} et~al.,}{{Tyul'bashev} et~al.}{2018}]{Tyul'bashev2018}
{Tyul'bashev} S.~A.,  et~al., 2018, \mn@doi [Astronomy Reports] {10.1134/S1063772918010079}, \href {https://ui.adsabs.harvard.edu/abs/2018ARep...62...63T} {62, 63}

\bibitem[\protect\citeauthoryear{{Wang} et~al.,}{{Wang} et~al.}{2024}]{Wang2024b}
{Wang} Z.,  et~al., 2024, \mn@doi [arXiv e-prints] {10.48550/arXiv.2409.10316}, \href {https://ui.adsabs.harvard.edu/abs/2024arXiv240910316W} {p. arXiv:2409.10316}

\bibitem[\protect\citeauthoryear{Weltevrede}{Weltevrede}{2016}]{Weltevrede2016}
Weltevrede P.,  2016, \mn@doi [\aap] {10.1051/0004-6361/201527950}, 590, A109

\bibitem[\protect\citeauthoryear{{Wenger} et~al.,}{{Wenger} et~al.}{2000}]{Wenger2000}
{Wenger} M.,  et~al., 2000, \mn@doi [\aaps] {10.1051/aas:2000332}, \href {https://ui.adsabs.harvard.edu/abs/2000A&AS..143....9W} {143, 9}

\bibitem[\protect\citeauthoryear{{Yamasaki} \& {Totani}}{{Yamasaki} \& {Totani}}{2020}]{Yamasaki2020}
{Yamasaki} S.,  {Totani} T.,  2020, \mn@doi [\apj] {10.3847/1538-4357/ab58c4}, \href {https://ui.adsabs.harvard.edu/abs/2020ApJ...888..105Y} {888, 105}

\bibitem[\protect\citeauthoryear{Yao, Manchester  \& Wang}{Yao et~al.}{2016}]{Yao2016}
Yao J.~M.,  Manchester R.~N.,   Wang N.,  2016, \mn@doi [\apj] {10.3847/1538-4357/835/1/29}, 835, 29

\bibitem[\protect\citeauthoryear{Zhang, Harding  \& Muslimov}{Zhang et~al.}{2000}]{Zhang2000}
Zhang B.,  Harding A.~K.,   Muslimov A.~G.,  2000, \mn@doi [\apj] {10.1086/312542}, 531, L135

\bibitem[\protect\citeauthoryear{{Zhou} et~al.,}{{Zhou} et~al.}{2023}]{Zhou2023}
{Zhou} D.~J.,  et~al., 2023, \mn@doi [Research in Astronomy and Astrophysics] {10.1088/1674-4527/accc76}, \href {https://ui.adsabs.harvard.edu/abs/2023RAA....23j4001Z} {23, 104001}

\bibitem[\protect\citeauthoryear{van Straten \& Bailes}{van Straten \& Bailes}{2011}]{VanStraten2011}
van Straten W.,  Bailes M.,  2011, \mn@doi [\pasa] {10.1071/AS10021}, 28, 1

\makeatother
\end{thebibliography}



\appendix
\section{Images and DM fit statistics}
\begin{figure*}
    \centering
    \includegraphics[width=0.68\linewidth]{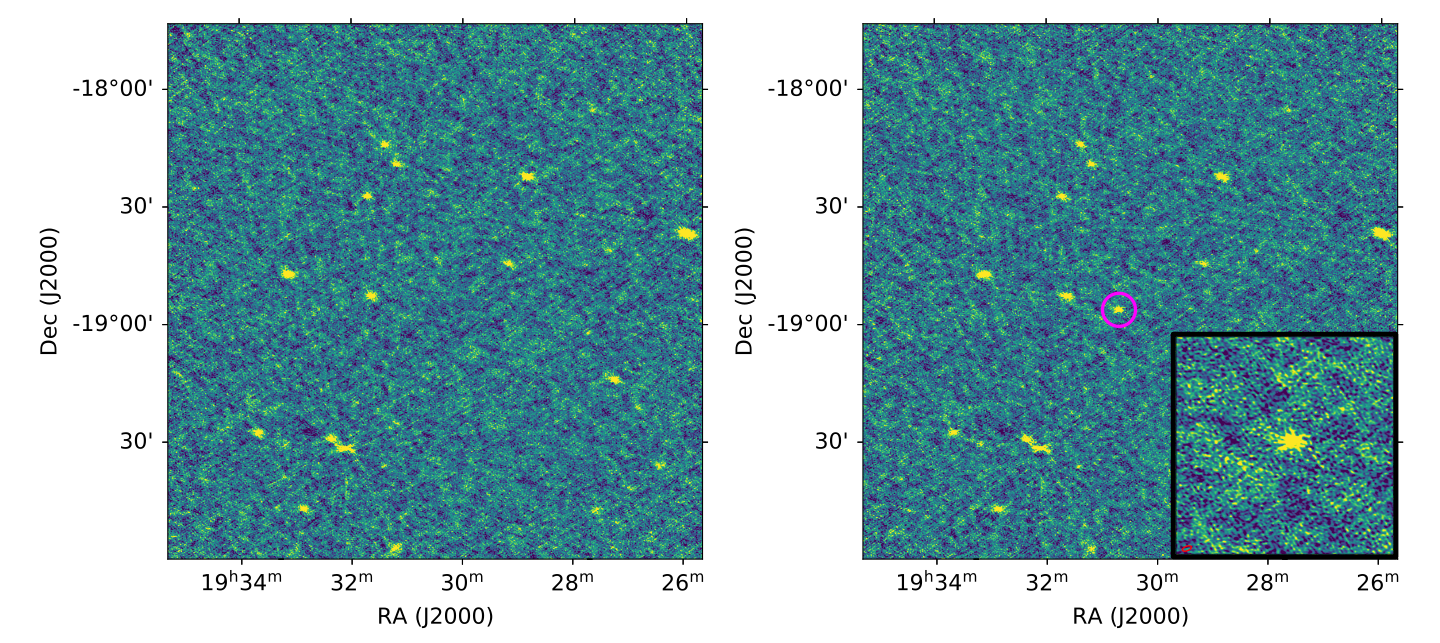}
    \includegraphics[width=0.68\linewidth]{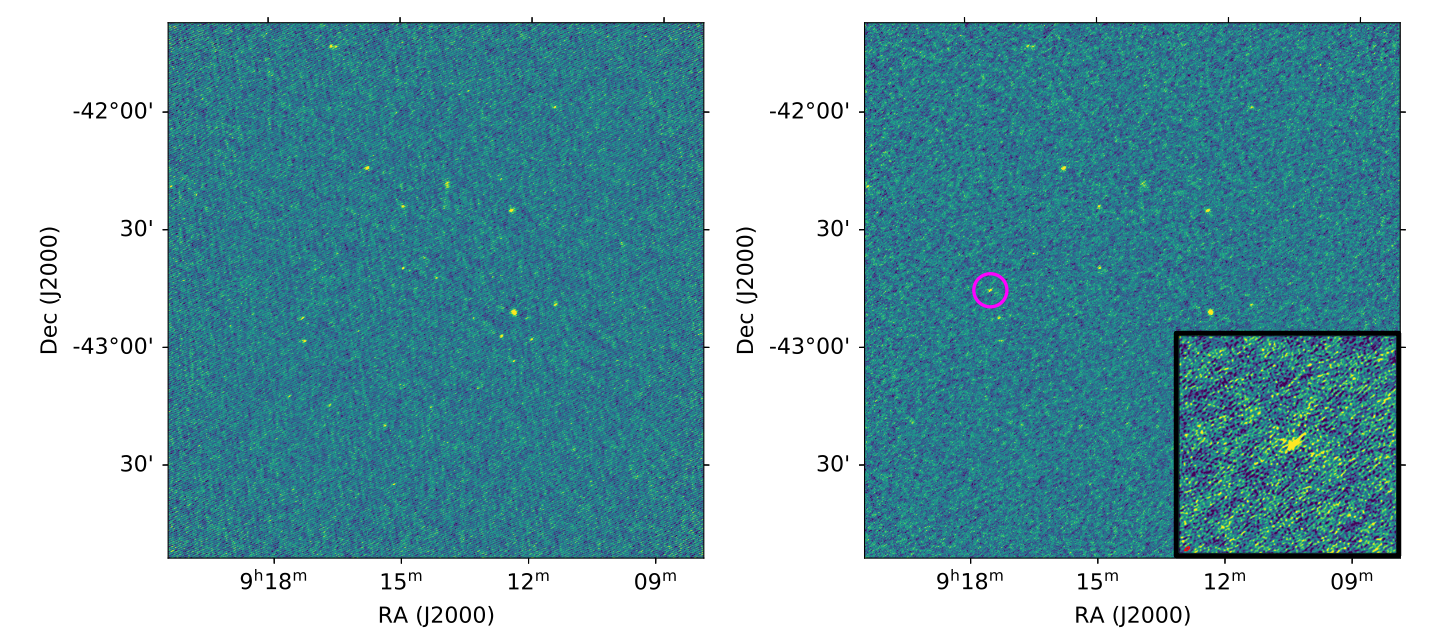}
    \includegraphics[width=0.68\linewidth]{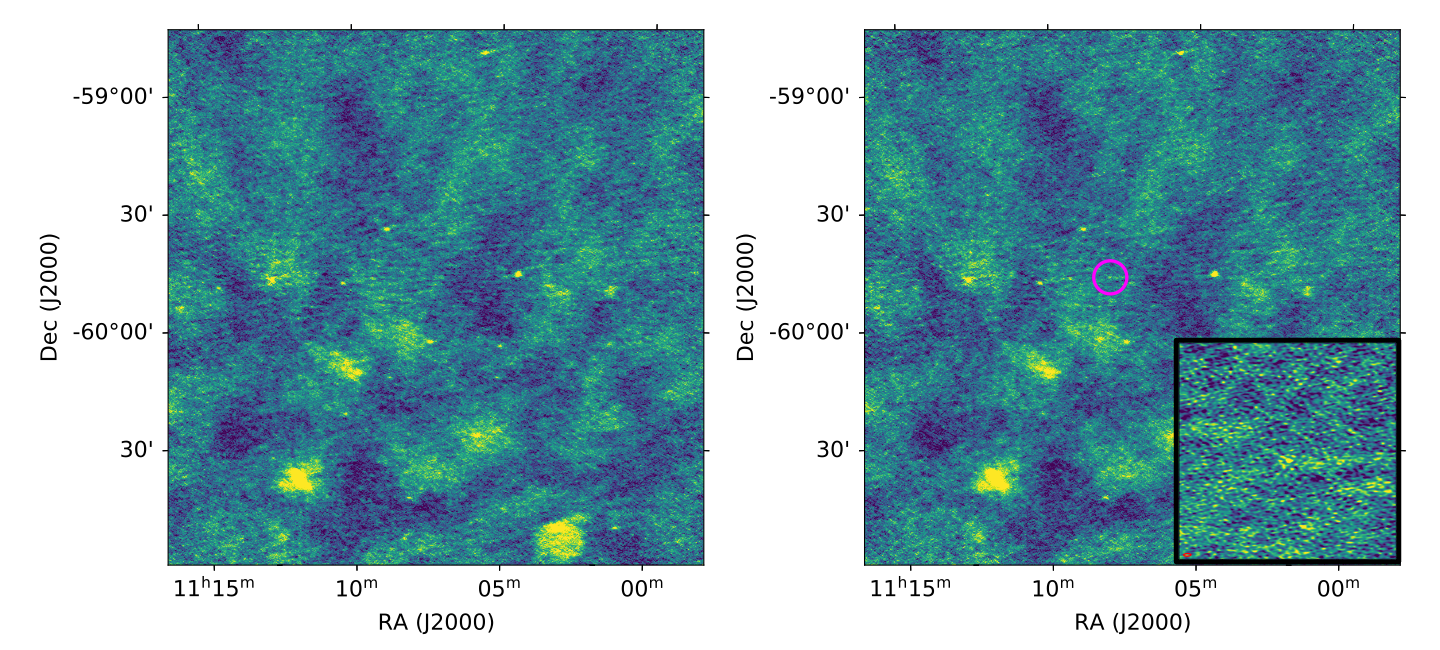}
    \includegraphics[width=0.68\linewidth]{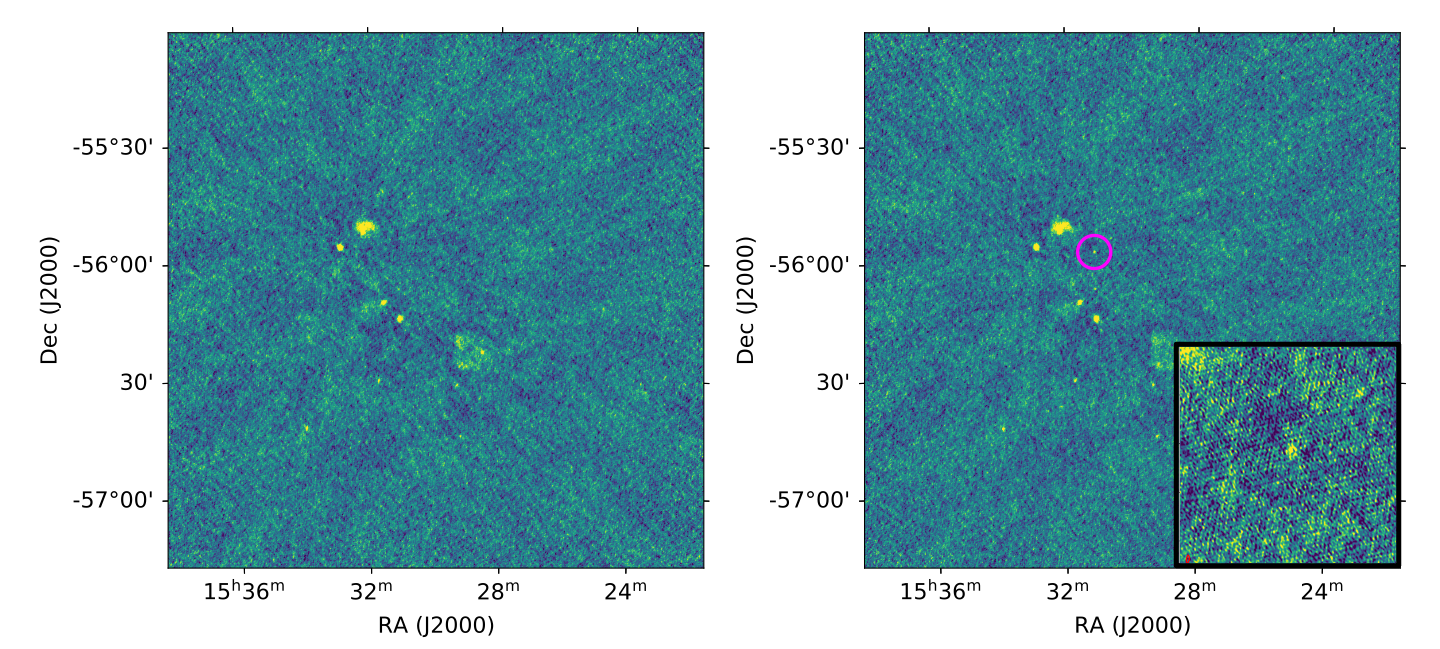}
    \caption{From top down: Images of the positions of MTP0020, MTP0031, MTP0034 and MTP0045, integrated over the duration of the pulse (right) and before the pulse detection (left). The magenta circle marks the transient source identified at the time of the pulse detection. The inset at the bottom right corner of the on-source image is a zoomed in view centred on the source. The red ellipse at the bottom left corner of the inset indicates the synthesised beam.}
    \label{fig:on-off}
\end{figure*}
\begin{table*}
\centering
\caption{Fit statistics from \scatfit{} for the sources we fit. The S/N is for the pulse dedispersed at the tabulated DM, and determined by \spyden.}
\label{tab:fitstats}
\begin{tabular}{lrrrllcr}
\hline
Source & DM (pc cm$^{-3}$) & Uncertainty & S/N & Pulse epoch & & Reduced-$\chi^{2}$ & BIC \\
\hline
MTP0014 & 71.299 & 0.031 & 53.8 & 2023 May 08 & UTC 23:32:16 & 1.3 & 4.1 \\
MTP0016 & 41.162 & 0.349 & 21.6 & 2020 Dec 25 & UTC 01:59:07 & 1.2 & 2.8 \\
MTP0017 & 31.517 & 0.210 & 18.8 & 2022 Jun 17 & UTC 08:42:19 & 1.2 & 2.9 \\
MTP0018 & 38.837 & 0.307 & 9.7 & 2021 May 10 & UTC 16:11:46 & 1.7 & 2.0 \\
MTP0020 & 63.143 & 0.009 & 74.9 & 2021 Jun 22 & UTC 03:41:45 & 2.1 & 7.9 \\
MTP0024 & 41.007 & 0.501 & 13.3 & 2021 Jun 07 & UTC 01:46:13 & 0.5 & $-$3.3 \\
MTP0026 & 206.828 & 0.253 & 10.0 & 2021 Jun 25 & UTC 23:05:53 & 0.5 & $-$3.0 \\
MTP0028 & 440.298 & 0.239 & 23.1 & 2021 Aug 10 & UTC 20:33:27 & 0.8 & $-$0.4 \\
MTP0029 & 201.422 & 0.222 & 17.3 & 2021 Aug 20 & UTC 21:46:32 & 0.6 & $-$2.1 \\
MTP0032 & 271.516 & 0.454 & 11.0 & 2021 Oct 12 & UTC 15:32:17 & 1.7 & 0.6 \\
MTP0034 & 92.736 & 0.447 & 27.8 & 2023 Feb 26 & UTC 04:38:30 & 0.9 & 0.6 \\
MTP0035 & 224.463 & 0.220 & 11.1 & 2021 Oct 26 & UTC 13:02:04 & 2.8 & 4.0 \\
MTP0036 & 128.894 & 0.716 & 9.4 & 2021 Oct 30 & UTC 01:20:14 & 0.4 & $-$3.8 \\
MTP0038 & 126.711 & 0.389 & 16.1 & 2021 Dec 09 & UTC 06:12:23 & 4.0 & 11.2 \\
MTP0039 & 95.313 & 0.091 & 36.5 & 2021 Dec 20 & UTC 10:40:29 & 0.2 & $-$11.6 \\
MTP0040 & 264.142 & 1.091 & 9.4 & 2021 Dec 16 & UTC 09:20:27 & 2.0 & 1.0 \\
MTP0042 & 250.353 & 0.656 & 12.9 & 2021 Oct 08 & UTC 18:56:32 & 0.5 & $-$2.4 \\
MTP0044 & 55.795 & 0.389 & 16.5 & 2022 Jan 02 & UTC 13:04:19 & 1.3 & 3.3 \\
MTP0045 & 56.602 & 0.563 & 13.8 & 2021 Dec 20 & UTC 03:55:39 & 1.8 & 5.6 \\
MTP0046 & 254.170 & 1.497 & 8.8 & 2022 Jan 15 & UTC 10:20:15 & 2.0 & 1.0 \\
MTP0047 & 152.396 & 0.357 & 10.7 & 2022 Feb 19 & UTC 10:13:36 & 2.0 & 0.1 \\
MTP0048 & 151.559 & 0.457 & 9.4 & 2022 Feb 03 & UTC 06:01:15 & 0.9 & 0.4 \\
MTP0049 & 346.506 & 0.678 & 9.8 & 2022 Feb 06 & UTC 06:06:28 & 0.9 & $-$1.3 \\
\hline
\end{tabular}
\end{table*}


\bsp
\label{lastpage}
\end{document}